\documentclass[a4paper, 11pt, twoside]{book}
\usepackage[latin1]{inputenc}
\usepackage{epsfig}
\usepackage{cite}
\usepackage{hyperref}
\usepackage{color}
\usepackage{amsmath}
\usepackage{graphicx}
\usepackage{appendix}
\usepackage{textcomp}
\usepackage{subfig}
\usepackage{float}
\usepackage{pdfpages}
\usepackage[absolute]{textpos}
\usepackage[top=3cm, bottom=3cm,inner=3.5cm,outer=2.5cm]{geometry}

\begin{document}
\pagestyle{empty}
\includepdf{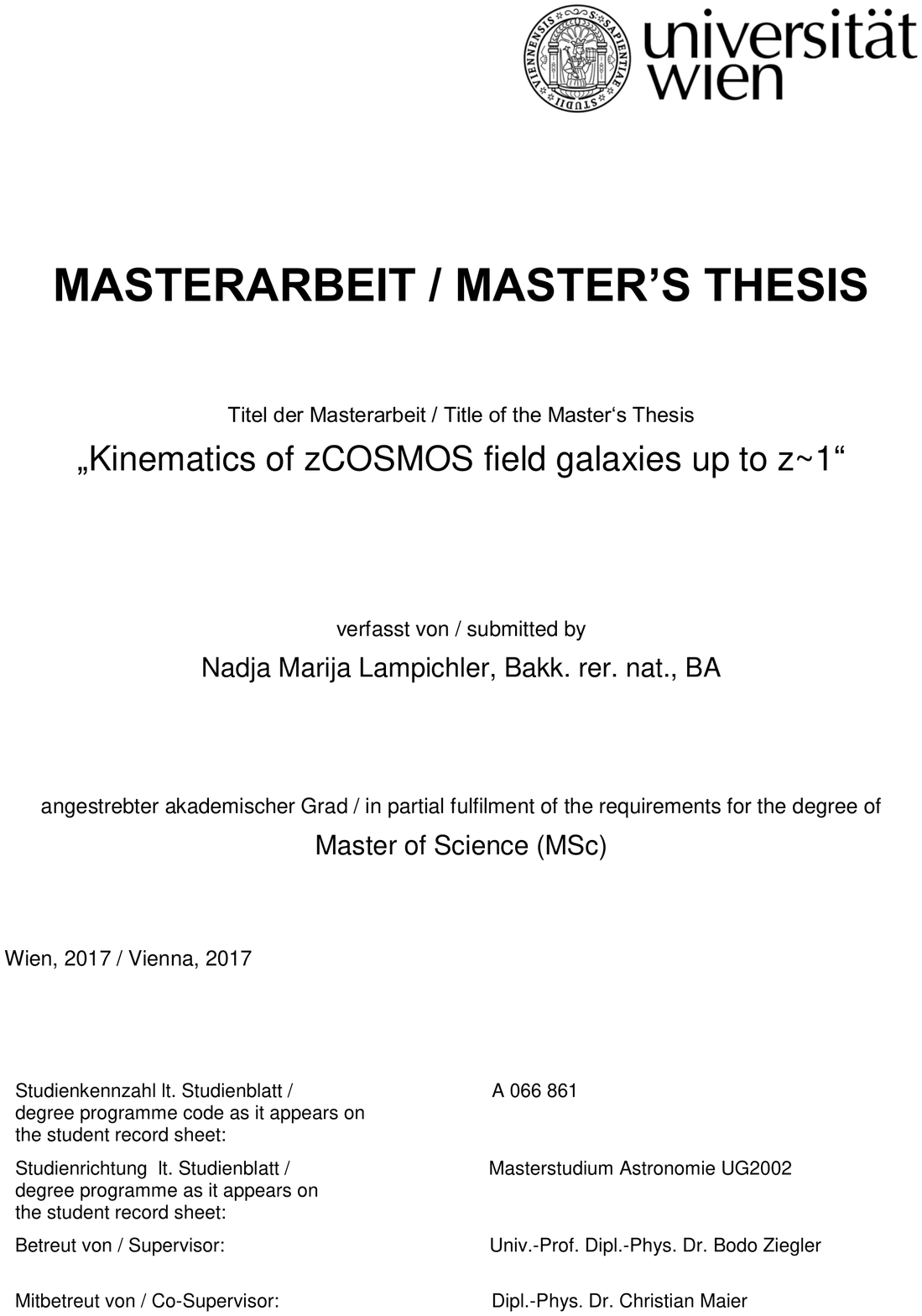}

\frontmatter
\pagestyle{headings}

\tableofcontents

\backmatter
\chapter{Acknowledgments}
First I would like my working group. Thank you for the opportunity to be part of this great group and work in the interesting field of extragalactic astronomy.\\
My thanks go especially to my supervisor Prof. Bodo Ziegler. Thank you for all your help, your suggestions, your expertise on the topic of the TFR and the VSR, and your patience. My special thanks go also to Dr. Christian Maier who co-supervised this master thesis. Thank you for all the time you invested in this thesis, for all your encouragements, advice and explanations which helped me a lot.\\
I would also like to thank the other members of my working group: Dr. Miguel Verdugo - thank you for your help regarding the python program and your improvements of the code; Jose Manuel Perez Martinez - thank you for sharing with me your findings and opinion regarding the extraction and simulation of the rotation curves, the Tully-Fisher diagram and velocity-size diagram, and for your advice; Lucas Ellmeier - thank you for helping me installing Ubuntu on my laptop.\\
A special thanks goes to Dr. Asmus B{\"o}hm. Thank you for your help, your explanations and your expertise on the topic of rotation curves and scaling relations, for answering my numerous questions and for your data of the FDF galaxies. I would also like to thank Dr. Benjamin B{\"o}sch who wrote the python-program I used for the rotation curve modelling.\\
In addition, I would like to thank my family: my parents Andrej and Heidi, my sister Mirjam and my brother Simon, and my grandfather Bartolomej. Hvala vam vsem za vse, kar ste zame naredili in za vso podporo v zadnjih letih! Brez vas to vse ne bi bilo mo\v{z}no. Predvsem vama hvala, mami in hati, da sta mi to vse omogo\v{c}ila. Mimi, hvala za vse pogovore in dobre besede, in za skupno pretrpene krize. In hvala dedi za vso podporo in za tedenske klice!\\
Finally, I would also like to thank my friends. My thanks go especially to Christoph: thank you for your support, your encouragement, your kind words and for listening. And thanks to Christoph K. (thank you for your corrections and comments!), Conny, Eva and Patrick, whom I met at the university and who all became very dear friends. Thank you for the past years we spent together, for all the experiences we shared in this time and for all your help!

\mainmatter

\chapter{Introduction}
\label{Intro}
This master thesis focuses on the kinematics and the evolution of spiral galaxies up to redshifts of $z\sim 1$. For this purpose two important scaling relations for disk galaxies, namely the Tully-Fisher relation and the velocity-size relation, are constructed for a sample of zCOSMOS galaxies and compared to the equivalent relations for local galaxies. Furthermore, the relation between stellar mass and halo mass is investigated.\\
Scaling relations are a very powerful tool to analyse the evolution of galaxies. Especially the Tully-Fisher relation has been the subject of many studies over the past couple of years. The results of these studies were often discrepant and contradictory. Where some publications conclude that there is a significant evolution with time of the Tully-Fisher relation, others did not find any evolution at all. The aim of this thesis is to analyse by means of the zCOSMOS sample, whether an evolution of the B-band and the stellar Tully-Fisher relation as well as the velocity-size relation with redshift can be found, and compare the results with the results of other works. Beside this, the stellar-to-halo mass ratio for the zCOSMOS sample is studied and compared to predictions from simulations.\\
This chapter gives at first a short introduction about the field of extragalactic astronomy and then focuses on spiral galaxies. They are the main objects of investigation in this thesis and it is thus important to have an overview of their main characteristics. Some important parameters of spiral galaxies like the mass, size and the large scale structures are described in more detail. The last part of this chapter deals with the rotation curves of spiral galaxies and their important role concerning the "discovery" of dark matter.\\
As scaling relations and their evolution with time are the main subject of this thesis, chapter \ref{scalrel} treats this topic in more detail. First a short overview is given on scaling relations in general and their significance for extragalactic astronomy and cosmology. Then some results from recent studies regarding this topic are presented. Chapter \ref{data} presents the data used in this thesis and describes VIMOS, the spectrograph that was used to record the spectra, and the zCOSMOS-survey. In chapter \ref{datared} the data reduction program VIPGI is presented and the individual data reduction steps are described in detail. The rotation curve extraction from emission lines is described in chapter \ref{extractRC}. Chapter \ref{simulations} deals with the modelling of these observed rotation curves in order to get values of the maximum rotation velocity $v_{max}$. The results are presented in chapter \ref{results}, and summarised and discussed in chapter \ref{discussion}.

\newpage
\section{The beginnings of extragalactic astronomy}
Astronomy is the most ancient natural science. It dates back thousands of years, as far as the Stone Age. At all times people were fascinated by the vastness and beauty of the sky and the celestial bodies, trying to unravel the mysteries connected to them. For peoples on all continents astronomy was inseparably connected to their religious, mythological and astrological beliefs. It played a major role in their lives. Many cultures were even able to use their findings about the movement of the celestial bodies to create calendars that, for these times, were unbelievably precise. Already in the antiquity there were incredibly many things known about astronomy and progress was made very fast.\\
But although people have engaged in astronomy for such a long time, the branch of this science concerned with objects that lie outside our galaxy, the extragalactic astronomy, is very young.\\
Already in the 10th century the Persian astronomer Al Sufi stated that Andromeda was different from the other objects in the sky, because of its blurred form~\cite[1]{Fridman}. But it took many centuries until it was known that Andromeda was a galaxy, considerably further away than the stars seen on the sky. A first step toward the understanding what a galaxy is and that we may as well be part of one, was made by the English Astronomer Thomas Wright. He assumed that the stars of the Milky Way he was observing, are distributed in a thick disk. The German Philosopher Immanuel Kant liked this concept and developed it further. He stated that the Milky Way was "only one of the many 'Island Universes', scattered in the infinite Universe". Another important point was the compilation of astronomical catalogues of celestial objects. They often included also galaxies, although these were not identified as such at the time of the publications of the calatogues. In this context, the Messier catalogue by Charles Messier from 1784, and the General Catalogue of Nebulae and Clusters of Stars by John Herschel from 1864 are of special significance. In the Messier catalogue, one 39 out of 108 objects were galaxies, as was discovered many years later.\\
One could say that the actual beginning of modern extragalactic astronomy as we know it was in the 1920s. In these years the extragalactic nature of Andromeda and other spiral nebulae was confirmed which was a huge scientific breakthrough that changed the way people looked at the universe. At the same time also our own galaxy was the subject of many investigations: a concept of its structure was developed, its centre was determined, there were the first estimations of its dimensions and more. An important point is also that the rotation of the Milky Way was confirmed. All these discoveries led to the - later confirmed - assumption that the extragalactic spiral objects are actually similar to our own galaxy, with similar properties like structure and rotation. So a good approach to understand galaxies properly, is to complement the "detailed local information that we can deduce about our own Galaxy [...] by the less detailed but global picture that we have of other galaxies"~\cite[17]{Binney}.\\
Soon after the discovery of the real nature of the spiral nebulae it was realised that the number of Milky Way-like objects was amazingly high. Already in the first years after the "birth" of extragalactic astronomy it was estimated that the number of galaxies that could actually be observed, was about 120.000~\cite[2]{Fridman}. Today it is assumed that there are about $\sim 10^{12}$ galaxies in the whole universe. Although extragalactic astronomy is only roughly one hundred years old, many achievements and discoveries have already been made in this field, especially after the 1940s. In these years huge progress was made in science and technology and this led to a rapid development of astronomy. Many new methods of astronomical research were developed that resulted in new discoveries and findings. E.g. the radio band became available and soon the whole wavelength range could be covered by astronomical observations.\\
Today our knowledge of the characteristics, formation and evolution of galaxies and the universe is enormous compared to what was known a hundred years ago. However, there are still a lot of things that we do not know and understand yet and that are waiting to be discovered.

\section{Spiral galaxies}
The main object of investigation in this thesis are spiral galaxies. Spiral galaxies are stellar systems, i.e. gravitationally bound assemblages of stars. In the universe there are many different kinds of stellar systems, varying over "some fourteen orders of magnitude in size and mass, from binary stars, to star clusters containing $10^{2}$ to $10^{6}$ stars, through galaxies containing $10^{10}$ to $10^{12}$ stars, to vast clusters containing thousands of galaxies"~\cite[3]{Binney}.\\
Galaxies can vary in sizes, shapes and masses. On the basis of those properties they can be divided into four main types: elliptical, lenticular, spiral and irregular galaxies. In this section some of the most important properties of spiral galaxies are explained.

\subsection{Size, mass and different types of spiral galaxies}
\begin{figure}[!h]
  \centering
  \includegraphics[height=4.5 cm]{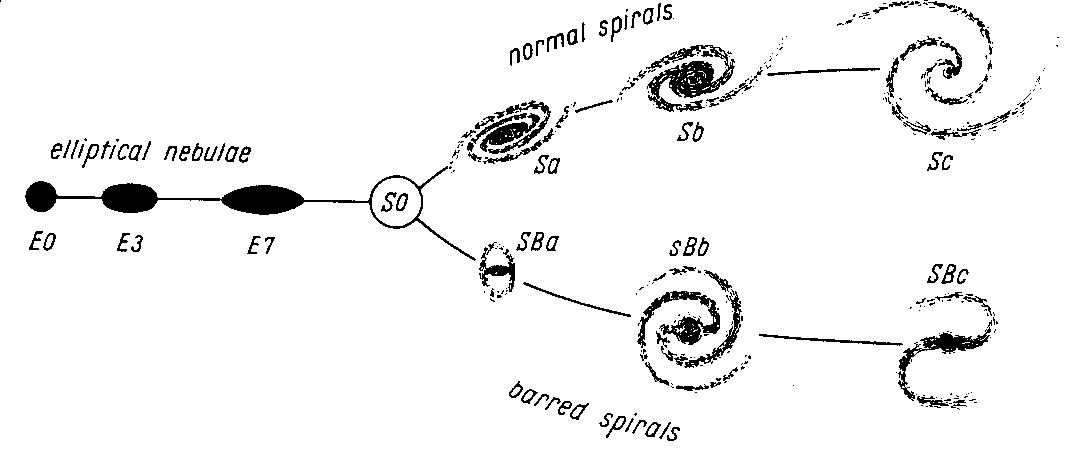}
   \caption{\small The Hubble classification system for galaxy morphologies ~\cite[45]{Hubble}.}
   \label{fig:hubble}
\end{figure}

Values for the size of spiral galaxies usually range between 1 and 50 kpc and for their masses between $10^{10}$ to $10^{12}$ solar masses ($M_{\odot}$)~\cite[29]{Bertin}. The velocities of the stars in those galaxies are typically 100-400 km/s. As an example, with about $10^{11}$ stars, $\sim10^{12} M_{\odot}$ of gas, a radius of roughly 10 kpc and a typical circular velocity of $\sim 200$ km/s, the Milky Way populates the area of the medium-sized galaxies~\cite[3-4]{Binney}.\\
Spiral galaxies can be found in all environments, as field galaxies or members of groups and clusters. The interesting thing here is that almost $80\%$ of all bright galaxies in the field are spiral galaxies, whereas in high-density regions the fraction is only about $10\%$~\cite[23]{Binney}.\\
Regarding the shape, one could say that spiral galaxies consist of two main components: a prominent bright and rather thin rotating disk, the most characteristic feature of this galaxy type, that is composed of Population I stars, gas and dust, and a nuclear spheroid of Population II stars, called the bulge~\cite[22-23]{Binney}. The luminosity of the bulge compared to the luminosity of the disk correlates very well with other properties of spiral galaxies and is one of the main criteria of Hubble's classification of spiral galaxies.\\
The so-called Hubble tuning fork diagram is a morphological classification system for normal galaxies developed by Edwin Hubble after the discovery of the nature of galaxies~\cite[21-28]{Bertin}. In this case the term normal refers to non-peculiar objects. In spite of the fact that this classification system is already about 90 years old and that huge progress has been made in extragalactic astronomy since its invention, it is still the main tool used for the classification of galaxy morphologies. It is based on the observed shape of the galaxies and can be divided into three main linear sequences: the elliptical galaxies, the normal and the barred spiral galaxies. The original tuning fork diagram made by Hubble in 1936 can be seen in figure \ref{fig:hubble}~\cite[45]{Hubble}.

\noindent
The elliptical galaxies are not important in this work, it should just be mentioned that they are arranged in the order of increasing flattening. Also the lenticular galaxies (S0, located between the elliptical and the spiral galaxies) and the irregular galaxies will not be described any further.
\newpage
\noindent
The spiral galaxies can first of all be divided into galaxies with or without a bar. In addition, on both linear spiral sequences there is a subdivision into a, b and c types. The most important criteria for the distinction between those types are~\cite[26-28]{Bertin}~\cite[5]{Fridman}: 
\begin{enumerate}
\item the already mentioned ratio between the bulge and the disk luminosity or rather the size of the bulge compared to the size of the disk (decreasing from a to c)
\item the gas content of the disk (increasing from a to c) which is also the primary parameter in this classification scheme
\item the galaxy's total mass (decreasing from a to c)
\item the nature of the spiral structure $\longrightarrow$ the spiral arms are tighter in a-type galaxies, whereas they are clumpier in c-type galaxies, which means that features like HII-regions tend to be more visible
\item the changing colour indices (c-type galaxies are bluer)
\item the dust content (increasing from a to c)
\item the maximum rotation velocity (decreasing from a to c)
\end{enumerate}

\subsection{Large scale structures in spiral galaxies}
In the previous section spiral galaxies have been described as consisting of two components, the disk and the bulge. But in fact there are more components that compile a spiral galaxy. Those large scale structures will be described briefly in the following.\\

\subsubsection{The galactic disk}
The disk is the most characteristic feature of spiral galaxies. It consists of a massive stellar disk (mainly Population I stars) and a much less massive gas disk. The main movement of both components of the disk structure is the rotation. This means that the velocities of the chaotic motion of the stars and gas clouds are significantly lower than the rotation velocities and can therefore usually be neglected~\cite[13]{Fridman}. It should also be mentioned that there is a difference between the rotation velocity of stars and gas, although it is usually rather small.\\
An important characteristic of the galactic disk is its radial surface brightness distribution which can be described very well by an exponential power law~\cite[31]{Bertin}:

\begin{equation}
I(r)\sim I_{0}e^{\dfrac{-r}{r_{d}}}
\end{equation}

\noindent
Here $I_{0}$ is the central surface brightness of the galaxy, \textit{r} is the radius and $r_{d}$ is the scale length. Interestingly, $I_{0}$ is often assumed to be more or less constant for spiral galaxies, with $I_{0}\simeq 140 L_{\odot}pc^{-2}$~\cite[23]{Binney}.\\
Also the vertical structure of the stellar disk is often parameterized by an exponential power law with the scale height $h_{z}$~\cite[11]{Fridman}:

\begin{equation}
I(z)\sim I_{0}e^{\dfrac{-\mid z\mid}{h_{z}}}
\end{equation}

\subsubsection{The bulge}
Another important feature of spiral galaxies is the spherical or flattened concentration of stars in the centre, called the bulge~\cite[37-39]{Fridman}. In contrast to the disk, the bulge mainly consists of old Population II stars. Its size compared to the size of the disk varies with the Hubble-type, being more prominent in early-type spiral galaxies. The dimensions and total mass of the bulge are generally very small and can be neglected in most considerations when compared to the radius and mass of the disk.\\
Bulges are sometimes described as a form of small ellipticals in the centre of spiral galaxies since their form and surface brightness distribution are similar to their larger counterparts. In contrast to the galactic disks, the surface brightness distribution of bulges is not described by an exponential power law, but by the so-called de Vaucouleurs law, also known as "the $r^{1/4}$ law":

\begin{equation}
log\left( \frac{I(r)}{I_{e}}\right) =-\beta\cdot \left[ \left( \dfrac{r}{r_{e}}\right) ^{1/4}-1\right] 
\end{equation}

\noindent
$r_{e}$ is the effective radius, defined as the radius whithin which half of the total light is emitted. $I_{e}$ is the effective surface brightness, thus the brightness within $r_{e}$. $\beta$ has the value 3.331.\\
This law is a special case of the Sersic equation, an empirical, but well probed law that describes how the intensity \textit{I(r)} of a galaxy varies with the radius:

\begin{equation}
I(r)=I_{e}\cdot exp\lbrace -\nu_{n}[(\dfrac{r}{r_{e}})^{1/n}-1]\rbrace
\end{equation}

\noindent
The advantage of this equation is that by means of it the surface brightness distribution of elliptical galaxies and bulges as well as of galactic disks can be described. If $n=4$ and $\nu_{n}=7.669$, the Sersic law transforms into the de Vaucouleurs law, whereas $n=1$ yields an exponential distribution describing the disks of spiral galaxies. The $r^{1/4}$ law is naturally relevant only in the centre of the spiral galaxy. The majority of the galaxy's radial luminosity profile is characterised by the exponential distribution as can be seen in figure \ref{fig:br_dist} from~\cite{Freeman70}:
\newpage

\begin{figure}[!h]
  \centering
  \includegraphics[height=7 cm]{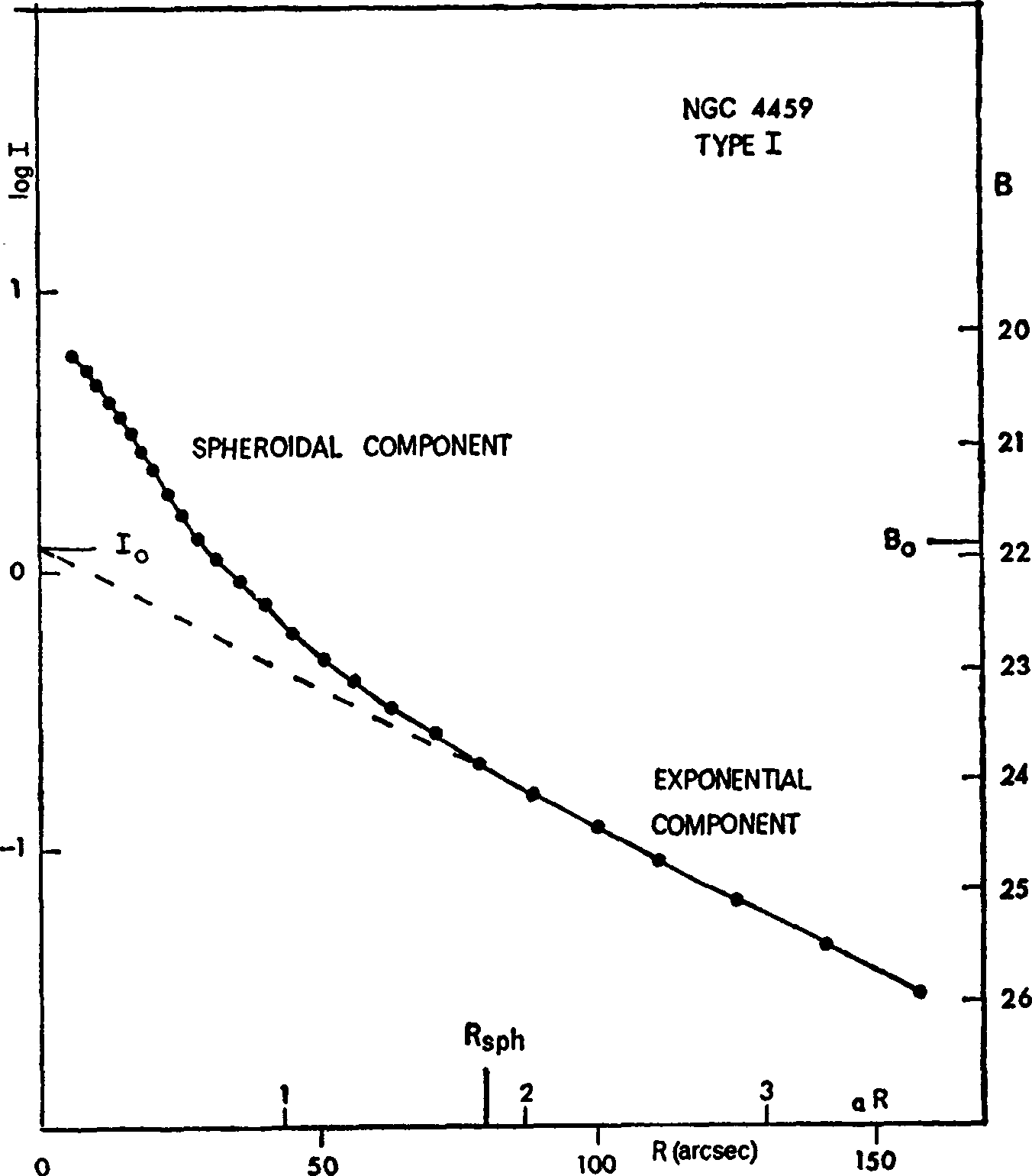}
   \caption{\small Radial brightness distribution of the spiral galaxy NGC 4459~\cite{Freeman70}.}
   \label{fig:br_dist}
\end{figure}

\noindent
An interesting fact that should be mentioned is the relation between the effective radius of the bulge $r_{e}$ and the scale length of the disk $r_{d}$ of a galaxy. This can be seen very well in figure \ref{fig:bulge_effr} from~\cite{Aguerri05}:

\begin{figure}[!h]
  \centering
  \includegraphics[height=5 cm]{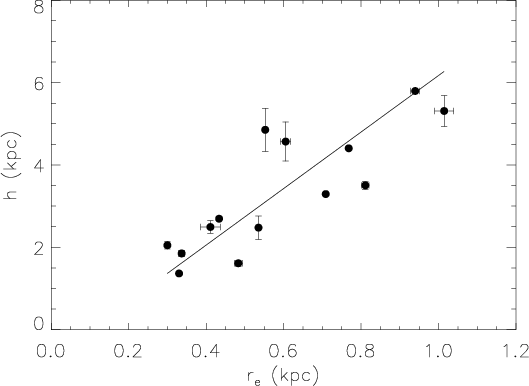}
   \caption{\small Relation between the bulge effective radius and the disk scale length of spiral galaxies~\cite{Aguerri05}.}
   \label{fig:bulge_effr}
\end{figure}

\subsubsection{The galactic bar}
As already mentioned, spiral galaxies can be divided into galaxies with or without a bar. Galaxies with a central bar, like the Milky Way, are called barred spirals (SB). The galactic bar is an oval-shaped elongated bright region in the centre of the galaxy with a length of typically 1-5 kpc~\cite[39-42]{Fridman}. It contains stars, but also large amounts of gas and dust, especially in late-type spiral galaxies. The most important property of galactic bars is their stellar solid-state rotation with the angular velocity $\Omega_{b}$. The stars move along the bar in complicated elongated orbits. The major axes of these stars' orbits form at the same time the major axis of the bar.\\
It is interesting that only thirty to forty years ago it was believed that galactic bars are a rather rare phenomenon. The fraction of barred spirals was estimated to be only about $10\%$. Today it is assumed that at least $50-70\%$ of all spiral galaxies are SB-galaxies. An interesting fact is also that the existence of a bar apparently does not depend on the Hubble type.\\
Galactic bars have several characteristic features. Two important ones should be presented here briefly: Firstly, the formation of radial flows of gas along the bar to the galactic centre was observed. These radial flows supply the central region with gas which leads to enhanced star formation. The second characteristic that was observed are long thin veins. Those are narrow regions where the insterstellar medium is strongly compressed. This means that also the concentration of dust is higher there which furthermore leads to an increased opacity of the medium, seen as the thin veins. Beside these two features, several structures can be observed in different bars, e.g. inner polar or nuclear rings, inner bars and so on.

\subsubsection{The halo}
The characteristic of spiral galaxies is, as explained above, their flat rotation disk that is very thin compared to the diameter of the galaxy. But in fact, there are individual stars and other objects like globular clusters, bound gravitationally to the galaxy, which do not lie in the galactic plane, but can straggle rather far from it~\cite[33]{Bertin}. The region they are located in, is called the halo. The stars in the halo are often called "high-velocity" stars. Because in contrast to the stars in the galactic disk, the velocity dispersion of the halo stars is comparable with their circular velocity~\cite[41]{Fridman}. For simplicity, the shape of the halo is usually described as spherical, although in reality it is more diffuse~\cite[33]{Bertin}. Still the spherical description is a very good approximation.\\
The number of stars in the halo is considerably lower than the number of stars in the disk. Also halo stars have typically lower luminosities. Therefore halos from distant galaxies cannot be observed as the light from the disk is too dominant~\cite[41-44]{Fridman}. That is one reason why the halos are called "dark halos". The other reason is the assumption that despite of their relatively low stellar mass the halos are in fact quite massive, having a mass between half to three times the mass of the disk. 
There are several indications for this mass, however since it has not been observed directly yet, the hypothesis of dark matter was introduced. The most important evidence for this kind of matter comes from the rotation curves (RCs) of spiral galaxies which will be described in more detail in Section \ref{RCs}. There are also several other features that support the presence of dark matter halos. To give one more example, the existence of polar rings perpendicular to the galactic disk in some galaxies is an indication for a dark matter halo and can even help to determine its form.\\
In figure \ref{fig:gal_sketch} a sketch of a spiral galaxy lying in a halo can be seen~\cite[42]{Fridman}:

\begin{figure}[!h]
  \centering
  \includegraphics[height=3 cm]{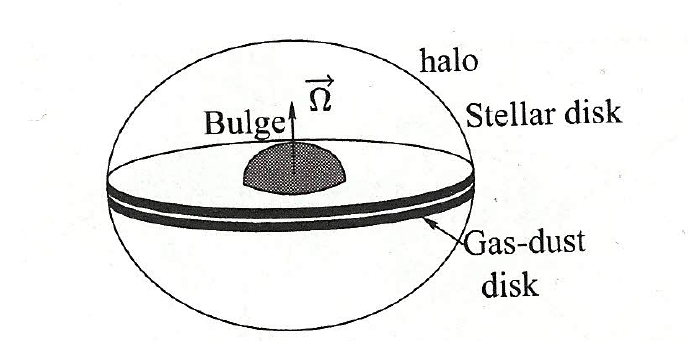}
   \caption{\small A sketch of the main structural elements of a spiral galaxy~\cite[42]{Fridman}.}
   \label{fig:gal_sketch}
\end{figure}

\subsubsection{The spiral structure}
Spiral galaxies get their name from their spiral arms, which are clearly the most noticeable feature of this galaxy type. As has already been mentioned, the nature of the spiral structure depends on the Hubble type: early-type spirals tend to have tighter spiral arms than late-type spirals, while spiral arms of the latter are clumpier. Also the length, the prominence or the geometry - e.g. the number - of spiral arms can vary strongly from galaxy to galaxy. What is important is that they are almost always existent~\cite[22]{Binney}. But what exactly are spirals arms? They are bright filaments composed of stars, gas and dust in the galaxy's disk. Numerous stars are forming in these regions and the high amount of young massive stars, especially O and B stars, is the reason for their brightness.\\
The Irish astronomer Lord Rosse was the first to describe the spiral structure of several nebulae~\cite[400]{Fridman}. In 1845, he built a 183 cm reflector making it the largest telescope in the world for many decades. This allowed him to discover spiral arms in individual galaxies. Of course at this time it was still believed that these nebuale belonged to our stellar system.\\
Nowadays we know quite a lot about spiral structure. Usually a distinction is made between two main types of spiral arms~\cite[31-32]{Fridman}: 

\begin{enumerate}
\item symmetrical and long arms that form a global spiral pattern (often called grand design spirals), and
\item fragmentary spiral patterns that is irregular, short pieces of arms (flocculent spirals).
\end {enumerate}

\noindent
Observations imply that the number of the second type is significantly larger than the number of galaxies with a symmetric and regular structure. However it should be mentioned that there is no clear distinction between the two types, as there is a continuous transition from grand design to flocculent spirals even to irregular galaxies.\\
The two types of spiral galaxies described above do not only have different appearances regarding their spiral structure, also the physical mechanisms of formation of the latter are believed to be different. In the case of the global spiral patterns, a density wave propagating through the whole galactic disk is believed to be the trigger. The fragmentary spiral patterns however are probably the result of differential galactic rotation that leads to the breaking of spiral arms and moreover to the stretching of areas of star formation. But also tidal interactions can in principle result in the formation of spiral arms. This means that there are several theories explaing mechanisms for the formation of spiral structures, although at the moment the density wave theory has the most support.\\
The question of the origin of the spiral arms was occupying scientists already in the late 1920ies, soon after the discovery of the extragalactic nature of galaxies. In 1929, the English astronomer James Hopwood Jeans assumed that the spiral pattern was generated completely by unknown forces~\cite[400-401]{Fridman}. Because soon a serious paradox regarding the formation of these structures was found: the galactic disks have a differential rotation which means that the angular velocity decreases with the radius. But if spiral arms would rotate differentially too, they should be stretched thin more and more over time which would lead to a disappearance of the spiral structure in only 1 to 2 rotation periods. However, observations of stars of the Milky Way have shown that they have made at least 50 rotations, without the spiral structure fading away. It took several decades and the research of many scientists to solve this paradox, but the foundation for the solution was laid already in 1938 by the Swedish astronomer Bertil Lindblad. He suggested that the spiral arms in the galaxies are density waves. He claimed that the wave fronts of these density waves rotate with a constant angular velocity and that also their phase velocities are constant. This theory was revived by Lin and Shu who developed it significantly. Today there are several observations that point to and reinforce the wave nature of the galactic spiral structure. The spiral structure of the velocity fields of stars and gas, the visibility of the spiral pattern in nearly all galactic components (e.g. HII-regions, molecular clouds, dust etc.) and a characteristic age gradient of stars across the spiral arms are only a few of them.\\
As has already been mentioned above, spiral arms are characterised by an increased concentration of young stars, which implies a higher star formation rate than in the remaining regions of the disk~\cite[32-35]{Fridman}. A galactic shock wave propagating through the disk can of course lead to the formation of star birth regions by compressing gas, but observations have also shown that the star formation in the regions between the spiral arms is often only $\sim30\%$ less than in the arms. It is proven that spiral arms have a much larger fraction of molecular gas, but still the star formation efficiency there is only about $\sim10-20\%$ greater than in the areas between the spirals.\\
Although there are galaxies with leading spiral arms, the clear majority of observed spiral galaxies have lagging arms. This means that they rotate with their ends backwards. In addition, branching spiral structures are often observed in galaxies. This means that, a galaxy with e.g. two spiral arms in the inner part can have four or more arms in its outer parts. A very interesting fact is that the length, contrast and also brightness of the spiral pattern depend strongly on the rotation velocity of the disk. In many cases of rather slow rotating galaxies ($v \leq 100$ km/s) spiral arms are very faint and chaotic or even not observable at all. What is also very important is that the observed parameters of the spiral structure depend strongly on the observed spectral range. In the ultraviolet range spiral arms are brighter and the contrast is higher, whereas in the visual range, e.g. in the V-band, and the near-infrared the structures are a lot less bright and also smoother. However, in the middle- and far-infrared the spiral pattern is again good visible. The last characteristic that should be mentioned is that the strength of the magnetic field in the spiral arms is considerably greater than in in the areas between them. This is primary due to the gas compression.

\section{RCs of spiral galaxies, mass distribution and dark matter}
\label{RCs}
As mentioned above, the disks of spiral galaxies are rotation-supported which means that the orbits of stars and gas are almost circular and the chaotic motions are negligible compared to the rotation velocity. This chapter deals with the characteristics of rotation curves (RCs) of spiral galaxies which are a very important property of this galaxy type and can tell us a lot about the nature of the galaxies.\\
First of all, what are RCs? A RC of a galaxy shows the rotation velocities of the galaxy's stars or gas in dependence of the distance from the galactic center.\\
RCs are essential for the understanding of spiral galaxies. The first things scientists want to know about a galaxy are usually its distance, size and mass. The latter can be estimated with the help of the galaxy's RC. The orbits of stars and gas and thus their rotation velocities are strongly dependent on the gravitational field of a galaxy, and the gravitational field is naturally connected to the mass distribution~\cite[35]{Bertin}. As stated by James Binney and Scott Tremaine, one could say that RCs "provide the most direct method of measuring the mass distribution of a galaxy"~\cite[598]{Binney}.\\
To derive the RC of a galaxy, the Doppler shift of its emission lines is used~\cite[35]{Bertin}. As the galaxy rotates, one side of it is basically moving toward and the other side away from the observer. This means that from the observer's point of view the velocities of both sides are different. This leads to a change of the wavelength of the light emitted from the galaxy due to the optical Doppler effect. And this change can be seen in the emission lines that appear tilted in the spectra. From this tilt velocities can be calculated. Typically, emission lines e.g. from HII regions, like $H\alpha$ or $H\beta$, are used to measure RCs~\cite[598]{Binney}. But it is also common to observe at radio wavelengths and use the 21 cm hyperfine transition emission line of atomic hydrogen (HI). The latter possibility is often preferred because the 21 cm emission line traces the cold gas, and radio observations of HI have shown that the gas disk extends a lot farther than the optical stellar disk~\cite[32]{Bertin}. This means that in most cases the RC can be determined better from radio observations of HI~\cite[35]{Bertin}.\\
In this context some differences between the stellar and the gas disk component of a galaxy should be mentioned. The shape of the RC derived from the gas component is normally very similar to the one from the stellar component, but there are some characteristic differences~\cite[28]{Fridman}. First, there is a small, but systematical difference between the velocities of the two components: on average, the gas is rotating faster, typically by $5-15\%$. In special cases it is also possible that the two disks rotate in opposite directions. Computing the mean value $\frac{(v_{gas}-v_{*})}{v_{gas}}$ for different galaxy types shows that this value increases from late-type to early-type galaxies which means that the velocity of the gas compared to the stellar velocity increases. Furthermore, in the RCs derived from the gas component more fine-scale inhomogeneities can be seen. The gas is also influenced more strongly by the chaotic, non-circular motions than the stellar disk. This is linked especially to the presence of the spiral pattern. These effects can lead to "wavelike" features in the RC of the gas, especially in the outer part.\\
As a first approximation, we can divide the RCs of spiral galaxies into two types: single hump and double hump galaxies~\cite[13-15]{Fridman}. This is valid for the RCs of stellar as well as gas disks. The single hump RCs are characterised by a nearly solid-state rotation in the central part ($v_{rot}(r)\propto r$). In this case, $v_{rot}(r)$ increases linearly with the distance from the centre until it reaches the so-called turn-over radius. From this point forward the value of $v_{rot}(r)$ stays more or less constant which is why this part of the RC is also called "the plateau". It is possible that for very large \textit{r} $v_{rot}(r)$ increases or decreases slowly, but usually only slightly.\\
The second type of RCs is characterised by a strong increase of $v_{rot}(r)$ with radius in the central part, however only on the range of about $0<r<0.3-1$ kpc. After that $v_{rot}(r)$ decreases, only to increase again afterwards and reach a plateau, just like the single hump RCs. Because of the decrease in $v_{rot}(r)$ and the subsequent increase a second "hump" can be seen in the RCs, yielding the name of double hump. Fridman and Khoperskov explain that this "internal 'hump' of the RCs of this type appears to be superposed on the section of the almost solid-state rotation of the galaxies with the single-hump curves"~\cite[13]{Fridman}. It is interesting that the maximum value of $v_{rot}$ of this inner hump is usually very similar to the value of $v_{rot}$ in the area of the plateau. However, the reason for the formation of double hump RCs is not completely understood yet. One possibility is the presence of a bulge in a not very dense disk that could lead to a decrease of $v_{rot}(r)$ in the transition zone between the bulge and the disk. Also a dense stellar disk with a "hole" in the central part observed in several galaxies could be an explanation for the double hump form. It should be mentioned that up to the middle of the 1980s, it was believed that almost all galaxies belong to the single hump type. But after the increase of resolution power more and more galaxies with a double hump RC were detected which implied that this group of galaxies is considerably larger than initially expected. In figure \ref{fig:hump} a sketch of the two RC types can be seen:

\newpage
\begin{figure}[!h]
  \centering
  \includegraphics[height=5 cm]{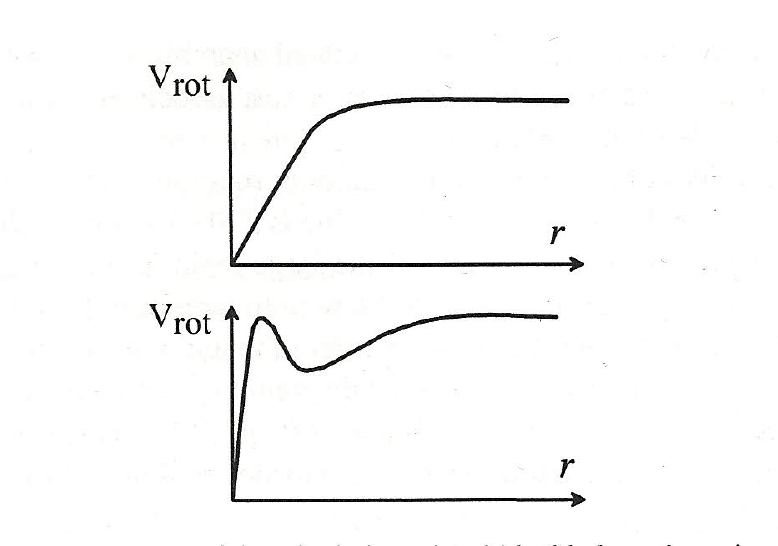}
   \caption{\small Typical form of the "single hump" (above) and the "double hump" RCs (below)~\cite[15]{Fridman}.}
   \label{fig:hump}
\end{figure}

\noindent
As explained above, the "plateau" of the RC was mentioned as the area where the value of $v_{rot}(r)$ remains more or less constant. Until the 1960s it was not possible to observe this flat part of the RC~\cite[598-599]{Binney}. Due to observational limits the measurements were restricted, so only the central parts of galaxies could be studied. This means that the RCs observable at this time mostly showed only a sharp rise of $v_{rot}(r)$ in the centre, at best followed by a short flat region. But there were not enough data points to know for sure how the RC continued after that. As this behaviour matched the expectations of the RC of an exponential disk, scientists assumed that after the last measured points, there was a Keplerian falloff of the velocity ($v_{rot}(r)\approx \sqrt{GM/r}$). This was reinforced by the fact that "the light of the galaxy was already mostly contained within the radius of the last measured point"~\cite[598]{Binney}. Based on the assumption that the unobserved region was Keplerian, until 1970 about thirty RCs and estimated total masses were published. However, at the same time the sensitivity in optical as well as radio observations improved significantly. Suddenly it was possible to measure RCs to larger radii. This showed that after the steep rise in the centre, the RCs stayed flat and that there was no sign of a Keplerian falloff which was a big surprise. Today we know that almost all RCs of spiral galaxies are more or less flat in the outer region. There are of course galaxies where the velocities decrease in the outer regions (but not as fast as in the Keplerian case). However this does not occur often and the reasons are mostly nearby companions that perturb the velocity field or very massive bulges.\\
But what does the flattening of the RCs mean for the mass distribution of the galaxies? It means that the mass does not decline beyond the optical radius as was assumed for many years. A RC with an almost constant velocity implies that the mass distribution extends to large radii and even increases linearly with radius. But as this mass cannot be seen, it is believed that spiral galaxies are surrounded by a massive and very large halo of dark matter, therefore called "dark halo". This was first postulated already in 1970 by Freeman. He claimed: "For NGC 300 and M33, the 21-cm data give turnover points near the photometric outer edges of theses systems. [...] if they are correct, then there must be in these galaxies additional matter which is undetected [...] Its mass must be at least as large as the mass of the detected galaxy [...]"~\cite{Freeman70}.\\
But still the problem of dark matter in spiral galaxies was not an important subject until the mid-1980s~\cite[37-38]{Bertin}. Today there are many observations that support the existence of dark matter. But RCs of spiral galaxies are by far the best and also most direct evidence for the presence of dark matter halos around galaxies. As Giuseppe Bertin and Chia-Ch'iao Lin express it, the halo mass is believed to be "at least comparable to the amount of mass in the visible form within a sphere of the size of the stellar disk ($r\sim4r_{d}$)"~\cite[38]{Bertin}. As already mentioned above, beyond this radius the mass should grow more or less linearly with the distance from the centre.\\
The shape of the RCs of spiral galaxies shows that the total circular velocity of a galaxy is the composition of the circular velocities of all its components~\cite[44-45]{Fridman}. Every component contributes to the total circular velocity which can be expressed as follows:

\begin{equation}
v_{c}=\sqrt{(v_{c}^{halo})^{2}+(v_{c}^{disc})^{2}+(v_{c}^{bulge})^{2}+...}
\end{equation}

\begin{figure}[h!]
  \centering
  \includegraphics[height=5.5 cm]{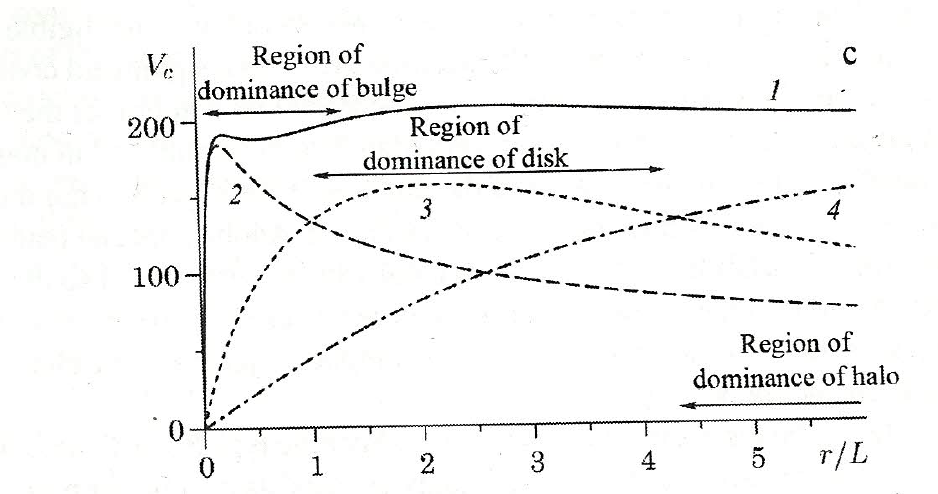}
   \caption{\small Radial dependence of the circular velocities $v_{c}(r)$ of indiviual components. (1) is the total circular velocity, (2), (3) and (4) are the circular velocities of the bulge, the disk and the halo, respectively~\cite[45]{Fridman}.}
   \label{fig:circvel}
\end{figure}

\noindent
The contributions of the individual components can be observed in figure \ref{fig:circvel}. It can be seen how the central part of the RC of a galaxy is dominated by the bulge (depending on how prominent and massive it is), followed by a region of dominance of the disk and finally the region dominated by the halo. This is why for the best results multicomponent models including gas as well as stellar subsystems are used to describe the mass distribution of a galaxy.

\chapter{Scaling Relations}
\label{scalrel}
\begin{figure}[!h]
  \centering
  \includegraphics[height=6 cm]{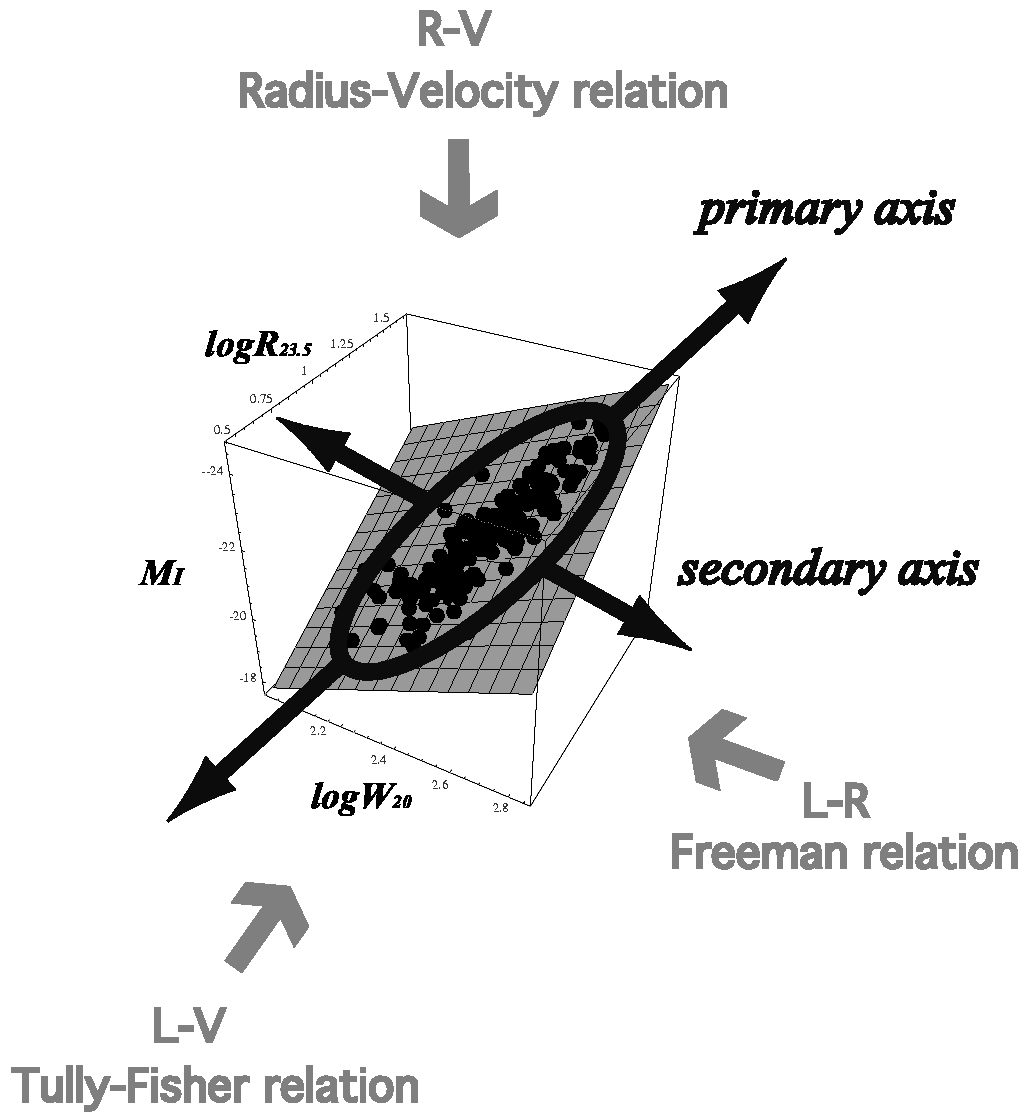}
  \caption{\small Spiral galaxies are distributed on a unique plane in the three-dimensional space of luminosity \textit{L}, radius \textit{r}, and rotation velocity \textit{v}. In this schematic figure, the I-band absolute magnitude $M_{I}$ [mag] represents \textit{L}, the face-on isophotal radius $R_{23.5}$ [kpc] represents \textit{r} and the HI line width $W_{20}$ [km/s] represents \textit{v}. Well-known scaling relations are indicated as projections of this scaling plane~\cite{Koda00a}.}
  \label{fig:fundplane}
\end{figure}

An important tool to study galaxies and their evolution, are the so-called scaling relations. Scaling relations connect fundamental characteristics of galaxies. This means that if a particular property of a galaxy is known, e.g. the luminosity or the size, in most cases another property, like the total mass, can be derived by means of scaling relations, assuming that the galaxy is a "typical" object (e.g. a main sequence galaxy). Apart from the fact that we can infer from scaling relations how galaxies evolve, some of them can also be used as a distance indicator, as elaborated below. The most important scaling relations for spiral galaxies, which are also of central importance for this master thesis, are the Tully-Fisher relation (TFR) and the velocity-size relation (VSR).\\
The origins of the famous TFR date back to the late 70s of the 20th century. In 1977 the astronomers Richard Brent Tully and James Richard Fisher published a paper in which they describe their observation of a tight correlation between the luminosity \textit{L} of a spiral galaxy and its maximum rotation velocity $v_{max}$ ($L\propto v_{max}^{a}$) for galaxies from the Local Group and the nearby M81 and M101 groups~\cite{Tully77}. Their purpose was to show that such a correlation could be used as a distance indicator by deriving the luminosity from $v_{max}$ and then comparing it to the apparent magnitude. To determine $v_{max}$ they used the global neutral hydrogen line profile width which is a distance-independent observable and can be measured directly. In their paper they used the sample of nearby galaxies to derive the distances to the Virgo cluster and the Ursa Major cluster, and also gave a preliminary estimate of the Hubble constant ($H_{0}$=80 km/s/Mpc).\\
Since the publication of this paper in 1977, the TFR has been the subject of many observational as well as theoretical studies. Besides trying to determine the origin of the slope and scatter of the relation or trying to find an evolution of the relation (which will be explained in more detail below), the TFR was put into the framework of a fundamental plane for spiral galaxies, similar to the fundamental plane for elliptical galaxies. The latter correlates the velocity dispersion $\sigma_{v}$, the effective radius $r_{e}$ and the surface brightness $\mu_{e}$ within $r_{e}$~\cite{Dressler87}. In the case of spiral galaxies, the disk scale length $r_{d}$ was introduced as the third parameter besides the luminosity and the maximum rotation velocity to form the fundamental plane~\cite{Burstein97}. Figure \ref{fig:fundplane} shows a sketch of the fundamental plane for spiral galaxies, taken from Koda et al. (2000)~\cite{Koda00a}:\\

\newpage

\noindent
Besides the luminosity \textit{L}, the maximum rotation velocity $v_{max}$ and the scale length $r_{d}$, the fundamental plane for spiral galaxies can also be constructed with other parameters which in principle exhibit the same characteristics. Fridman and Khoperskov (2013) state that to construct the three-dimensional plane, an energy characteristic, a spatial scale and a kinematic characteristic are necessary~\cite[50]{Fridman}. The energy can be represented by the absolute magnitude, e.g. $M_{I}$, or the central surface brightness $I_{0}$. For the spatial scale besides the disk scale length $r_{d}$ also radii of different isophotes, e.g. $R_{23.5}$ or $R_{25}$ (at 23.5 and 25 mag/arcsec$^{2}$, respectively), can be used. And instead of $v_{max}$ the HI line width can stand for the kinematic characteristic. This can be observed in figure \ref{fig:fundplane}. In this figure one can see that well-known scaling relations as the TFR and VSR (referred to as R-V) are simply projections of this fundamental plane. The third well-known scaling relation indicated in this figure is the Freeman relation that links the luminosity and the disk size. However, the Freeman relation is not relevant for this master thesis and will not be discussed further. Interestingly, the distribution of spiral galaxies on the fundamental plane seems to have an elongated "surfboard" shape. According to Koda et al. (2000) this can be understood as being the result of the existence of two dominant physical factors, for example the mass and the angular momentum~\cite{Koda00a}.\\
Today the TFR is a very important tool in extragalactic astronomy not only because of its purpose as a distance indicator, but primarily because it provides a link between the visible and the dark matter components of a galaxy. As has already been discussed in the previous section, this is due to the fact that the $v_{max}$ of a galaxy is proportional to its total mass, and the luminosity, or absolute magnitude, correlates to the visible mass. The TFR predicts that galaxies that are slow rotators also have lower luminosities than fast-rotating galaxies. The reason for this can be explained with a combination of the virial theorem and a nearly constant mass-to-light ratio~\cite{Asmusthesis}~\cite{Koda00b}.\\
Scaling relations like the TFR and the VSR are also a very significant tool to test the validity of the hierarchical scenario of the galaxy formation. This scenario predicts a hierarchical growth of galaxies with time. It states that the majority of the matter in the universe is composed of nonbaryonic and nonluminous matter that interacts only gravitationally, and stars and gas are embedded in halos of this dark matter~\cite{Boehm07}. According to this scenario, the low-mass systems were the first virialized structures that, after an epoch of numerous merger and accretion events, grew into larger systems. If this scenario is true, it should effect the evolution of scaling relations with time and thus be visible by means of differences between scaling relations of nearby and distant galaxies. For example, the evolution of the disk sizes can be traced with the VSR. In the case of a hierarchical growth scenario, the disk sizes, and thus the radii, should be smaller for distant galaxies.\\
The TFR can be used to derive the evolution of galaxies in luminosity. It is assumed that e.g. in the B-band, distant galaxies should be more luminous for a given total mass than local galaxies. The reason for this is that galaxies observed at higher redshifts have on average younger stellar populations~\cite{Asmusthesis}. Or with other words, the fraction of luminous, hot and massive stars is considerably higher than in the local universe.\\
This topic - the evolution of scaling relations with look-back time - was the subject of numerous investigations and publications in the past years. In the first place, scientists investigate the change of the zeropoints of the scaling relations, thus whether and by what amount the luminosity or the size of an object of a given mass (i.e. a given $v_{max}$) change with time. But another parameter is of significant interest, especially in the case of the TFR: its slope. The investigation of a possible change of the slope has been a main issue of many studies. An evolution of the slope of the TFR would be especially remarkable because it would not only indicate an evolution in luminosity, but even a mass-dependent evolution.\\
About twenty years ago the first attempts to construct an optical TFR for distant galaxies were made. Two examples for this are the works of Vogt et al. (1996) and Rix et al. (1997). In both cases, the number of galaxies used to construct a distant TFR is rather low. Vogt et al. (1996) determined the rotation velocity of nine faint field galaxies with redshifts between $0.1\leq z\leq 1$ and found a modest increase in luminosity ($\Delta M_{B}\leq 0.6$ mag)~\cite{Vogt96}. Rix et al. (1997) measured the [OII] emission linewidths of 24 blue field galaxies, but only up to redshifts of about $z\sim 0.35$~\cite{Rix97}. They conclude that distant blue, sub-$L^{*}$ galaxies are even about 1.5 mag brighter than local galaxies.\\
In the past two decades extensive surveys, such as the zCOSMOS-survey, provided data for thousands of galaxies that made it possible to construct distant scaling relations for far greater samples. Still, the results from studies that investigated a possible evolution in time of zero-point - or even slope - of scaling relations were rather discrepant. For example, Vogt (2001)~\cite{Vogt01} and Flores et al. (2006)~\cite{Flores06} found nearly no evolution up to redshifts of $z\simeq 1$, while e.g. B\"ohm et al. (2004)~\cite{Boehm04} found that spiral galaxies at $z\simeq 1$ were approximately by $\Delta M\simeq -1$ mag brighter than local galaxies with similar $v_{max}$. Also Bamfort et al. (2006)~\cite{Bamfort06} and Fernandez Lorenzo et al. (2010)~\cite{Fernandez10} got similar results regarding the change in luminosity. B\"ohm et al. (2004)~\cite{Boehm04} focused their work also on a possible slope change of the TFR, discussing a potential stronger luminosity evolution for low-mass disk galaxies. In contrast to their findings, Weiner et al. (2006)~\cite{Weiner06} found a stronger brightening in high-mass spirals.\\
In 2007, B\"ohm and Ziegler showed that a strong enough evolution of the TFR scatter with time could imitate an evolution of the slope through selection effects~\cite{Boehm07}. On the assumption of an incompleteness effect because of the magnitude limit, they stated that a scatter that is three times higher than the local scatter could account for the observed slope change of the distant TFR. And since the last few years, the assumption predominates that the slope of the local TFR holds up to at least $z\sim 1$, for all TFR variants~\cite{Boehm16}.\\
Although the VSR is also a very important scaling relation for spiral galaxies, there are considerably less observational studies that deal with its evolution with redshift. Two examples for this are the works from Puech et al. (2007)~\cite{Puech07} and Vergani et al. (2012)~\cite{Vergani12}. The first did not find differences in disk sizes at a given $v_{max}$ between z=0 and z=0.6, whereas the latter found a small increase of $\sim 0.12$ dex since $z\simeq 1.2$. In their work they looked at the half-light radii of the galaxies observed in the I-band.\\
Besides the growing amount of data from large surveys available to scientists, another development of the recent years made it possible to analyse and understand the evolution of scaling relations better: numerical simulations. Simulating the evolution of galaxies is not an easy task because many physical processes and factors have to be taken into account. For a long time, a major problem was that simulated spiral galaxies had a too low angular momentum~\cite{Boehm16}. Including internal and also external processes, e.g. different kinds of feedback etc., yielded simulated galaxies that had properties similar to observed galaxies and thus could be used to reproduce the scaling relations. A frequently used work to compare theoretical predictions with observational results is from Dutton et al. (2011)~\cite{Dutton11}. They combined N-body simulations and semi-analytic models in order to predict the evolution of scaling relations up to high redshifts. Regarding the luminosity and disk size evolution, their simulations predict that at given $v_{max}$, disk galaxies at z=1 are by about $\Delta M_{B}\sim -0.9$ mag brighter and by about $\Delta r_{d}\sim 0.2$ dex smaller than local galaxies. This seems to be in quite good agreement with the above-mentioned results from observational studies.\\
One more thing should be briefly mentioned here: the potential dependence of scaling relations on the environment. Many studies have tried to determine whether there is a difference between scaling relations of field and cluster galaxies, especially with regard to the TFR. Currently it is believed that field and cluster galaxies at a certain redshift have the same TFR slope, but a different scatter which is higher in denser environments (e.g. B\"osch et al. 2013~\cite{Boesch13}). The reason for this could be interaction processes that are typical for cluster environments, like e.g. tidal interactions that lead to an increased star formation or the so-called ram-pressure stripping which pushes gas out of a spiral galaxy and thus restrains or even stops the star formation~\cite{Boehm16}. The consequences of this can be a broader range in luminosities for a given $v_{max}$ and also a larger fraction of perturbed gas kinematics than in the case of field galaxies~\cite{Boesch13a}. All these factors can lead to an increased scatter of the TFR. However, if the cluster and field samples are matched considering the RC quality - which means that galaxies that have rather perturbed RCs because of interactions typical for cluster environments are rejected - their TFRs are very similar~\cite{Ziegler03}.

\chapter{Data}
\label{data}
The galaxies investigated in this master thesis are zCOSMOS galaxies. Their spectra were recorded with the HR-red grism of the VIMOS spectrograph between December 2009 and April 2010. The data were public in 2014 when I started with this master thesis.\\
A fundamental point regarding the spectra is that they were recorded with tilted slits. This means that the slits were not all aligned in the same direction, but for each object the slit was aligned with the semi-major axis of the galaxy. If the slits were tilted, the slit width was automatically adjusted to keep the width along the dispersion direction constant (1")~\cite{Pelliccia17}. In this way also the spectral resolution was kept constant, regardless of the position angle of the slit.\\
In this section the spectrograph VIMOS and the zCOSMOS survey are presented.

\section{VIMOS}
The name VIMOS is an acronym for "Visible Multi-Object Spectrograph". As the name suggests, the instrument is a visible wide field imager and multi-object spectrograph~\cite{vimos}. It is mounted on the Nasmyth focus B of UT3 Melipal, one of the four unit telescopes of the Very Large Telescope operated by the European Southern Observatory on Cerro Paranal in the Atacama Desert in Chile. VIMOS consists of four identical arms, called quadrants. Each quadrant has a field of view of $7'\times8'$ with a pixel size of 0.205"/px. The quadrants are separated by a gap of $\sim2'$. The wavelength range covered by the spectrograph goes from 360 to 1000 nm and the obtainable spectral resolution \textit{R} ranges from $R\sim200$ to $R\sim2500$.\\
The instrument can be operated in three different modes: Imaging (IMG), Multi-Object Spectroscopy (MOS) and as an Integral Field Unit (IFU). The mode that is relevant in this work is the MOS mode. With MOS it is possible to place multiple slits on the field of view and thus obtain simultaneous spectra of many objects. On the VIMOS instrument MOS is carried out using one mask per quadrant~\cite{mos}. Prior to the observation, the masks have to be prepared with the VIMOS Mask Preparation Software (VMMPS). For each quadrant there are six grisms available in MOS mode: LR blue, LR red, MR, HR blue, HR orange and HR red, where LR, MR and HR stand for low, medium and high resolution, respectively. Depending on which grism is used, the spectral resolution and the wavelength coverage vary. For example, the data used in this work were obtained with the HR red grism that has the highest possible spectral resolution of 2500 and a wavelength coverage of 650-875 nm.\\
It is also important that the maximum number of slits per quadrant depends on the spectral resolution. It varies from $\sim150-200$ at R=200 to $\sim40$ at R=2500. 

\section{COSMOS and the COSMOS field}
COSMOS stands for "Cosmic Evolution Survey". It was the first survey large enough to address the evolution of different aspects of observational cosmology and their connection~\cite{Scoville07}. It was designed to investigate the coupled evolution of galaxies, active galactic nuclei (AGNs), star formation and dark matter with large-scale structure out to $z\sim6$, which is the majority of the Hubble time. Some examples of more specified goals of the COSMOS survey are the evolution of galaxy morphology, the assembly of galaxies, clusters, and dark matter on scales up to $>2\times10^{14} M_{\odot}$ and the mass and luminosity distribution of the earliest galaxies. To probe the correlated evolution of all those diverse aspects, the survey included spectroscopy from X-ray- to radio wavelengths and multiwavelength imaging, including HST (Hubble Space Telescope) imaging. The field that was selected for the survey is a $\sim2$ deg$^{2}$ ($1.4^{\circ}\times1.4^{\circ}$) area that is located near the celestial equator, aligned north-south, east-west. It is centered at R.A.$=10^{h}00^{m}28.6^{s}$ and decl.$=+02^{\circ}12' 21.0''$ (J2000.0). The location was chosen to guarantee the visibility by all astronomical facilities. Decisive was also that there are no bright X-ray, UV or radio sources in this area and that - in comparison to other equatorial fields - the Galactic extinction in this field is remarkably low and uniform ($\langle E_{B-V}\rangle \sim0.02$ mag).\\

\section{zCOSMOS}
\label{zCOSMOS}
zCOSMOS is a large redshift survey that was carried out in the COSMOS field using the VIMOS spectrograph with observation time of about 600 hours~\cite{Lilly07}. It was undertaken between 2005 and 2010. The purpose of this survey was to characterise the environments of the COSMOS galaxies out to $z\sim3$, "from the 100 kpc scales of galaxy groups up to the 100 Mpc scale of the cosmic web"~\cite{Lilly07}. About the decision to undertake such a survey, Lilly et al. (2007) say: "The power of spectroscopic redshifts in delineating and characterising the environments of galaxies motivates a major redshift survey of galaxies in the COSMOS field"~\cite{Lilly07}.\\
The specific scientific goals of zCOSMOS can be subdivided into three categories:

\begin{enumerate}
\item The spectroscopic redshifts should be used to create maps of the large-scale structure and to measure the density field up to $z\sim3$. For these purposes, spectroscopic redshifts can yield results that are considerably more precise than those from photometric redshifts.
\item The spectra obtained with zCOSMOS should be used to diagnose the galaxies, i.e. provide informations about e.g. the star formation rates, the reddening by dust, the metallicities, AGN classification etc.
\item The third aim was to replace and complement the photometrically estimated redshifts by reliable accurate spectroscopic redshifts.
\end{enumerate}

\noindent
The spectrograph VIMOS was selected for this survey amongst others because its quadrant design enables to cover a large area. In order to cover the whole COSMOS field, simply the so-called "pointings" have to be moved across the larger survey field.\\
Another important thing to mention is that the survey consists of two parts: zCOSMOS-bright and zCOSMOS-deep. zCOSMOS-bright is the brighter, lower-redshift component of the survey. It is magnitude-limited with the constraint $I_{AB}<22.5$ and yields redshifts between $0.1<z<1.2$. This part of the survey covers the whole 1.7 $deg^{2}$ COSMOS ACS field. The resulting sample consists of about 20.000 galaxies. In constrast, zCOSMOS-deep is a sample of about 10.000 galaxies with $B_{AB}<25$ that have been selected through two color-selection criteria (BzK and UGR) and that have redshifts between $1.4<z<3.0$. This part of the survey covers only the central 1 $deg^{2}$ of the COSMOS field.\\
In figure \ref{fig:cosmos} the COSMOS field with the pointings and the total coverage of both components of the zCOSMOS survey can be seen:

\begin{figure}[!ht]
  \centering
  \includegraphics[height=12 cm]{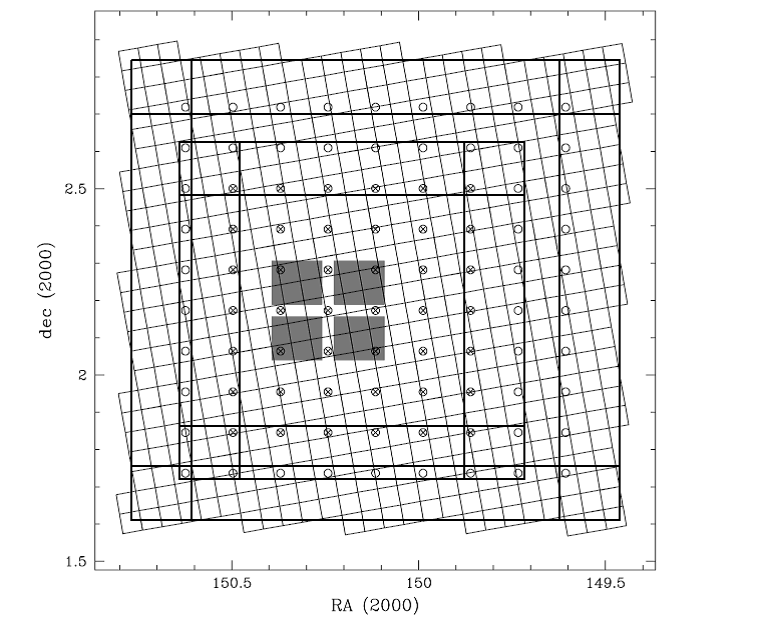}
  \caption{\small Pointing centers and total coverage of the two components of zCOSMOS compared with the mosaic of 600 COSMOS ACS images. The field centers of the 90 pointings of the zCOSMOS survey are indicated by the small circles. They are uniformly spaced in declination and R.A. which assures a constant sampling rate over a large area. zCOSMOS-bright uses all 90 pointings, whereas zCOSMOS-deep uses a mosaic of 42 central pointings (indicated by crosses). The four shaded areas show the VIMOS quadrant pattern for one pointing. Figure from~\cite{Lilly07}.}
  \label{fig:cosmos}
\end{figure}

\chapter{Data reduction}
\label{datared}
The data used in this master thesis were recorded by Laurence Tresse from the Laboratoire d'Astrophysique de Marseille. In the observation period between December 2009 and April 2010, a set of seven pointings, or rather 28 masks, were observed, each mask containing approximately 30 galaxy spectra.\\
For my master thesis I used five out of the 28 masks. As a first step, those five masks, containing altogether 160 spectra, were reduced with the software package VIPGI. To do this, I had science, bias and lamp frames at my disposal. In this chapter VIPGI as well as the main steps of data reduction carried out will be presented in more detail.

\section{VIPGI}
VIPGI is a software package that was designed with the objective of simplifying the reduction of data obtained with the spectrograph VIMOS. The name stands for "VIMOS Interactive Pipeline and Graphical Interface".\\
After VIMOS started operating in 2002, it was suddenly possible to obtain up to 1000 spectra in MOS mode during a single exposure~\cite{Scodeggio05}. This was of course challenging for the data reduction. In order to make it easier for the astronomical community to cope with and process the huge amount of spectroscopic data, the development of a data reduction and analysis pipeline was commissioned. This led to the development of VIPGI which has been publicly available to astronomers all over the world since June 2005.\\
The functions of VIPGI can be divided into two main parts. First, the data reductions recipes, which are the main purpose of this software package. The data reduction steps include wavelength calibration, creation of master bias, master flat and master lamp files, the preliminary reduction of each individual science file and finally the combination of several science files into one file. Those steps will be described in more detail in the next section. What is important here, is that VIPGI is a semi-automatic data reduction pipeline, which means that, as described in~\cite{Scodeggio05}, "there is no single 'do it all' recipe". It is not possible to input multiple raw data into the program and receive completely reduced files as the output. In fact, it is possible, and also recommended, to check the intermediate data reduction steps. In this way, one can understand what the program is doing, and also at which step something might have gone wrong if the result does not look good. It should also be mentioned that all slits in a given mask are always being reduced simultaneously.\\
All the recipes of the data reduction pipeline are written in the C programming language, whereas the second main part of VIPGI, the graphical interface, is written mostly in Python. The graphical interface offers many different possibilites to analyse the data. As a start, it is possible to look at and check the data at the various stages of the data reduction. After this it is also possible to analyse the data further, for example by plotting individual spectra. The tool to do this is called "slit summary" and is one of the most important tools of the graphical interface. It plots each one-dimensional spectrum of the mask with the corresponding two-dimensional spectrum as well as the one-dimensional sky spectrum. This allows to "visually check the reality of spectral features that are present in the one-dimensional spectrum"~\cite{Scodeggio05} which could also be due to sky or other residuals. It is also possible to get redshift estimates with this tool. This is achieved by manually indicating the position of several spectral lines within the spectrum. The tool yields a list of possible redshifts depending on known absorption and emission lines. After choosing a redshift from the list, all spectral lines that should be visible at this redshift are indicated on the plot and one can check if the redshift estimate is good or completely wrong.\\
Beside the mentioned points, there was another motivation to design a graphical interface. Marco Scodeggio and other scientists involved in this project wanted to provide the possibility to organize the large amount of data in a reasonable way, so astronomers working with the data would have a better overview.\\
This can be achieved by creating an \textit{organizer table} with all the data that belong together at the very beginning of the reduction procedure. The data in the organizer table are then automatically grouped by category and observation date. This is possible because the information needed for this classification is normally already included in the names of the data files. In my case, the observation date was not part of the file names, so I had to look into the header files in order to find out which files were recorded in the same night. However, the file type was included in the file names. For example, a typical name of a science file is: \textit{sc$\_$preimg83$\_$P1Q1$\_$vm$\_$HR$\_$Red$\_$M1Q1$\_$Q1$\_$1b.fits}. Here, "sc" stands for science and "P1Q1" for the observed pointing and the quadrant. In the case of a lamp or a bias file, "sc" in the file name would be replaced with "lp" or "bs". The appendix "1b" is simply used to number the several files from the same pointing and quadrant. "$HR\_Red$" means that the file was taken with high resolution and that the wavelength range is approximately from 630 to 870 nm. All the files used in this master thesis were recorded with 1"-slits in the HR red mode, so the spectral resolution is always 2500 and the dispersion is 0.6 \AA/px.\\
So depending on how the file names begin, the files are automatically placed into substructures. Now it is easy to choose between the individual quadrants and find the required bias, lamp or science files. When a master bias or a master lamp is created, the resulting files are named accordingly, starting with \textit{msbias} or \textit{mslamp}, and are put into the correct subdirectory.\\
This system of organizing the data automatically into a "rigidly predefined directory structure", as phrased by Scodeggio et al. (2005), helps the users of VIPGI to easily and quickly select the correct data~\cite{Scodeggio05}. In this way mistakes, like selecting the wrong input files, can be prevented and also the time required for the data reduction process is reduced significantly.

\newpage
\section{Main steps of data reduction}
The masks I worked with are P4Q1, P6Q1, P7Q1, P4Q2 and P5Q2. Thus three masks from the first quadrant and two from the second quadrant were used, but none from the third and the fourth quadrant. The masks were chosen based on the assumption that there are multiple bright emission lines visible in their spectra.\\
In table \ref{tab:pointings} the most important information about these five masks are shown: the number of single frames that were combined after the reduction process, the total integration time of all science frames in seconds, the dates of the nights when a certain pointing was observed along with the "names" of the exposures taken in a certain night, and the number of spectra contained in one mask.

\begin{table} [h!]
\begin{center}
\small
  \begin{tabular}{ l | c | c | c }
    
    Mask & P4Q1 & P6Q1 & P7Q1 \\ \hline
    Exposures & 7 & 6 & 6 \\ 
    Integration time [s] & 6920 & 6045 & 6045\\ 
    Observation dates & 16.03.2010 (1/2) & 14.01.2010 (1/2) & 25.01.2010 (1/2) \\ 
                      & 17.03.2010 (1a/1b/2a/2b) & 24.01.2010 (1b/1c/2b/2c) & 11.02.2010 (1a/2a) \\
                      & 05.04.2010 (1d) &  & 21.02.2010 (1b/2b) \\
    Spectra & 35 & 29 & 30 \\ 
  \end{tabular}
\end{center}

\begin{center}
\small
  \begin{tabular}{ l | c | c }
  
    Mask & P4Q2 & P5Q2 \\ \hline
    Exposures & 6 & 6 \\ 
    Integration time [s] & 6045 & 6045\\ 
    Observation dates & 16.03.2010 (1/2) & 22.02.2010 (1/1a/2/2a)\\ 
                      & 17.03.2010 (1a/1b/2a/2b) & 12.03.2010 (1b/2b)\\ 
    Spectra & 33 & 33 \\
  \end{tabular}
  \caption{\small Number of exposures, observation dates and number of spectra per mask.}
  \label{tab:pointings}
\end{center}
\end{table}

\noindent
The exposure time of a single frame was always about 1007 seconds, except in the case of frame \textit{1d} from P4Q1, where it was 875 seconds. It should be mentioned that not all existent frames of those five masks were reduced and used for the combined science files. For P4Q1 one exposure from 02.04.2010, for P6Q1 two exposures from 20.01.2010 and for P4Q2 one exposure from 02.04.2010 and one from 05.04.2010 were not used. In all these cases the seeing value did not meet the requirements. The seeing constraint was 0.80", but e.g. in the case of the two exposures from 20.01.2010 (P6Q1) that were dismissed, the seeing was 1.40" and 1.89", respectively. Additionally, in the case of P4Q1 and P4Q2, the exposure times of the discarded frames were considerably shorter (twice 350 and once 874 seconds).\\
In table \ref{tab:seeing} the seeing value of the individual exposures as well as the mean value for each mask are listed. The mean seeing values are mostly around 0.7", which implies good observing conditions. In the case of P7Q1 the mean value is somewhat higher because the individual values of two exposures were above 1". Nevertheless these exposures were included because otherwise only four exposures would be available for the combined science frame, and thus the total integration time would be considerably lower than in the case of the other four masks. The resulting mean value of 0.95" still implies reasonable seeing conditions.

\newpage
\begin{table} [h!]
\begin{center}
\small
	\begin{tabular}{ l | c  c  c  c  c }

& P4Q1 & P6Q1 & P7Q1 & P4Q2 & P5Q2 \\ \hline
1 & 0.57 & 0.82 & 1.30 & 0.57 & 0.82 \\ 
2 & 0.59 & 0.89 & 1.08 & 0.59 & 0.73 \\ 
1a & 1.00 & / & 0.81 & 1.00 & 0.79 \\ 
2a & 0.66 & / & 0.73 & 0.66 & 0.75 \\ 
1b & 0.70 & 0.77 & 0.82 & 0.70 & 0.86 \\ 
2b & 0.64 & 0.74 & 0.93 & 0.64 & 0.75 \\ 
1c & / & 0.46 & / & / & / \\ 
2c & / & 0.58 & / & / & / \\ 
1d & 0.88 & / & / & / & / \\ \hline
\hline
Mean & 0.72 & 0.71 & 0.95 & 0.69 & 0.78 \\ 
	\end{tabular}
	\caption{\small The seeing values of the indiviual exposures and the mean seeing value per mask in ["].}
	\label{tab:seeing}
\end{center}
\end{table}

\begin{figure}[!h]
  \centering
  \includegraphics[height=8 cm]{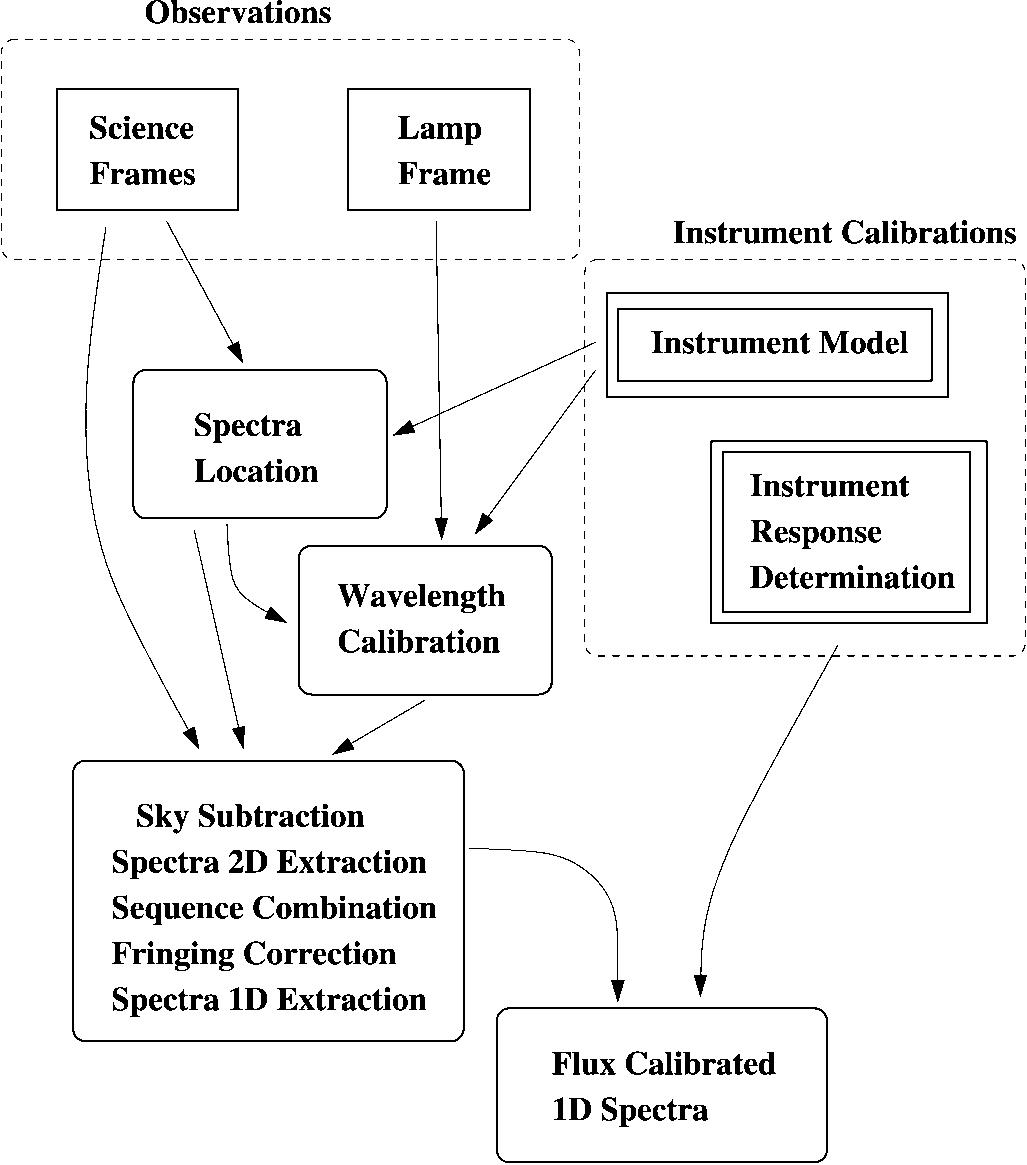}
   \caption{\small Block diagram of the main steps of the VIPGI data reduction process~\cite{Scodeggio05}.}
   \label{fig:Vipgi}
\end{figure}

\noindent
Figure \ref{fig:Vipgi} shows a diagram from~\cite{Scodeggio05} where the main steps of the data reduction process of VIPGI are outlined. In the following part of this section I will describe those main steps. Hereby, I will concentrate on the left side of the diagram, that is the procedures conducted by me. I will not go into detail regarding the instrument calibrations.

\subsection{Lamp line catalogue}
\begin{figure}[!h]
  \centering
  \includegraphics[height=6 cm]{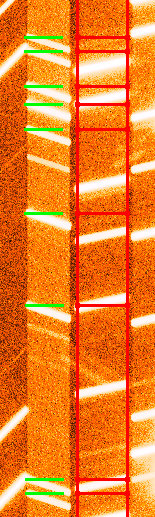}
   \caption{\small Lamp image (red) and corresponding lines of a spectrum (green) shown with the tool "Check Line Catalogue".}
   \label{fig:checklinecat}
\end{figure}

Before starting with the first step of the data reduction process - the first guesses adjustment - the lamp line catalogue needs to be checked which is later attached to the data and used for the wavelength calibration. There are line catalogues available for all the different observing modes. Depending on the observing mode, a standard catalogue with main lines for all lamps is appended to the data, different only from the low to the high resolution observations~\cite{Garilli09}. It is however probable that there are lines missing in the required catalogue or that the catalogue even contains lines that are not needed.\\
As my spectra were observed in the HR red mode, the file \textit{$lineCatalog\_HR\_red.fits$} should be appended to the data. To do that, the command \textit{VmAppTable [name of the lamp frame] [name of the line catalogue]} is used.\\
As already mentioned, the wavelength range of the HR red mode is approximately from 630 to 870 nm (=240 nm). However, this is valid only for slits located in the middle of the mask. The wavelength range of the spectra I used extends further in both directions. Depending on the position of the slit on the mask, the spectra can range from about 570 to 920 nm. The provided HR red line catalogue contained only lines between 630 to 870 nm, so it was necessary to manually add lines at wavelengths smaller than 630 nm and larger than 870 nm. To check which additional lines are needed, I used the tool \textbf{Check Line Catalogue} at \textit{Adjust first guesses}. Therein the lamp image is superimposed on an input, e.g. a science frame. An example can be seen in figure \ref{fig:checklinecat}. The red horizontal lines show all the lines contained in the line catalogue, and the green lines are the corresponding lines in a certain spectrum. The red lines are used as a reference and can be moved as a whole in order to easier identify the green lines. The green lines are always shown for two different spectra. This makes it easier to compare which required lines are already present in the lamp catalogue, which should be added and which should be deleted. It should be mentioned that sometimes it is problematic, if two lines are too close together. In these cases it can be better if one of the two lines is deleted.

\noindent
After comparing the lines of the lamp line catalogue with the spectra, I realised that not only lines below 630 and above 870 nm were missing, but also some lines between 630 and 870 nm. The catalogue contained only helium and argon lines, but no neon lines. To find out the wavelengths of the missing lines, one can find the plots of the lines that are used for the wavelength calibration for each grism on the VIMOS homepage from ESO~\cite{lineplots}. 
In figure \ref{fig:lamp_plots} the plots used to complement my line catalogue are shown:
\newpage

\begin{figure} [H]
\centering
\subfloat{\includegraphics[width= 17 cm]{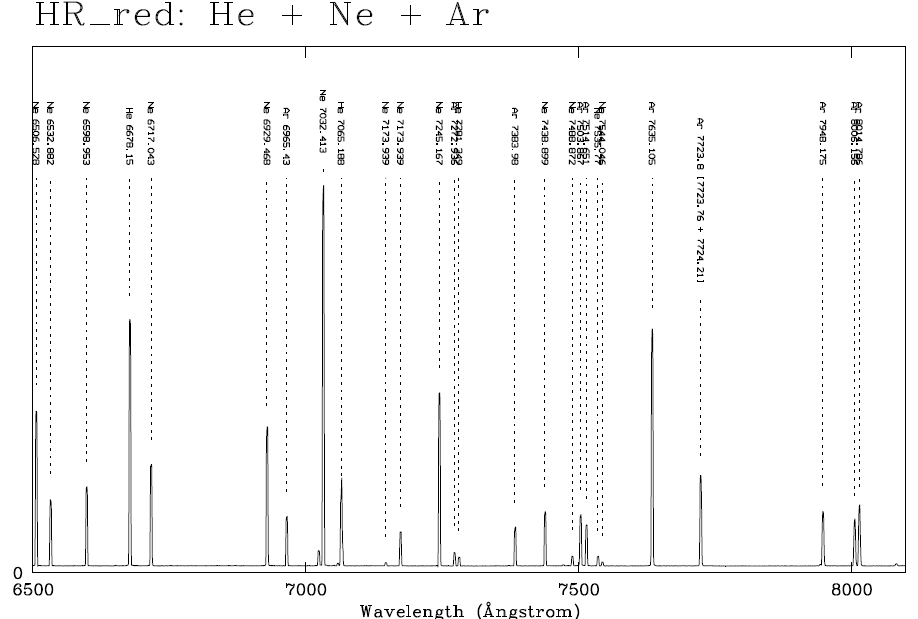}}\\
\subfloat{\includegraphics[width= 17 cm]{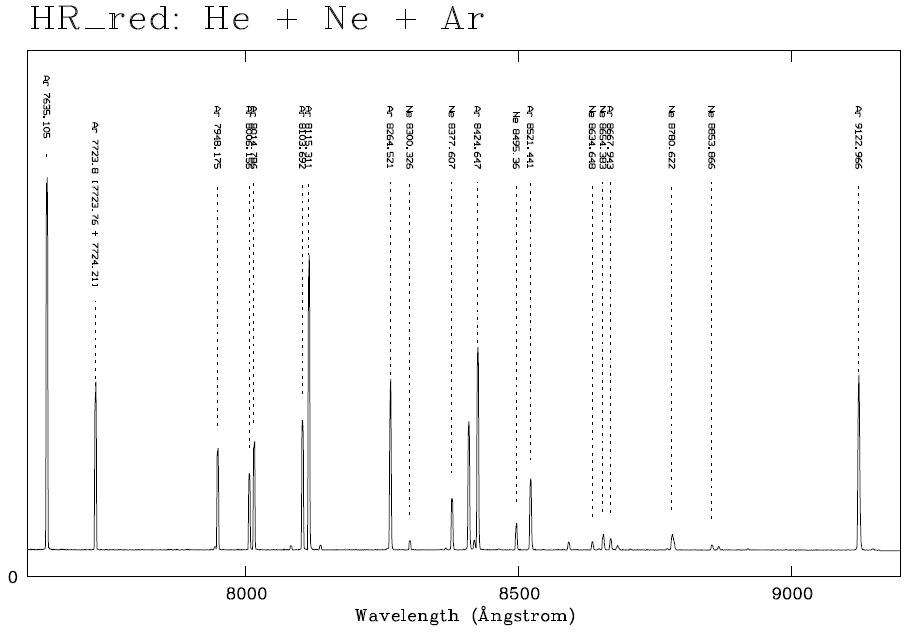}}
\caption{\small Lines used for the wavelength calibration for the HR red grism~\cite{lineplots}.}
\label{fig:lamp_plots}
\end{figure}

\noindent
Table \ref{tab:primarycat} lists the lines that were already included in the line catalogue at the beginning, whereas table \ref{tab:complcat} lists the lines of the final complemented line catalogue.

\begin{table} [h!]
\begin{center}
\small
	\begin{tabular}{ c | c  c }

 & Line & Wavelength [\AA] \\ \hline
1 & He & 6678.200 \\ 
2 & Ar & 6965.431 \\ 
3 & He & 7065.188 \\ 
4 & Ar & 7147.040 \\ 
5 & Ar & 7272.936 \\ 
6 & He & 7281.349 \\ 
7 & Ar & 7383.980 \\ 
8 & Ar & 7503.867 \\ 
9 & Ar & 7514.651 \\ 
10 & Ar & 7635.105 \\ 
	\end{tabular}
\quad
	\begin{tabular}{ c | c  c }

 & Line & Wavelength [\AA] \\ \hline
11 & Ar & 7724.206 \\ 
12 & Ar & 7948.176 \\ 
13 & Ar & 8006.157 \\ 
14 & Ar & 8014.786 \\ 
15 & Ar & 8103.693 \\ 
16 & Ar & 8115.311 \\ 
17 & Ar & 8264.523 \\ 
18 & Ar & 8424.647 \\ 
19 & Ar & 8521.442 \\ 
  &	   &   \\
	\end{tabular}
\caption{The 19 lines of the primary line catalogue.}
\label{tab:primarycat}
\end{center}
\end{table}

\begin{table} [h!]
\begin{center}
\small
	\begin{tabular}{ c | c  c }

 & Line & Wavelength [\AA] \\ \hline
1 & He & 5852.488 \\ 
2 & Ne & 5944.834 \\ 
3 & Ne & 5975.534 \\ 
4 & Ne & 6029.997 \\ 
5 & Ne & 6074.338 \\ 
6 & Ne & 6096.163\\ 
7 & Ne & 6128.16 \\ 
8 & Ne & 6143.062 \\ 
9 & Ne & 6163.594 \\ 
10 & Ne & 6217.281 \\ 
11 & Ne & 6266.495 \\ 
12 & Ne & 6304.789 \\ 
13 & Ne & 6334.428 \\ 
14& Ne & 6382.991 \\ 
15 & Ne & 6402.246 \\ 
16 & Ne & 6506.528 \\ 
17 & Ne & 6532.882 \\ 
18 & Ne & 6598.953 \\ 
\textbf{19} & \textbf{He} & \textbf{6678.200} \\ 
20 & Ne & 6717.043 \\ 
21 & Ne & 6929.468 \\ 
\textbf{22} & \textbf{Ar} & \textbf{6965.431} \\ 
23 & Ne & 7032.413 \\ 
\textbf{24} & \textbf{He} & \textbf{7065.188} \\ 
\textbf{25} & \textbf{Ar} & \textbf{7147.040} \\ 
26 & Ne & 7173.939 \\ 
27 & Ne & 7245.167 \\ 
	\end{tabular}
\quad
	\begin{tabular}{ c | c  c }

 & Line & Wavelength [\AA] \\ \hline
\textbf{28} & \textbf{Ar} & \textbf{7272.936} \\ 
\textbf{29} & \textbf{He} & \textbf{7281.349} \\ 
\textbf{30} & \textbf{Ar} & \textbf{7383.980} \\ 
31 & Ne & 7438.899 \\ 
32 & Ne & 7488.872 \\ 
\textbf{33} & \textbf{Ar} & \textbf{7503.867} \\
\textbf{34} & \textbf{Ar} & \textbf{7514.651} \\ 
35 & Ne & 7535.77 \\ 
36 & Ne & 7544.046 \\ \
\textbf{37} & \textbf{Ar} & \textbf{7635.105} \\ 
\textbf{38} & \textbf{Ar} & \textbf{7724.206} \\ 
\textbf{39} & \textbf{Ar} & \textbf{7948.176} \\ 
\textbf{40} & \textbf{Ar} & \textbf{8006.157} \\ 
\textbf{41} & \textbf{Ar} & \textbf{8014.786} \\ 
\textbf{42} & \textbf{Ar} & \textbf{8103.693} \\ 
\textbf{43} & \textbf{Ar} & \textbf{8115.311} \\ 
\textbf{44} & \textbf{Ar} & \textbf{8264.523} \\ 
45 & Ne & 8377.607 \\ 
\textbf{46} & \textbf{Ar} & \textbf{8424.647} \\ 
47 & Ne & 8495.36 \\ 
\textbf{48} & \textbf{Ar} & \textbf{8521.442} \\ 
49 & Ne & 8634.648 \\ 
50 & Ne & 8654.383 \\ 
51 & Ar & 8667.943 \\ 
52 & Ne & 8780.622 \\ 
53 & Ne & 8853.866 \\ 
54 & Ar & 9122.966\\ 
	\end{tabular}
\caption{The 54 lines of the final complemented line catalogue. The 19 lines of the primary line catalogue (table \ref{tab:primarycat}) are printed in bold.}
\label{tab:complcat}
\end{center}
\end{table}
\newpage

\noindent
To distinguish the lines from the primary line catalogue from the later added lines, the former are written in bold in table \ref{tab:complcat}. It can be seen that the primary line catalogue contained only 19 lines, whereas for the final line catalogue 35 new lines were added. Almost all of them are neon lines (32), as they were completely missing in the primary catalogue. One of the added lines is a helium line and two are argon lines.\\
Adding these 35 lines to the lamp catalogue provides a more accurate wavelength calibration and thus more correct results from the data reduction.

\subsection{First guesses}
The first step of the data reduction are the first guesses. The first guesses are a "preexisting calibration of the instrument properties"~\cite{Scodeggio05}. This information is appended to the raw data fits header of each file mostly already at observation time. It contains the MOS slit positions which are later on used for a first guess of the positions of all slit spectra on the CCD. Furthermore, it contains information for the inverse dispersion solution which is fundamental for the wavelength calibration. For this the positions of several strong emission lines on the CCD are measured. Line catalogues with known emission lines are necessary for this. They can be obtained either from calibration lamp exposures (as explained above) or the night sky lines present in the science frames. With the help of the line catalogue, as a first guess the positions of the lamp lines are measured in each spectrum and fitted against the known line wavelengths.\\
The first guesses from the header files can save a lot of time. But although they may give a very good estimate in some cases, the opposite may be the case in others. That is why the first thing to do is to check the first guesses and, if necessary, adjust them. This is one of the most important steps in the data reduction, as Scodeggio et al. (2005) state, "the accuracy and stability of the wavelength calibrations are obviously of greatest importance for spectroscopic surveys"~\cite{Scodeggio05}.\\
To check the quality of the first guesses regarding the wavelength calibration and the spectra location, several data browsing and plotting tools are available in VIPGI. The first thing that to be done is to click on the field \textit{Browsing $\rightarrow$ Adjust First Guesses}. Here one can choose between five different options. It is important that for every option always one science (or flat field) and one lamp file from the same night are selected.\\

\noindent
The first option is \textbf{Shift Only}. Once run, the lamp image is displayed with the imaging and data visualization program \textit{ds9}, "with some regions superimposed, as computed from the flat field or science image given in input"~\cite{Garilli09}. The first guesses of the slit positions are shown as green vertical lines. These lines show where the left edge of each slit should be. Also the supposed positions of the arc lines within each spectrum are plotted as green horizontal lines. Both, the vertical as well as the horizontal lines, can be displaced to a greater or lesser extent from their correct position. However, the positions of the arc lines are not really relevant in this step. It can be assumed that the positions of the most central lines are correct, but the important thing here is that the green vertical lines should be moved in such a way that they really coincide with the left edges of the spectra, at best within 3-4 pixels. This can be achieved by clicking on the blue rectangles shown on the image and moving them left-right or up-down, respectively. After this has been performed successfully, one has to click \textit{yes} which will lead to the program computing a shift of the first guesses. The result are shown on the image. If the newly computed first guesses are satisfactory, the tool can be closed and the changes saved. In case of the results not being good enough, one can simply repeat the \textbf{Shift Only}-procedure and compute the first guesses again. This can be repeated as many times as necessary. It should be mentioned here that this applies to all the tools that are part of \textit{Adjust First Guesses}. The procedure is always the same: after changes have been made, the program computes new first guesses and depending on how well they fit, one can exit and save the changes or iterate the procedure.\\
Figure \ref{fig:shiftonly} shows examples of good and bad performances of the \textbf{Shift Only}-procedure:

\begin{figure}[H]
  \centering
  \subfloat[Good example]{\label{main:a}\includegraphics[height= 4 cm]{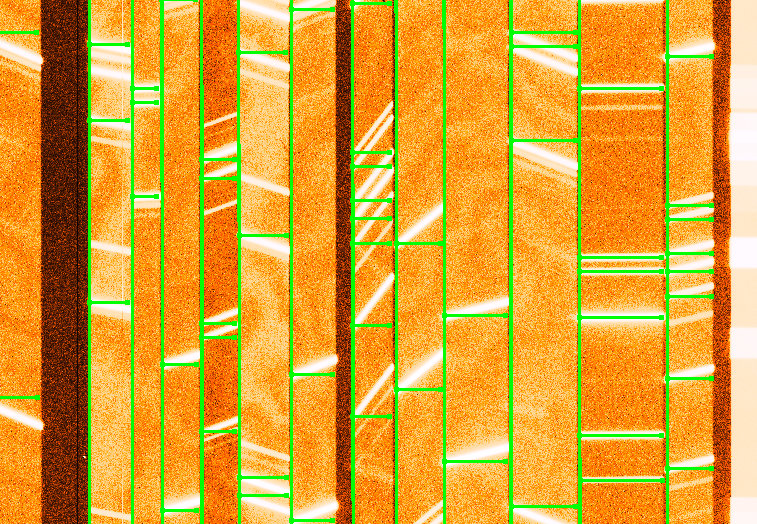}}\\
  \subfloat[Bad example]{\label{main:b}\includegraphics[height= 4 cm]{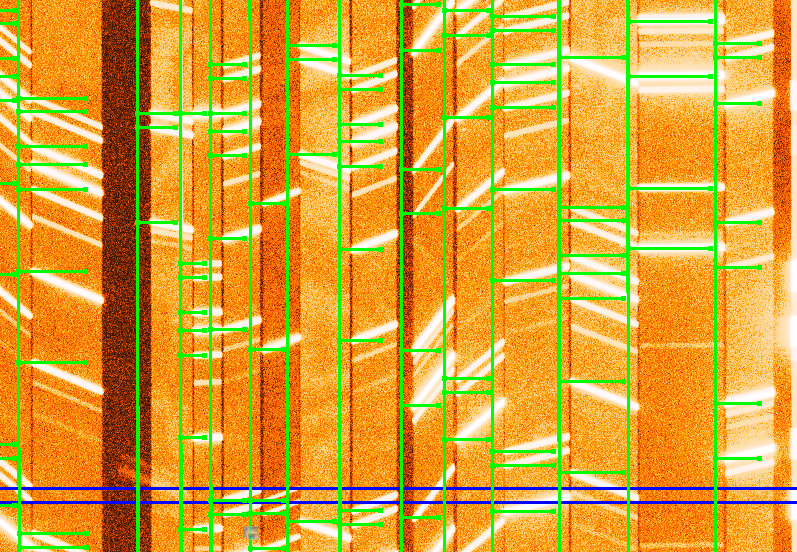}}
  \caption{\small Examples of good and bad performances of the \textbf{Shift Only}-procedure. It is well visible that in the case of the bad example the green vertical lines do not coincide with the left edges of the spectra. On the lower picture also the blue rectangle can be seen that is used to move the green lines.}
   \label{fig:shiftonly}
\end{figure}

\noindent
\textbf{Check Lamp Catalogue} is another tool selectable at \textit{Adjust First Guesses} and has already been described in the previous subsection. This step was not carried out for every night separately, but only once at the beginning to help complete the line catalogue.\\

\noindent
A tool that was not used is \textbf{Rotate}. In the multiobject spectroscopy it can happen that the grisms are not perfectly aligned and that the spectra are therefore a little bit tilted. With the \textbf{Rotate}-procedure one can check if this is the case and, if necessary, correct it. This problem did not come up in my data, so this step was left out.\\

\noindent
An important tool of the \textit{Adjust First Guesses} is \textbf{Choose Lines}. With this tool representative lines can be chosen that are used for a global dispersion model. This step is not necessary. If it is omitted, VIPGI automatically chooses the first and the last as well as the median and the quartile line from the lamp catalogue~\cite{Garilli09}. However, the more lines are chosen, the more precise the wavelength calibration can be carried out. For this reason I decided to use all the lines in my lamp catalogue.\\

\noindent
The last and most time-consuming, but also most important step regarding the first guesses is the procedure \textbf{Complete Adjustment}. It is similar to \textbf{Shift Only}, however now the correct position of the arc lines is of importance. As always, after selecting a science and a lamp file, the lamp image is displayed on \textit{ds9}, with regions superimposed. At this step only green horizontal lines are shown, indicating - as in \textbf{Shift Only} - the presumable positions of the arc lines within the individual spectra. The number of lines that is shown depends on how many lines have been chosen at the step \textbf{Choose Lines}. It should be noted that in the case of high resolution grisms all slits are shown, whereas in the case of low resolution only a part is shown. This can of course be changed, but it is of no importance here because the data used in this thesis are taken with high resolution. In addition to the green lines, a red region with all lines from the line catalogue is shown on the image. This red region can be moved around in order to help identify the individual lines in a certain spectrum and also distinguish between very close lines. In this step, the green lines have to be moved individually in such a way, that they coincide exactly with the corresponding real arc line. Sometimes it is not necessary to move a line at all, but in some cases it is indispensable. In order to examine all the lines, the tool \textit{View$\rightarrow$Panner} from \textit{ds9} can be used.\\
This last step of the first guesses was the most time-consuming. In most cases the first newly computed first guesses for the dispersion solution were not satisfactory. After the first run many green lines were moved back to their previous position. So in the majority of cases this step had to be repeated at least once, or even twice, until the results were good enough.\\
A very important point regarding this step is that - as mentioned in chapter \ref{data} - the spectra were recorded with tilted slits. This means that although the green lines superimposed on the lamp image are horizontal, the arc lines are not. This can be seen in figure \ref{fig:compladj}:

\begin{figure}[H]
  \centering
  \includegraphics[height= 5 cm]{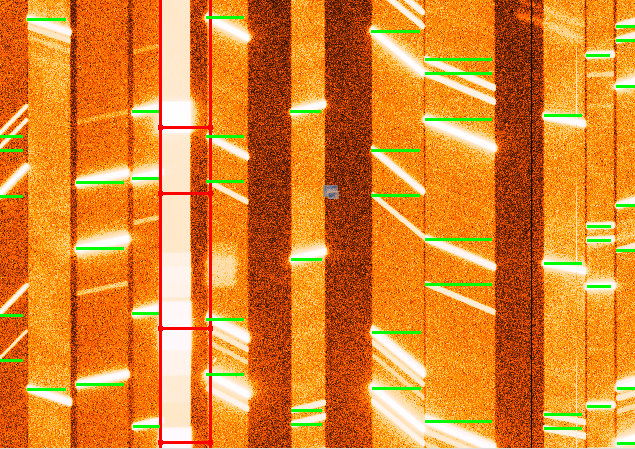}
  \caption{\small An example for the \textbf{Complete Adjustment}. It can be seen that the green lines are horizontal, the arc lines however are tilted. The green lines are moved in such a way that they coincide with the left ends of the arc lines. On the left side of the picture the red region with the lines from the lamp catalogue can be seen.}
   \label{fig:compladj}
\end{figure}

\noindent
This makes it somewhat more difficult to decide how and where to move the green lines. It would be possible to match them with the left or right side, or even the middle. Here, the lines were moved in such a way that the left end of the green line coincides with the left end of the arc line. This can be observed in figure \ref{fig:compladj}. Of course it is crucial that all the lines are moved in the same manner.\\
As previously stated, at all the steps of the first guesses a science (or, if existent, a flat field) and a lamp file from the same night have to be selected. However, this procedure of the first guesses does not have to be performed with all the science files. It is sufficient if the first guesses are done only for one science frame per night and then later on applied to the remaining exposures via the master flat from the same night. This approach is only valid for files from the same night. The first guesses from one night cannot be applied to another night because the observing conditions may vary greatly and cause wrong results. Furthermore, in most cases two lamp files from the same night were available. Normally, calibration lamp exposures for a night are taken right after all the science frames have been observed, but sometimes also during the following day~\cite{Scodeggio05}. In the case of the data used in this thesis, always the first lamp file was used.\\
Table \ref{tab:fg_files} lists which science and which lamp file were used for the first guesses of the individual observation nights:

\begin{table} [h!]
\begin{center}
\small
	\begin{tabular}{ c | c  c }

Obs. night & Science & Lamp \\ \hline
16.03.2010 (P4Q1) & 1 & 4 \\ 
17.03.2010 (P4Q1) & 1a & 4a \\ 
05.04.2010 (P4Q1) & 1d & 4d \\ 
14.01.2010 (P6Q1) & 1 & 4 \\ 
24.01.2010 (P6Q1) & 1b & 4b \\
25.01.2010 (P7Q1) & 1 & 4 \\ 
11.02.2010 (P7Q1) & 1a & 4a \\ 
21.02.2010 (P7Q1) & 1b & 4b \\ 
16.03.2010 (P4Q2) & 1 & 4 \\ 
17.03.2010 (P4Q2) & 1a & 4a \\ 
22.02.2010 (P5Q2) & 1 & 4 \\
12.03.2010 (P5Q2) & 1b & 4b \\ 
	\end{tabular}
	\caption{\small The science and the lamp files that were used for the first guesses of the individual observation nights.}
	\label{tab:fg_files}
\end{center}
\end{table}

\subsection{Master bias, master flat and master lamp}
\label{Master frames}
\begin{figure}[H]
  \centering
  \includegraphics[height= 4.5 cm]{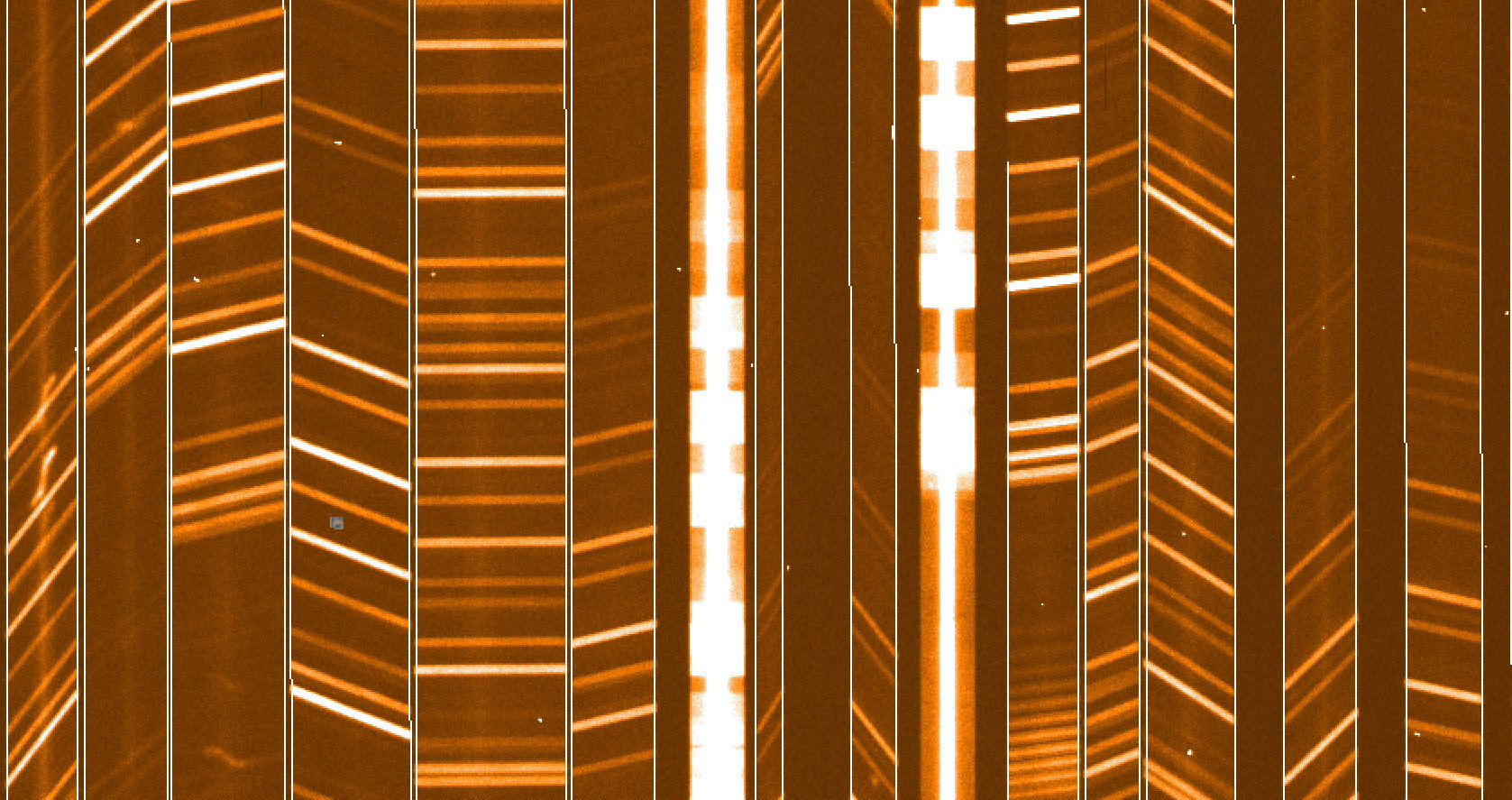}
  \caption{\small An example for the \textbf{Check Spectra Location}-tool. It can be seen that the white lines are in good agreement with the edges of the spectra.}
   \label{fig:checksploc}
\end{figure}

If the first guesses are satisfactory, the next step is to create the master files required for the data reduction. First of all, the master bias is created. Bias frames are zero seconds-exposures that comprise only the readout noise created during the recording of the spectra, that is the noise of the electronics. By subtracting the master bias from a science frame, this undesirable noise is eliminated. To create the master bias in VIPGI, one has to select a couple of bias frames and then click on the button \textbf{Create Master Bias} in the \textit{Reduction} menu which starts the reduction recipe \textit{pipeMasterBias}. The number of selected bias frames is arbitrary. Here, five bias frames from one night were used to create a master bias. Also, only one master bias was created for one quadrant. As three of the five masks used in this thesis are from the first quadrant and two from the second quadrant, only two master biases had to be created. It is not necessary to create one for each night or mask individually, because as already stated, the bias frames contain only the readout noise and are thus not dependent on the observing conditions. This does not apply in the case of the master flat and the master lamp frame as they contain the information about the first guesses. These have to be created for each night separately.\\
After creating the master bias, the next step is to run the program \textbf{Locate Spectra}. There are two purposes of this program: to trace the spectra on the CCD and to create a spectroscopic master flat file~\cite{Garilli09}. The first is the main purpose of this step because the location of the spectra is fundamental for the following steps of the data reduction. In contrast, the second purpose is not really required. In fact, Garilli et al. (2009 ) even say that "experience has shown that to correct for a flat field makes fringing corrections very bad"~\cite{Garilli09}. To avoid this, it is possible to instead of that create a mock master flat. For this, simply the option "Locate Spectra" in the parameter files has to be set to "yes". Consequently, by running the program \textbf{Locate Spectra} only the spectra are located and all pixel values of the resulting mock master flat are set to 1. The mock master flat was also applied in the case of the data used in this thesis. In order to run the program \textbf{Locate Spectra} and create the mock master flat, the master bias as well as a science file have to be selected. For each night the same science frame was selected that has already been used for the first guesses.\\
To check how well the spectra have been located, the program \textbf{Check Spectra Location} from the \textit{Browsing} menu can be run. After selecting the created (mock) master flat and a science frame from the same night, an image is displayed with \textit{ds9} that shows the spectra as well as superimposed white lines that indicate the supposed contours of regions containing a certain spectrum. By zooming into the image, one can check if the white lines coincide with the edges of the spectra. An example can be seen in figure \ref{fig:checksploc}

\noindent
The last master file that is required for the reduction is the master lamp. It is created by selecting a lamp file as well as the previously created master bias and the (mock) master flat for this particular night, and clicking on the \textbf{Create Master Lamp} button from the \textit{Reduction} menu which starts the recipe \textit{pipeSpMasterLamp}. Another possibility is to select the corresponding entry in the \textbf{Parameter Files}-menu which displays the parameter file editor for this recipe. This option has the advantage that it allows to check and edit the parameter file before executing it.\\
At this step, the master flat is used to find the spectra on the CCD. The first guesses that are stored in the header are used to find the expected positions of all the lines in the line catalogue for each slit. For every arc line the program tries to determine the peak and after doing so, fits all the peaks with a $3^{rd}$ degree polynomium~\cite{Garilli09}. The results are then stored in the resulting master lamp.\\
The master lamp is of great importance for the following reduction steps, therefore it is essential that it is done properly. There are two possibilities to check the quality of the master lamp frame. The first is the \textbf{Check Master Lamp} tool in the \textit{Browsing} menu. It is similar to the tool \textbf{Check Spectra Location} and displays a selected raw lamp frame with \textit{ds9}, with superimposed positions af arc lines. This can be seen on the following picture \ref{fig:checkmslamp}:

\begin{figure}[H]
  \centering
  \includegraphics[height= 4.5 cm]{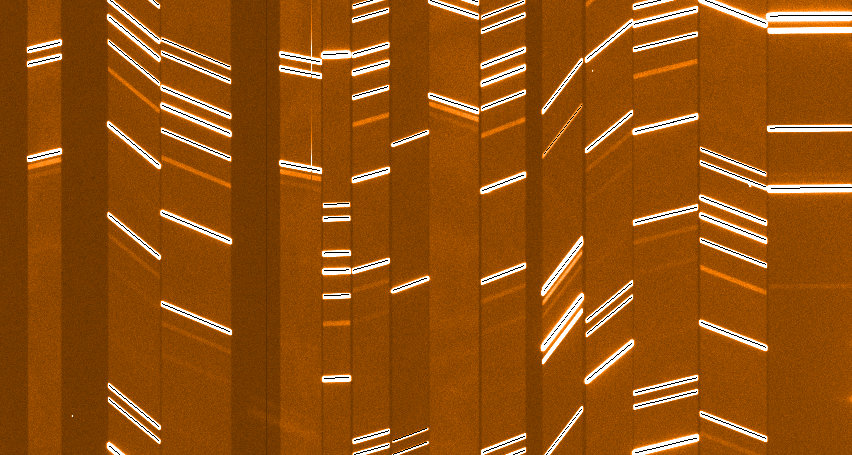}
  \caption{\small An example for the \textbf{Check Master Lamp} tool. It can be seen that the black thin lines match the arc lines very good.}
   \label{fig:checkmslamp}
\end{figure}

\noindent
With \textbf{Check Master Lamp} one can check the quality of the master lamp by how well the black lines coincide with the real arc lines from the raw lamp file. If the result is not good, in the worst case the first guesses have to be repeated, especially the \textbf{Complete Adjustment}. With this tool it can also be seen if certain lines tend to cause problems in several spectra or if the wavelength calibration in individual slits is not good enough yet. If this is the case, it is sufficient to adjust only these slits (or lines) with the \textbf{Complete Adjustment}.\\
The second tool to analyse the master lamp is \textbf{Show Lambda Calibration} in the \textit{Plotting} menu. It is a very useful tool because it offers several options to check the quality of the wavelength calibration. It is possible to inspect and analyse each slit and also each line individually, by selecting either \textit{One slit} or \textit{One line}. In both cases, individual lines that seem to be not good enough can be selected and removed. The first option shows the precision of all lines for a certain slit, whereas the second option shows the precision of one certain line in all slits. As a measure for the precision the rms-values are used. Both options distinguish between four kinds of lines (displayed with different symbols or colours): good lines (actually used; blue circles), bad lines (dismissed already in advance because they are too far away from their expected position; red circles), lost lines (for which it was no possible to find the peak; magenta triangles) and rejected lines (black triangles). The rejected lines are those removed manually during this process.\\
There is also a third option: besides individual lines and slits one can look at a histogram with the rms values for all slits with the \textit{Summary}-option. Below the histogram the slits are listed in order of decreasing rms. Here one can see which slits are rather bad and subsequently take a closer look at those with the option \textit{One slit}. If at \textit{One slit} or \textit{One line} certain lines are removed manually, the summary histogram changes consequently. In the case that a removed line worsens the rms values and thus the quality of the master lamp, the removal can easily be undone with \textit{Restore original fit}. Regarding the rms values it should be mentioned that the mean rms of all slits should be around 1/7 of the pixel size~\cite{Garilli09}. Thus in the case of the HR red grism which has a dispersion of 0.6\AA/px, the mean rms should be around $\sim 0.09$\AA. This means that the rms of the individual slits should be of the same order. If the peak of the rms distribution and several rms values are considerably larger, this can e.g. be a sign that there is a problem with a misidentified line. A rms of only 1\AA is already too large.\\
After implementing changes with the tool \textbf{Show Lambda Calibration}, one can analyse the master lamp again with \textbf{Check Master Lamp} and look if the black lines match the arc lines now better or not. The three options of the tool \textbf{Show Lambda Calibration} can be seen in figure \ref{fig:lambdacal}:

\begin{figure}[H]
  \centering
  \subfloat[One line]{\label{main:a}\includegraphics[height= 6.5 cm]{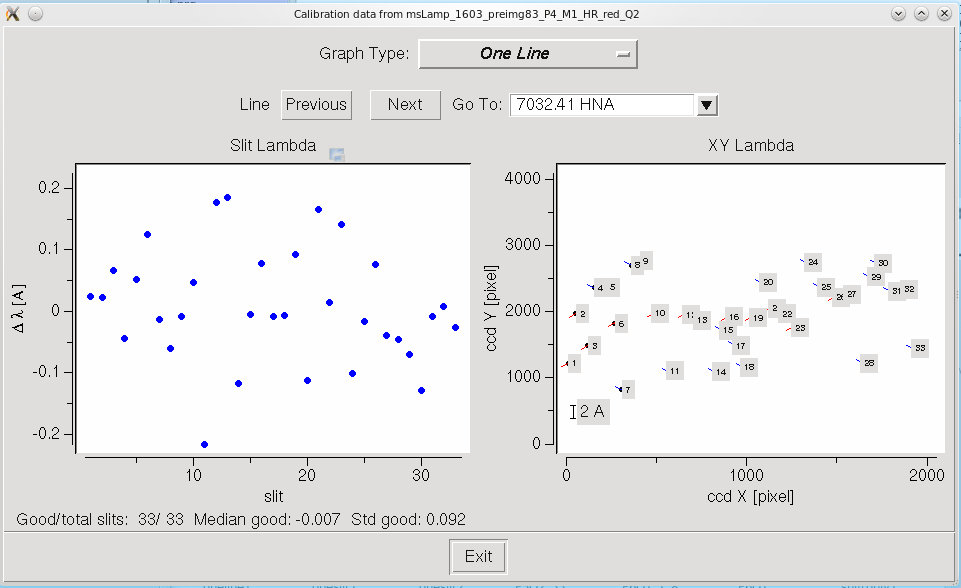}}\\
  \subfloat[One slit]{\label{main:b}\includegraphics[height= 7.5 cm]{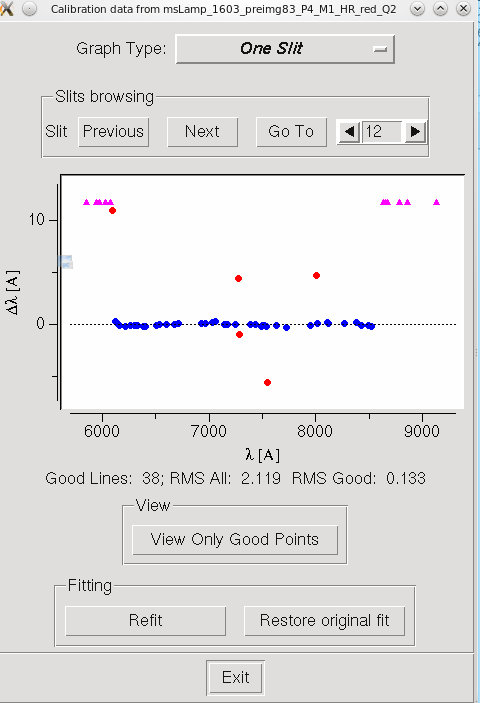}}
  \subfloat[Summary]{\label{main:c}\includegraphics[height= 7.5 cm]{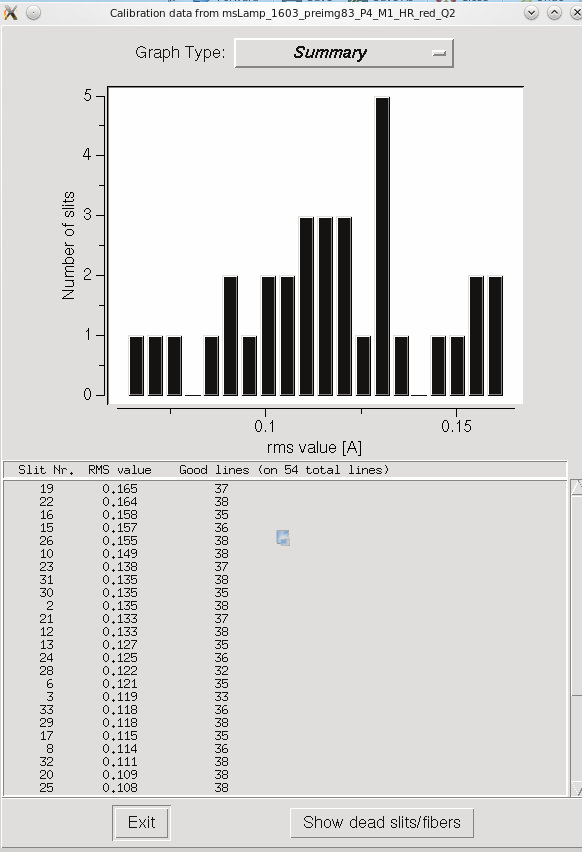}}
  \caption{\small The three options of the tool \textbf{Show Lambda Calibration} that are used to check the quality of the master lamp: \textit{One line} (\textit{a}), \textit{One slit} (\textit{b}) and \textit{Summary} (\textit{c}).}
   \label{fig:lambdacal}
\end{figure}

\subsection{Preliminary Reduction, Reduce Single Observation and Reduce Sequence}
\label{reduction}
The final step of the data reduction process is the combination of the individual exposures into one file to increase the signal-to-noise ratio. This step consists roughly said of three parts. If the quality of the first guesses of the individual nights is satisfactory, the \textbf{Preliminary Reduction} is performed. This step is implemented for each observation night separately. For this, all science frames from one night and the corresponding master bias and master flat are selected. The main purpose of the preliminary reduction is to subtract the master bias and to append the extraction table from the master flat to the science frames~\cite{Garilli09}. Furthermore, trimming of the science frames to "eliminate prescan and overscan areas" as well as interpolation to remove bad pixels are performed at this step~\cite{Scodeggio05}. The output files of the \textbf{Preliminary Reduction} are named the same as the input science files, only with the suffix BFC. Those BFC-files are needed for the next step, the \textbf{Reduce Single Observation}-tool. This program is very important as it provides fully reduced science frames. During this step, which requires the BFC-files and the master lamps as the input, amongst others the wavelength calibration is refined, the sky background is subtracted, objects are detected (as groups of contiguous pixels above a certain threshold) and the 2D spectra are extracted~\cite{Garilli09}~\cite{Scodeggio05}. Those reduction steps are carried out on all slits individually, slit after slit. Of course also at this step only files from the same night are selected.\\
As mentioned above, the output files, that now have the suffix BFCR, are fully reduced and could in principle be already used for scientific analysis. But as they are still individual exposures, the signal-to-noise ratio (S/N) is mostly not good enough. To increase it, the single exposures of a certain night have to be added. If \textit{n} exposures are combined, the exposure time, that is the signal, is \textit{n}-times higher, whereas the noise increases only with $\sqrt{n}$~\cite{im_comb}. E.g., if 4 exposures are combined, the S/N is doubled.\\
To combine all the individual exposures of a certain mask, the program \textbf{Reduce Sequence} is used. For this tool, the BFCR-files of one mask as well as the corresponding master lamp need to be selected. The BFCR-files are already reduced, so in the parameter file of this recipe \textit{Use already extracted spectra} and \textit{BFCR only} have to be set to \textit{yes}. In this case, the program knows that the science frames are already reduced and the master lamp is not used. If those parameters were set to \textit{no}, the program would reduce the single exposures before combing them. This can be done, if only exposures from the same night are combined. In the case that exposures from different nights should be combined, the first option is recommended so that each exposure is reduced with its own master lamp.\\
If already reduced BFCR-files are used, the main purpose of the \textbf{Reduce Sequence}-tool it to first detect if there is an offset between the single frames, and then combine them. If the parameter \textit{Use header information for shifts computing} in the parameter file is set to \textit{yes}, the program automatically computes the gobal offset from one image to another on the basis of the telescope pointing coordinates saved in the header file~\cite{Garilli09}. The individual slits are then shifted according to the global offset and finally added. The output of this program are as many BFCS-Files as there were input BFCR-Files, and a file with the suffix \textit{seq}. This \textit{seq}-file contains several extensions. The most important is the \textit{EXR2D}-extension which shows the combination of the individual reduced science frames, that is the file that is subsequently used for the analysis. The second very important extension is \textit{SKY2D}. This extension shows the computed residual sky. Depending on how well the sky lines of the individual slits match, one can see if the data reduction has been done good or bad. Another practical extension is \textit{WIN}. In this extension the information is stored where exactly a certain slit starts and ends and also where the objects detected during the \textbf{Reduce Sequence} are located, all written in pixels. This is useful if one wants to cut the individual slits to use it for further analysis. Figure \ref{fig:exr2d} shows an example of an \textit{EXR2D}-file and the corresponding \textit{SKY2D}-file:

\begin{figure}[H]
  \centering
  \subfloat[EXR2D]{\label{main:d}\includegraphics[height= 9 cm]{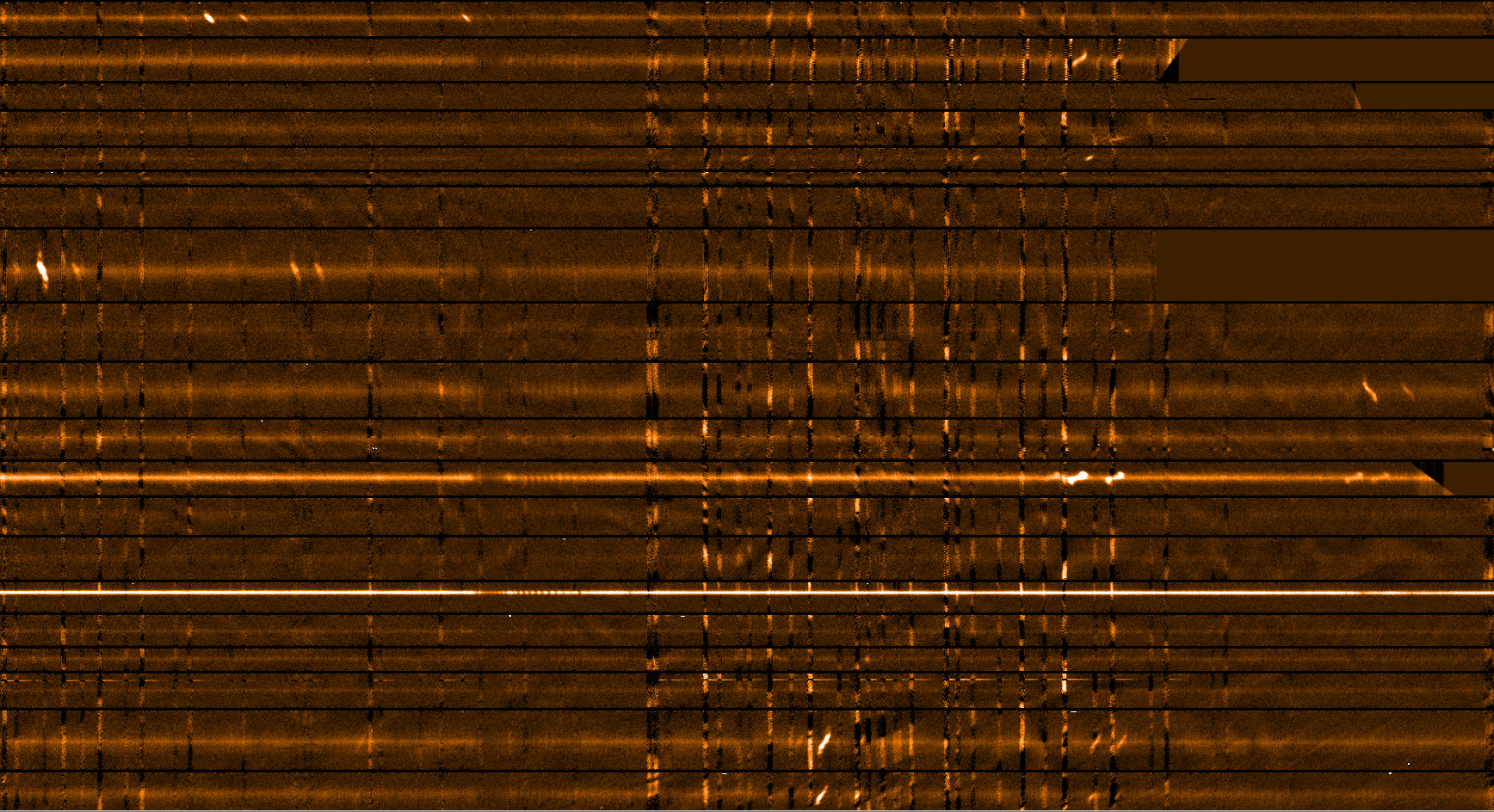}}\\
  \subfloat[SKY2D]{\label{main:e}\includegraphics[height= 9 cm]{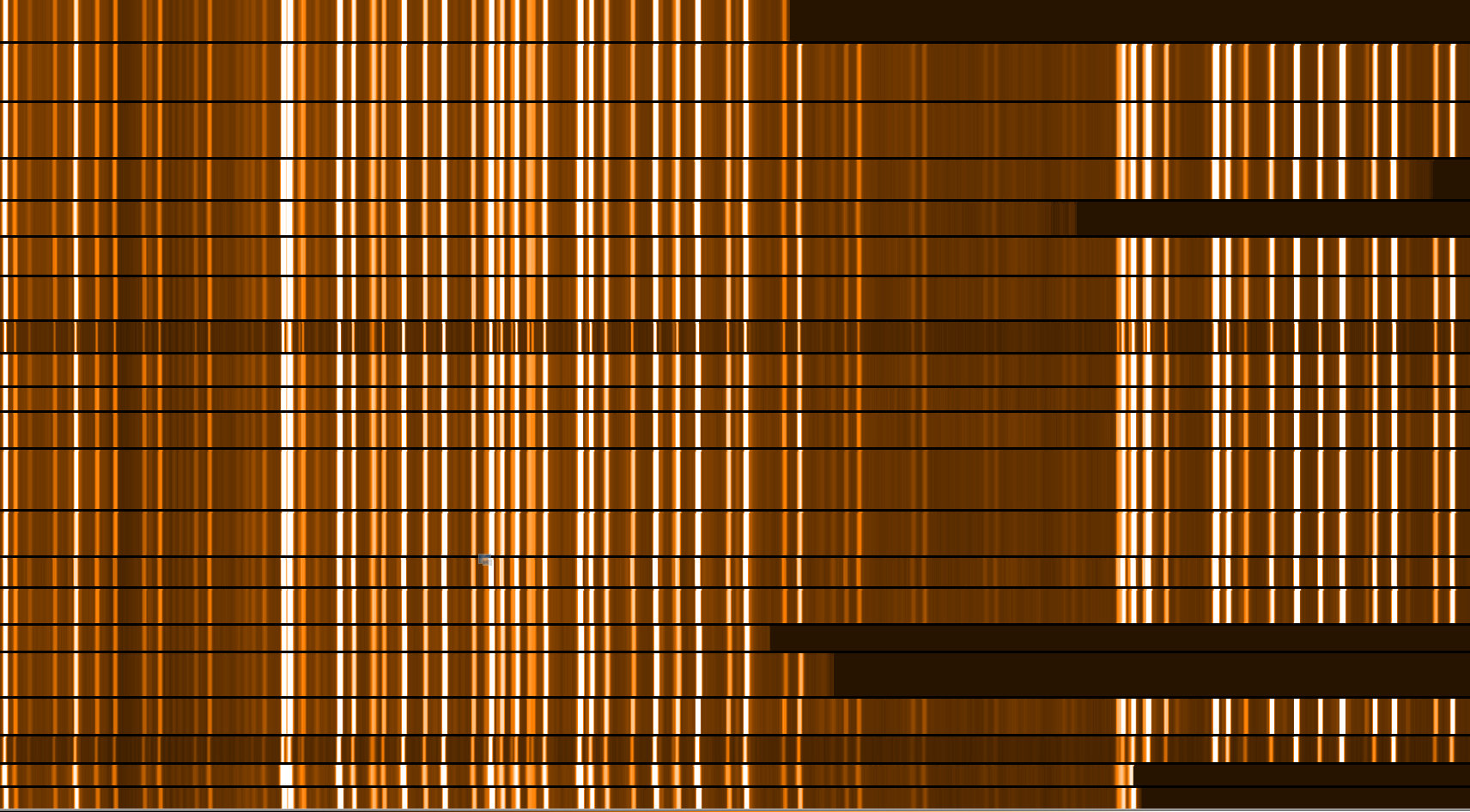}}
  \caption{\small An example of the \textit{EXR2D} and the \textit{SKY2D} extension of a \textit{seq}-file. Panel \textit{a} shows the reduced 2D spectra. Multiple bright emission lines can be seen as well as several sky lines. Panel \textit{b} shows the residual sky. The sky lines in the different slits coincide mostly very good which means that the data reduction has been done satisfactorily. It can be also seen very well how the spatial extensions of the slits differ depending on the redshift of the objects.}
   \label{fig:exr2d}
\end{figure}

\newpage
\noindent
A good tool to look at and analyse the combined \textit{seq}-file is \textbf{Show Slit Summary}. This feature contains several programs, for example the \textbf{Show Single Spectrum}-tool and the \textbf{Show Lambda Calibration}-tool that has already been described previously. The program \textbf{Show Single Spectrum} contains some very useful features. For one, it displays the 2D image of a certain slit upon the corresponding 1D spectrum. This makes it easier to get an impression of the quality of the combined file. Moreover, also the sky can be displayed below in order to see how well the background has been subtracted. It is also possible to zoom in and hence look at certain regions from the spectrum in more detail. Another very useful feature is the tool \textbf{Compute Redshift} which will be described in section \ref{redshift}.\\
Figure \ref{fig:slitsumm} in the following section shows an example for the \textbf{Show Single Spectrum}-tool.

\section{Determination of the redshift}
\label{redshift}
When astronomers are talking about redshift, it needs to be considered that they may refer to various types of redshift that all play an important role in astronomy: the cosmological redshift, the redshift due to the movement or rotation of objects and the gravitational redshift. The latter is not relevant for this master thesis and will not be described further, but the first two are both very important. This section is about the cosmological redshift. The redshift due to the rotation of galaxies and the Doppler effect will be described in chapter \ref{extractRC}.\\
The nature of the cosmological redshift was discovered in the 1920s by Edwin Hubble~\cite[383]{Weigert}. He examined the spectra of several galaxies with the objective of determining their radial velocities relative to our Galaxy. At that time it was expected that there would be a more or less uniform distribution of red- and blueshifted galaxies. However, Hubble found something completely different. Using the Doppler effect on characteristic spectral lines, he discovered that all galaxies were redshifted. As he had beforehand estimated the distances of those galaxies, he found furthermore that the redshift of the galaxies increased with distance. With other words: the farther away a galaxy is, the faster it seems to be "moving away". Today we know that this is due to the expansion of the universe which results in each point in the universe moving away from every other point. The consequence of this is that from our perspective, it seems that our Galaxy is in the center and all the other galaxies are moving away from us. This is of course not the case.\\ 
Hubble's discovery is expressed in the famous law, named after him:

\begin{equation}
v_{r}=\frac{\Delta\lambda}{\lambda_{0}}\cdot c=H_{0}\cdot d
\end{equation}

\noindent
Here $v_{r}$ is the radial velocity, $\lambda_{0}$ is the rest frame wavelength and $\Delta\lambda$ the shift of the observed relative to the rest frame wavelength. $c$ is the speed of light, $d$ the distance of the galaxy and $H_{0}$ is the so-called Hubble parameter which defines the rate of expansion and is today estimated to be $H_{0}\simeq 70 km s^{-1} Mpc^{-1}$. $\Delta\lambda/\lambda_{0}$ corresponds to the redshift \textit{z} of the galaxy. Figure \ref{fig:Hubblelaw} from Freedman et al. (2001) shows an example of Hubble's law~\cite{Freedman01}:

\begin{figure}[H]
	\centering
		\includegraphics[height=7 cm]{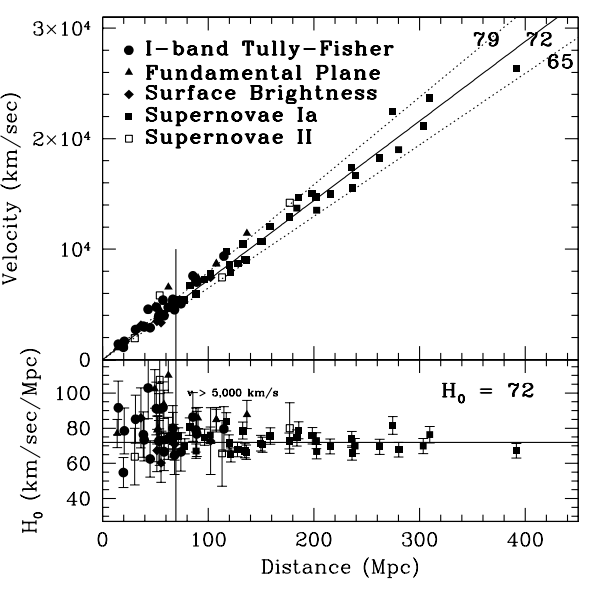}
	\caption{\small \textit{Top:} Hubble diagram of galaxies. The radial velocity $v_{r}=c\cdot z$ is plotted against the distance \textit{d}. The different symbols mark different methods for distance determinations: from Type I and Type II supernovae (filled and open squares, respectively), Tully-Fisher relation (filled cricles), the fundamental plane (triangles) and surface brightness fluctuations (diamonds). The lines represent the Hubble law $v_{r}=H_{0}\cdot d$ for three values of $H_{0}$. \textit{Bottom:} The value of $H_{0}$ as a function of distance~\cite{Freedman01}.}
	\label{fig:Hubblelaw}
\end{figure}

\begin{figure}[H]
  \centering
  \includegraphics[height= 8 cm]{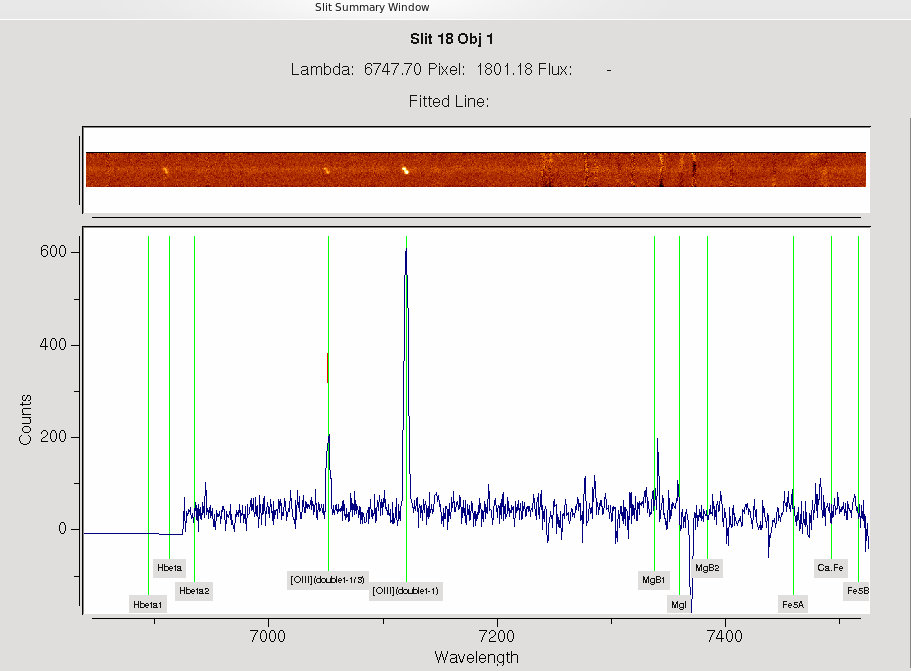}
  \caption{\small An example for the \textbf{Show Single Slit}-tool. Line positions expected at the computed redshift are plotted as green lines.}
   \label{fig:slitsumm}
\end{figure}

\noindent
In principle one could say that the Hubble law can be used as an ideal indicator for the distance: to determine the distance, only one value - the redshift - is needed~\cite[384-385]{Weigert}. Furthermore this value can be obtained easily from the spectrum of a galaxy. Of course in reality, it is not as easy as it seems. Especially in the local environment this relation cannot be applied because peculiar velocities of neighbouring galaxies, caused by mutual attraction, dominate over Hubble's law. A good example for this is the Andromeda galaxy which is moving with $\sim-130 kms^{-1}$ towards our Galaxy. As its radial velocity is negative, Hubble's law would imply a negative distance which is of course not possible. The escape velocity of a galaxy has to be around $v_{r}\geq3000kms^{-1}$ in order to clearly dominate over the local effects. This is the case at distances larger than 40 Mpc. However, for galaxies with $z\geq1$ Hubble's law would imply relativistic escape velocities and thus can not be applied anymore. But for galaxies with $z<1$ Hubble's law delivers a good estimation of the galaxy's distance.\\
Furthermore, the cosmological redshift of a galaxy contains information about the age of the galaxy~\cite[404]{Weigert}. The redshift gives us information about the time light takes to travel from a galaxy to us. Observing objects at high redshifts means looking into the past, and the redshift tells us how far into the past we are looking. This allows us to study the evolution of the universe over billions of years.\\
Summed up, the cosmological redshift yields fundamental informations about a galaxy. Thus it is crucial when dealing with extragalactic objects.\\
\newpage
\noindent
VIPGI offers an easy way to determine the redshifts of individual objects manually. This is done with the feature \textbf{Compute Redshift} from the tool \textbf{Show Single Spectrum}. To determine the redshift of a certain object, one has to choose one or more lines in the corresponding 1D spectrum. There are two possibilities to choose lines, either by "marking" or "fitting" them. After the lines have been selected, one has to click on \textbf{Compute Redshift} - here it is also possible to select which of the marked lines should be used for the computation. Depending on how many lines have been chosen, the program suggests several values for the redshift based on which emission or absorption lines they could be. If of course only one line has been chosen, multiple solutions will be proposed. In this case it is rather difficult to guess the correct emission line, and thus the correct redshift.\\
If several solutions are proposed, one can double-click on a certain redshift value, and after clicking on \textit{Show lines}, the expected locations of all the lines that should be visible at this redshift are plotted on the 1D spectrum. This makes it easy to verify if the redshift is correct by checking if there are really emission lines in the spectrum at the expected positions.\\
As the objects used in this thesis are zCOSMOS-galaxies, the spectroscopic redshifts for the majority of them were already available. For the ones without spectroscopic redshift information, the photometric redshifts from catalogues were used. As it is a good exercise to determine the redshift on one's own, I used the \textbf{Compute Redshift}-tool on the objects where I could identify emission lines. In the cases where only one emission line was visible, the provided spectroscopic and photometric redshifts helped to choose the correct line.\\
Figure \ref{fig:slitsumm} shows an example of the \textbf{Show Single Spectrum}-tool. The expected positions of emission lines from the computed redshift are plotted as green lines.

\newpage
\begin{table} [h!]
\begin{center}
\small
\begin{tabular}{l c c | l c c | l c c}
ID & $z_{computed}$ & $z_{zcosmos}$ & ID & $z_{computed}$ & $z_{zcosmos}$ & ID & $z_{computed}$ & $z_{zcosmos}$\\
\hline
700257 & 0.0789 & 0.0793 & 818788 & 0.335 & 0.3357 & 837598 & 0.6063 & 0.6063\\
700468 & 0.3315 & 0.3315 & 818816 & 0.5271 & 0.5409$^{a}$ & 837599 & 0.3758 & 0.3759\\
701290 & 0.8802 & 0.8813 & 819052 & 0.5987 & 0.5998 & 837603 & 0.2494 & 0.2499\\
810793 & 0.4325 & 0.4332 & 819177 & 0.3059 & / & 837607 & 0.1928 & 0.193\\
810929 & 0.3493 & 0.3502 & 819178 & 0.424 & 0.1242 & 837612 & 0.7485 & 0.7492\\
810934 & 0.4401 & 0.4332$^{a}$ & 830271 & 0.3497 & 0.3442$^{a}$ & 837613 & 0.8189 & 0.8205\\
810957 & 0.3112 & 0.3119$^{a}$ & 830319 & 0.5772 & 0.5874$^{a}$ & 837629 & 0.1934 & 0.1937\\
811012 & 0.8384 & 0.8386 & 830321 & 0.85 & 0.8512 & 837660 & 0.357 & 0.358\\
811102 & 0.4404 & 0.4408 & 830375 & 0.3728 & 0.3734 & 837661 & 0.6048 & 0.6047\\
811108 & 0.8958 & 0.8965 & 830408 & 0.2475 & 0.2476 & 837678 & 0.6168 & 0.606\\
811115 & 0.248 & 0.2483 & 830414 & 0.9238 & 0.9247 & 837768 & 0.3604 & 0.3611\\
811183 & 0.2221 & 0.2228 & 830461 & 0.2215 & 0.2217 & 837846 & 0.421 & 0.4218\\
811189 & 0.3549 & 0.3547 & 830462 & 0.7496 & 0.7499 & 837865 & 0.4221 & 0.4223\\
811224 & 0.9142 & 0.9171$^{a}$ & 830551 & 0.5182 & 0.519 & 837872 & 0.3738 & 0.3791$^{a}$\\
811233 & 1.0105 & 1.0116 & 830564 & 0.2917 & 0.2921 & 837887 & 0.2202 & 0.2202\\
811245 & 0.3908 & 0.3915 & 830569 & 0.7013 & 0.702 & 837889 & 0.3483 & 0.3483\\
812168 & 0.7419 & 0.6069 & 830590 & 0.6974 & 0.6868$^{a}$ & 837931 & 0.9022 & 0.0927\\
812243 & 0.4991 & 0.4995 & 830598 & 0.5188 & 0.5188 & 839042 & 0.2466 & 0.2472\\
812329 & 0.2195 & 0.2195 & 830618 & 0.7383 & 0.7389 & 839142 & 0.7498 & 0.7019$^{a}$\\
812388 & 0.6215 & 0.6226 & 830641 & 0.0928 & 0.0931 & 839178 & 0.6795 & 0.6621$^{a}$\\
812392 & 0.6704 & 0.6716 & 830731 & 0.741 & 0.7486$^{a}$ & 839193 & 0.8887 & 0.8819$^{a}$\\
812396 & 0.5839 & 0.5846 & 830732 & 0.3304 & 0.3306 & 839253 & 0.3322 & 0.3297$^{a}$\\
812429 & 0.2067 & 0.2064 & 830748 & 0.216 & 0.216 & 839295 & 0.5485 & 0.5493\\
812432 & 0.66 & 0.6611 & 830797 & 0.1238 & 0.1242 & 839304 & 0.6025 & 0.6024\\
812439 & 0.4235 & 0.4235$^{a}$ & 830822 & 0.1242 & 0.1242 & 839352 & 0.7331 & 0.7339\\
812455 & 0.7362 & 0.7362 & 837355 & 0.8259 & 0.8259 & 839379 & 0.7472 & 0.753$^{a}$\\
812457 & 0.1238 & 0.1243 & 837366 & 0.1127 & 0.1127 & 839384 & 0.6963 & 0.6966\\
812516 & 0.3706 & 0.3704 & 837367 & 0.2193 & 0.2193 & 839392 & 0.3478 & 0.3476\\
812519 & 0.4628 & 0.452$^{a}$ & 837379 & 0.2599 & 0.2597 & 839410 & 0.438 & 0.4379\\
812525 & 0.5387 & 0.5389 & 837433 & 0.7961 & 0.7961 & 839413 & 0.6752 & 0.6755\\
812636 & 0.3099 & 0.3106 & 837461 & 0.2199 & 0.2198 & 839423 & 0.3612 & 0.3606\\
812669 & 0.1855 & 0.1857 & 837485 & 0.4334 & 0.4339 & 839439 & 0.2495 & 0.2496\\
812675 & 0.0922 & 0.4726$^{a}$ & 837486 & 0.6164 & 0.6168 & 839440 & 0.7478 & 0.7471\\
817201 & 0.5298 & 0.5304 & 837487 & 0.3608 & 0.36 & 839451 & 0.3054 & 0.3058\\
817346 & 0.3434 & 0.3441 & 837491 & 0.9947 & 0.9944 & 839481 & 0.4292 & 0.4248$^{a}$\\
817351 & 0.1663 & 0.1608$^{a}$ & 837500 & 0.1298 & 0.1298 & 839599 & 0.3436 & 0.3369$^{a}$\\
817366 & 0.6408 & 0.6412 & 837511 & 0.2026 & 0.2001$^{a}$ & 839684 & 0.4388 & 0.3093\\
817640 & 0.9358 & 0.9363 & 837517 & 0.5999 & 0.5987 & 1470801 & 0.8008 & 0.7304\\
817647 & 0.219 & 0.214 & 837577 & 0.2746 & 0.275 & 1482419 & 0.3122 & /\\
\end{tabular}
\caption{117 objects for which it was possible to identify emission lines in the spectra and compute the redshift with VIPGI. The second column shows the computed redshift and the third column the spectroscopic or the photometric redshift$^{a}$, respectively.}
\label{tab:zvalues}
\end{center}
\end{table}

\noindent
Table \ref{tab:zvalues} shows the redshift values computed with VIPGI as well as the already provided spectroscopic and photometric redshifts for the objects where emission lines could be identified. Emission lines were found in 117 out of 160 spectra and used for the computation of the redshift. This means, for 43 objects it was not possible to identify emission lines, not even with the help of the provided zCOSMOS redshifts, as the signal for those objects was apparently too weak. For two of the 117 objects there was no redshift value available, neither from zCOSMOS nor from photometric measurements. However, it was possible to compute their redshifts with VIPGI. Therefore the redshift values of those two objects cannot be compared to already provided values. The computed redshifts are believed to be quite reliable though, because in both spectra three emission lines were clearly visible which makes the redshift computation quite unambiguous. Regarding the other 115 objects, one can see that in most cases the computed redshifts are rather similar to the spectroscopic and photometric redshifts. In the majority of cases the difference is smaller than 0.002. Table \ref{tab:zdiff} lists the objects for which the computed redshifts differ by more than 0.002 from the provided values:

\begin{table} [h!]
\begin{center}
\small
\begin{tabular}{l c}
ID & $\Delta z$ \\
\hline
810934 & 0.0069\\
812168 & 0.135\\
812519 & 0.0108\\
812675 & -0.3806\\
817351 & 0.0055\\
818816 & -0.0138\\
819178 & 0.2998\\
830271 & 0.0055\\
830319 & -0.0102\\
830590 & 0.0106\\
830731 & -0.0075\\
837511 & 0.0025\\
839142 & 0.0478\\
839178 & 0.0174\\
839193 & 0.0068\\
839253 & 0.0025\\
839379 & -0.0058\\
839599 & 0.0067\\
839684 & 0.1295\\
1470801 & 0.0704\\
\end{tabular}
\caption{20 objects for which the difference between the computed and the spectroscopic (or photometric) redshift is larger than 0.002.}
\label{tab:zdiff}
\end{center}
\end{table}

\noindent
Here, $\Delta z$ is defined as:

\begin{equation}
	\Delta z=z_{computed}-z_{cosmos}
\end{equation}

\noindent
A positive value of $\Delta z$ thus means that the redshift computed with VIPGI is higher than the spectroscopic redshift from zCOSMOS or the photometric redshift. One can see that for the majority of the objects in table \ref{tab:zdiff} this is the case. Only in five cases the computed redshift is smaller. Furthermore, in the case of almost half of the objects (9) the values of the computed and the provided redshifts differ only on the third and fourth decimal place. For four objects (812168, 812675, 819178, 839684) the difference is more considerable - the values differ already at the first decimal place. Another thing that stands out when looking at the two tables above is that for almost all objects with larger $\Delta z$ only less accurate photometric redshifts and no spectroscopic redshifts from zCOSMOS are available. Exceptions are 812168, 819178, 839684 and 1470801. This could explain the high $\Delta z$-values as photometric redshifts are less accurate than spectroscopic redshifts in most cases. However it is interesting that for three of the four objects with the largest $\Delta z$ spectroscopic redshifts are available. Why the difference between the zCOSMOS-redshifts and the computed redshifts of these objects is so large is not really clear. A possible reason could be the higher resolution of the spectra used in this master thesis.

\chapter{Extraction of rotation curves}
\label{extractRC}
For the extraction of RCs, the so-called Doppler effect, caused by the rotation of disk galaxies, plays a crucial role. The Doppler effect, named after the Austrian mathematician and physicist Christian Doppler, states that if a source of radiation is moving relative to an observer, the wavelength (or frequency) of the received radiation is changed compared to the wavelength (or frequency) of the emitted radiation~\cite[61]{Weigert}. In the case of a movement towards the observer the wavelength becomes smaller (\textit{blueshift}), if the source of radiation is moving away, the wavelength becomes larger (\textit{redshift}). This effect can be expressed with the following equation:

\begin{equation}
\frac{\Delta \lambda}{\lambda_{0}}=\frac{\lambda-\lambda_{0}}{\lambda_{0}}=\frac{v_{r}}{c}
\end{equation}

\noindent
Here, $\lambda_{0}$ and $\lambda$ are the rest frame and the measured wavelength, respectively, $\Delta \lambda$ is the difference between $\lambda$ and $\lambda_{0}$. $c$ is the speed of light and $v_{r}$ is the radial velocity with which the source of radiation is moving away from or towards the observer. However, this equation is only valid for the case $v\ll c$. If $v$ gets comparable to the speed of light, the relativistic Doppler effect has to be used:

\begin{equation}
\frac{\Delta \lambda}{\lambda_{0}}=\frac{\sqrt{1-(v_{r}/c)^{2}}}{1-(v_{r}/c)}-1
\end{equation}

\noindent
So by measuring $\lambda$ and knowing $\lambda_{0}$, it is possible to calculate how fast an object is approaching or moving away from us.\\

\noindent
Spiral galaxies are dominated by their rotating disk. This affects the emission lines in the spectrum of an object by redshifting one part and blueshifting the other part. If the cosmological redshift of an object - and thus its systemic velocity due to the expansion of the universe - is known, the Doppler effect can be used to convert these wavelength shifts due to the rotation into velocities and thus extract RC from the emission lines. In this master thesis, this has been done with a program written in the ESO system MIDAS.\\
Before describing what has been done in order to get the RCs, it is important to shortly explain which emission lines are most suitable for the extraction of RCs.\\
Emission lines in galaxy spectra stem from regions that contain excited interstellar gas~\cite[463]{Binney2}. There are several reasons for the excitation. Mostly the reason is the ionising radiation from a hot star or an AGN, but also strong shock waves can lead to the excitation of the gas by compressing it. Depending on what the interstellar gas is composed of, different emission lines can be observed. One of the best emission lines to trace the velocity field of a galaxy, is the emission line of atomic hydrogen HI, the 21 cm hyperfine transition emission line~\cite[32-35]{Bertin}. This emission line traces the distribution of the cold gas. Observations in the radio range, at which the 21 cm line can be observed, showed that the cold gas extends in general a lot farther than the stellar optical disk. Therefore with the HI line it is possible to get a RC that extends to larger radii than with optical emission lines.\\
In this master thesis optical emission lines are used for the RC extraction. The most prominent optical emission lines are usually the lines of ionised hydrogen, the Balmer lines, especially H$\alpha$ (656.3 nm), H$\beta$ (486.1 nm) and H$\gamma$ (434.1 nm)~\cite[463]{Binney2}. The origin of the Balmer lines are HII regions - star-forming regions where the ultraviolet radiation from hot young stars ionises the surrounding gas.\\
Besides the Balmer lines, the forbidden emission lines are also very important. The term forbidden comes from the fact that these lines were not known when they were observed for the first time. It was believed that the transitions causing these lines should not exist. Later it was discovered that they can occur if the particle density is low enough. As James Binney and Michael Merrifield write: "[...] in a low-density medium a forbidden transition can cause significant emission simply because once an ion is knocked into an excited state, it will hang there until it decays radiatively"~\cite[465]{Binney2}. Due to the low density, the time between the collisions of the particles is large enough so that the ions are not de-excited again. The forbidden lines that are most important and that also appear in the spectra used in this thesis, are the [OII]-doublet (372.7 nm), the two [OIII]-lines (495.9 and 500.7 nm), [NII] (658.3 nm) and the two [SII]-lines (671.7 and 673.1 nm). Forbidden lines trace especially emission nebulae and planetary nebulae.\\
As can be seen, different emission lines can trace different regions in a galaxy. Hence RCs derived from different emission lines can differ more or less.

\section{rcvimos}
\label{rcvimos}
For the extraction of RCs from emission lines, a MIDAS-program called \textit{rcvimos} was used. MIDAS is a package from the European Southern Observatory (ESO) that provides several tools for data reduction and image processing. This chapter describes the program in more detail: how it works and what it delivers as results.\\
To extract RCs from emission lines, two-dimensional (2D) spectra from the objects are needed as input. The 2D spectra were extracted from the reduced and combined \textit{seq}-file (see subsection \ref{reduction}) with the command \textit{cut2DSlit [name of seq-file] [new name of the cut 2D slit] [start of the slit] [end of the slit]}. The start and the end of the slit have to be written in pixel values.\\	
Before executing the program, two things have to be considered. First, a table called "default.tbl" has to be created and saved in the same directory as the \textit{rcvimos}-program before running the program for the first time. This table has to contain the following columns: x$\_$axis, y$\_$axis, line$\_$no, pixel$\_$no, x$\_$position, y$\_$position and pixel$\_$value. Furthermore, the command \textit{set/context long} has to be used each time after opening MIDAS. This command enables the use of commands and keywords pertaining to the reduction of long slit spectra~\cite{setcontext}.\\
Besides the extracted 2D slit three parameters are needed as input for the \textit{rcvimos}-program: the redshift of the object in this spectrum, the emission line from which the RC should be extracted and a boxcar value. The boxcar value defines the number of pixel rows that are averaged during the RC extraction process. The default value is 3 pixels.\\
In the program different numbers are attributed to different emission lines. In order to choose a certain emission line for the RC extraction, the corresponding number has to be inserted. Another important thing is that the program requires bdf-files as an input and not fits-files. To obtain the cut 2D slits as a bdf-file, the command \textit{indisk/fits [name of the fits-file] [name of the bdf-file]} in MIDAS was used. Finally, the program is run using the command \textit{@@rcvimos [name of the bdf-file] [redshift] [emission line] [boxcar]}.\\

\noindent
After starting the program, first the redshift-dependent expected position of the chosen emission line is calculated:\\

\begin{equation}
\lambda=\lambda_{0}\cdot (1+z)
\end{equation}

\noindent
Here \textit{z} is the object's redshift, and $\lambda_{0}$ and $\lambda$ are the rest frame wavelength and the expected redshifted wavelength of the selected emission line, respectively.\\
Taking the calculated spectral position of the emission line as the center, the program averages 100 columns (50 on each side of the center) in order to get a line profile as well as a part of the continuum along the spatial axis. After marking the edges of the line profile, the program derives the optical centre by fitting a Gaussian to the profile. If the centre is satisfactory, the emission line is subsequently fitted by the program row by row with a Gaussian. At this step the earlier mentioned boxcar comes into play. It defines how many neighbouring pixel rows are averaged at each step. E.g., in the case of the default value for the boxcar, always three rows are averaged. This is done to enhance the signal-to-noise ratio. If an emission line is rather faint, the boxcar can be enlarged to 5 rows. If a line is very strong, it is also possible to use only one row as the boxcar. It is recommended to not use a boxcar higher than 5 pixels because otherwise it is probable that the signal and thus the resulting RC are too smeared and imprecise.\\
The program starts at the derived optical centre - which is assumed to be also the kinematic centre - and fits the emission line row by row downwards. If no signal is found anymore, the procedure is repeated in the other direction. This yields blue- and redshifts along the spectral axis relative to the kinematic (optical) centre. By taking into account the systemic velocity of the object, these spectral shifts are converted into velocity shifts - that is observed line-of-sight rotation velocities - by means of the Doppler effect:\\

\begin{equation}
v_{r}=\frac{\lambda-\lambda_{0}}{\lambda_{0}}\cdot c
\end{equation}

\noindent
Here, $\lambda_{0}$ is not the rest frame wavelength of the emission line, but the already shifted wavelength due to the cosmological redshift of the object.\\

\noindent
In the original program six emission lines could be used for the extraction of RCs: H$\alpha$, H$\beta$, H$\gamma$, [OII] and the two [OIII]-lines. As in some of the spectra used in this thesis other emission lines were quite strong as well, the program was extended for four more emission lines: H$\delta$, [NII] and the two [SII]-lines. It should also be mentioned that for the [OII]-doublet a slightly changed program was executed that uses two Gaussians instead of one for the row-by-row fits.

\newpage
\begin{figure} [H]
\centering
\subfloat[Line profile along the spatial axis]{\label{main:f}\includegraphics[height= 7 cm]{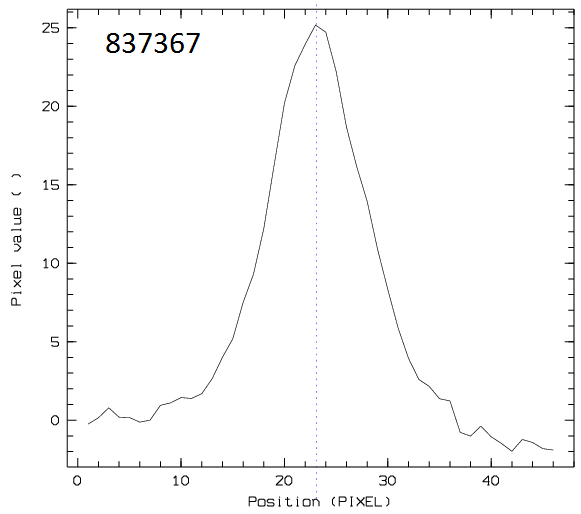}}
\subfloat[Two-Gaussian-fit for a OII-line]{\label{main:g}\includegraphics[height= 7 cm]{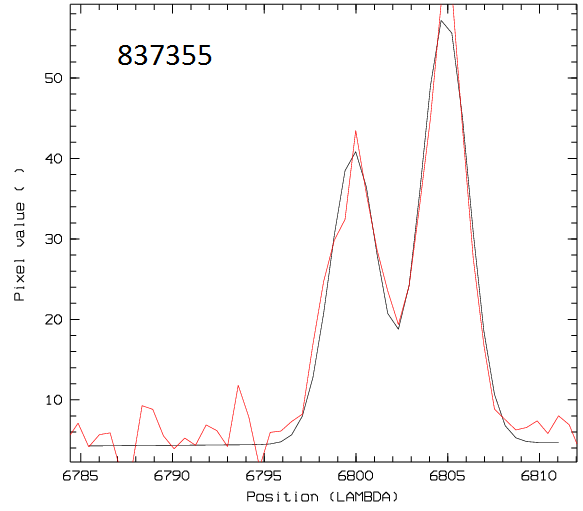}}\\
\subfloat[Extracted observed RC]{\label{main:h}\includegraphics[height= 7 cm]{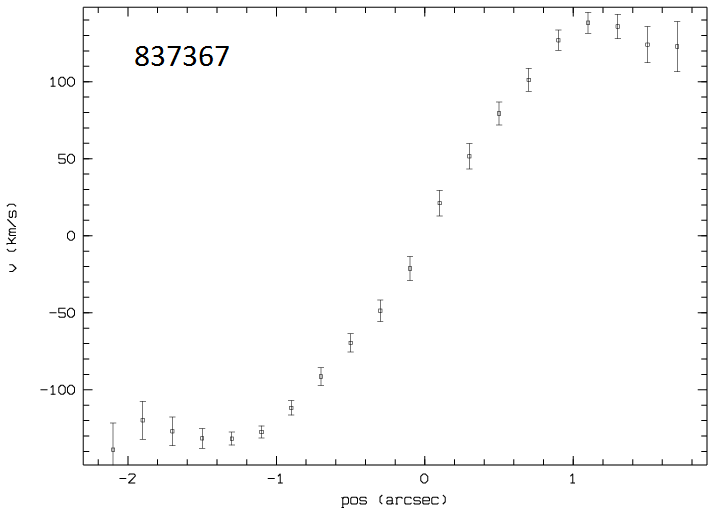}}
\caption{\small Examples for different stages of the rcvimos-program: the averaged line profile along the spatial axis with the marked optical centre (dotted vertical line), a two-Gaussian-fit for a [OII]-line and an extracted observed RC. Panels \textit{a} and \textit{c} are from the object 837367 (z=0.2193), panel \textit{b} is from the object 837355 (z=0.8248).}
\label{fig:rcvimos}
\end{figure}

\noindent
Figure \ref{fig:rcvimos} shows an example for the starting configuration, that is the line profile after the averaging of 100 columns of the spectra, with the marked optical centre (a), an example for two Gaussians for a [OII]-line (b) and an example for an extracted observed RC (c). The first and the third plots are both from the object 837367 (z=0.2193), the second plot is from the object 837355 (z=0.8248). The difference between the x-axes of the first two figures should be noted (pixels along the spatial axis vs. wavelength).\\

\begin{figure}[H]
  \centering
  \includegraphics[height= 7.5 cm]{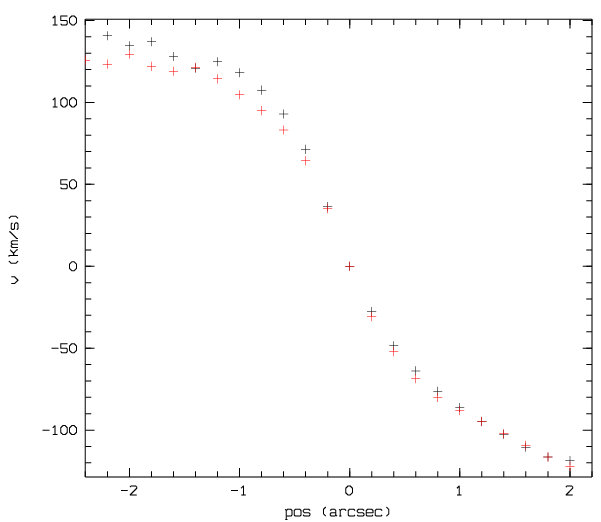}
  \caption{\small Two observed RCs of the object 839439 (z=0.2495), extracted with a boxcar of 3 pixels (black) and a boxcar of 5 pixels (red).}
   \label{fig:boxcar}
\end{figure}

\noindent
The program for the RC extraction was executed for 110 out of 160 objects. Regarding the remaining 50 objects, in 40 spectra no emission lines were found, and in the other 10 cases the emission lines were either too weak, too compact and/or directly behind a sky line which made the extraction of a RC impossible.\\
It should be mentioned that the \textit{rcvimos}-program is very sensitive to the redshift. If the redshift value is not accurate enough, the program either does not find the emission line at all or stops after a few iterations. To ensure that the redshifts are accurate enough, the values computed with VIPGI (see section \ref{redshift}) were used because they were computed directly from the spectra. Still, in the majority of cases the program was run several times with slighlty changed redshifts in order to find the redshift value at which the program works best and provides the best RC. Furthermore, for every object different boxcars were applied to see to which extent the resulting RCs differ. In many cases a boxcar of three pixel rows produced nice looking RCs that were not really different from the curves produced by a boxcar of five pixels.\\
Figure \ref{fig:boxcar} shows an example of the RCs of the object 839439 (z=0.2495) with boxcar=3 (black) and 5 (red). It can be seen that in this case both RCs are very similar. There are small differences on the receding side of the RC (positive velocities) which is supposably due to the averaging of different numbers of pixels.

\noindent
If the difference between the RCs extracted with boxcar=3 and boxcar=5 is as small as in the upper case, the one with boxcar=3 should be used for further analysis. However, in several cases averaging 3 pixels was not enough because the emission line was rather weak.\\

\noindent
Table \ref{tab:number_slits} shows a summary of the number of slits per mask in which emission lines could be used to extract RCs. It also lists in how many slits the lines were too weak, too compact or behind a sky line as well as in how many slits no emission lines were found at all.

\newpage

\begin{table} [h!]
\begin{center}
\small
	\begin{tabular}{ c | c  c  c  c  c }

& P4Q1 & P6Q1 & P7Q1 & P4Q2 & P5Q2 \\ \hline
Slits with extracted RCs & 24 & 20 & 21 & 25 & 20 \\
EL too weak/compact/behind SL & 2 & / & 2 & 3 & 3 \\
No EL found & 9 & 9 & 7 & 5 & 10 \\ \hline
Total & 35 & 29 & 30 & 33 & 33\\
	\end{tabular}
	\caption{\small Number of slits per mask in which emission lines were used for the RC extraction as well as number of slits with emission lines that were too weak, too compact or/and behind sky lines and number of slits with no emission lines found at all.}
	\label{tab:number_slits}
\end{center}
\end{table}

\begin{figure}[H]
  \centering
  \subfloat[812432]{\label{main:i}\includegraphics[width= 8 cm]{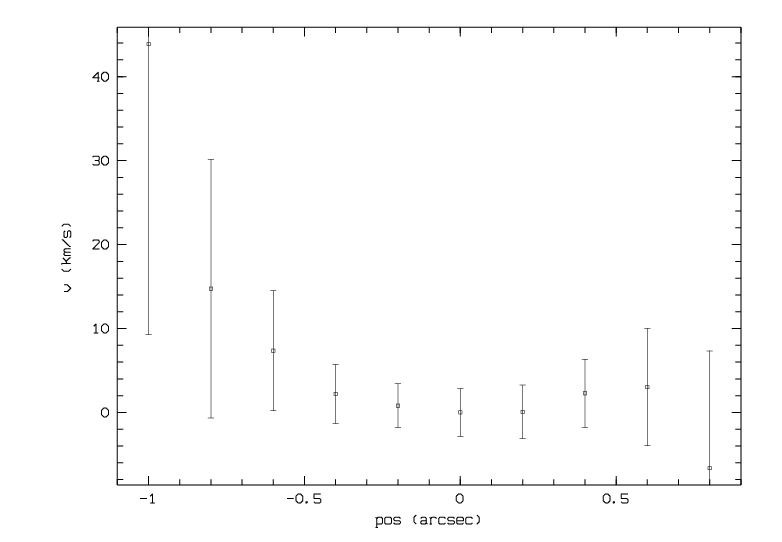}}
  \subfloat[819178]{\label{main:j}\includegraphics[width= 8 cm]{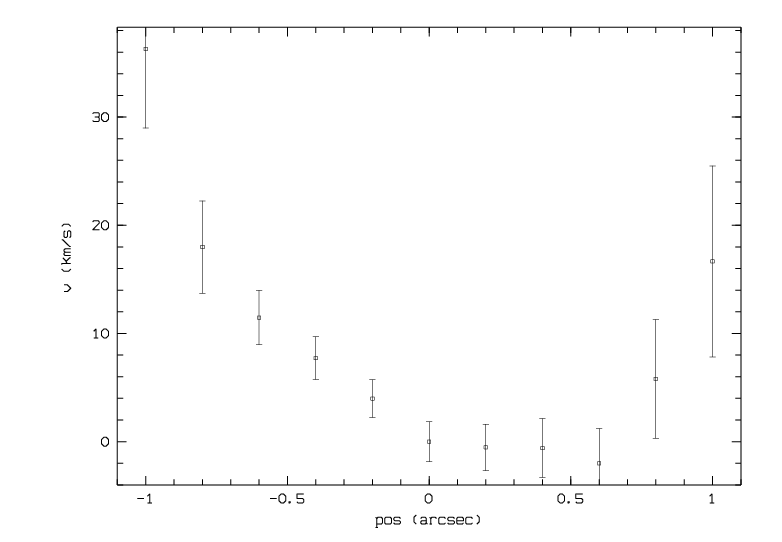}}\\
  \subfloat[OIII]{\label{main:k}\includegraphics[width= 7.5 cm]{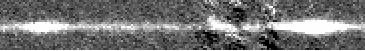}}
   \subfloat[H$\beta$]{\label{main:l}\includegraphics[width= 7.5 cm]{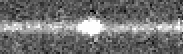}}
  \caption{\small Two examples for extracted RCs that do not look symmetric: the upper two panels show the RCs for the objects 812432 (z=0.66) (\textit{a}) and 819178 (z=0.424) (\textit{b}). The lower two panels show the corresponding emission lines that were used for the extraction: [OIII](2.) (\textit{c}) and H$\beta$ (\textit{d}).}
   \label{fig:badobj}
\end{figure}

\noindent
It can be seen that in each mask there are slits in which no emission lines are found. In percentage, the masks with the most "empty" slits are P6Q1 and P5Q2 - here, with 31.03$\%$ and 30.3$\%$ respectively, in almost a third of the slits no emission lines were found. In contrast, in P4Q2 only in 15.15$\%$ of the slits no emission lines could be found. Of course, these slits are not really empty. Spectra of objects were recorded, but obviously the signal was too low, so that no emission lines could be found, not even with the help of the provided zCOSMOS-redshifts. Moreover, in every mask, except for P6Q1, there were 2 or 3 slits in which emission lines were found, however they were too weak and/or too compact, or located directly behind a sky line, so that it was not possible to extract a RC.\\
Although it was possible to derive RCs from 110 objects, this does not mean that the extracted curves are all usable. In some cases the signal was too low or the lines were too compact, and the resulting RCs did not look symmetric as would be expected from regular rotating spiral galaxies. For example, sometimes the RCs looked kind of parabolic in shape which is probably due to the compact form of the emission lines. Regarding the number of objects with emission lines that did not produce usable RCs, more will be said in chapter \ref{simulations}.\\
Figure \ref{fig:badobj} shows two examples for not usable RCs. The RCs of two objects, 812432 (z=0.66) and 819178 (z=0.424), are shown as well as the corresponding emission lines that were used for the RC extraction. The emission lines that were used are [OIII](2.) for 812432 and H$\beta$ for 819178. In both cases the emission lines are very compact. In the case of 812432, the two [OIII]-lines visible are extended horizontally, but not vertically. This is of course a problem as the program fits Gaussians row by row to get a RC and this requires a sufficient vertical extension of the emission line. Although in the case of both objects the program gets measurements for 10 and 11 pixels, respectively, the resulting RCs are far from symmetric. There is also no distinction between a receding and an approaching side. The RC of 812432 is almost completely flat, with a slight rise on the left side. But also the error bars get larger on the left side. The RC of 819178 has a somewhat parabolic shape. In both cases the velocity values are rather low, ranging only to about 40 km/s.
\newpage

\noindent
In many cases more than one emission line was visible in the spectra. In these cases it was tried to extract RCs from as many emission lines as possible. Comparing the form and the velocity range covered by the different RCs is a good way to test their quality and reliability. However, as already mentioned, one should always keep in mind that different emission lines can trace different regions of the galaxy.\\
The following two tables show in how many slits how many emission lines per slit were used for the RC extraction (\ref{tab:elperslit}) and which emission lines were used how often (\ref{tab:elfreq}):

\begin{table} [h!]
\begin{center}
\small
	\begin{tabular}{ c | c  c  c  c  c | c}

 &  &  & $\#$ Slits &  &  & \\ 
$\#$ EL & P4Q1 & P6Q1 & P7Q1 & P4Q2 & P5Q2 & Total\\ \hline
1 & 14 & 15 & 11 & 9 & 9 & 58\\
2 & 4 & 2 & 6 & 10 & 6 & 28\\
3 & 6 & 2 & 3 & 5 & 4 & 20\\
4 & 0 & 1 & 1 & 1 & 0 & 3\\
5 & 0 & 0 & 0 & 0 & 1 & 1\\
	\end{tabular}
	\caption{\small Number of slits per mask in which 1, 2, 3, 4 or 5 emission lines were used for the RC extraction.}
	\label{tab:elperslit}
\end{center}
\end{table}

\begin{table} [h!]
\begin{center}
\small
    \begin{tabular}{ c | c  c  c  c  c | c}
EL & P4Q1 & P6Q1 & P7Q1 & P4Q2 & P5Q2 & Total\\ \hline
H$\alpha$ & 9 & 6 & 5 & 9 & 2 & 31\\
H$\beta$ & 7 & 6 & 10 & 13 & 11 & 47\\
H$\gamma$ & 0 & 1 & 1 & 0 & 0 & 2\\
OIII(1.) & 4 & 3 & 4 & 2 & 4 & 17\\
OIII(2.) & 9 & 8 & 13 & 14 & 10 & 54\\
OII & 9 & 6 & 2 & 7 & 7 & 31\\
NII(1.) & 0 & 0 & 0 & 0 & 1 & 1\\
NII(2.) & 0 & 0 & 0 & 0 & 1 & 1\\
SII(1.) & 2 & 0 & 1 & 2 & 0 & 5\\
    \end{tabular}
    \caption{\small Frequency of individual emission lines used for the RC extraction.}
    \label{tab:elfreq}
\end{center}
\end{table}

\begin{figure}[H]
  \centering
  \includegraphics[height= 6.5 cm]{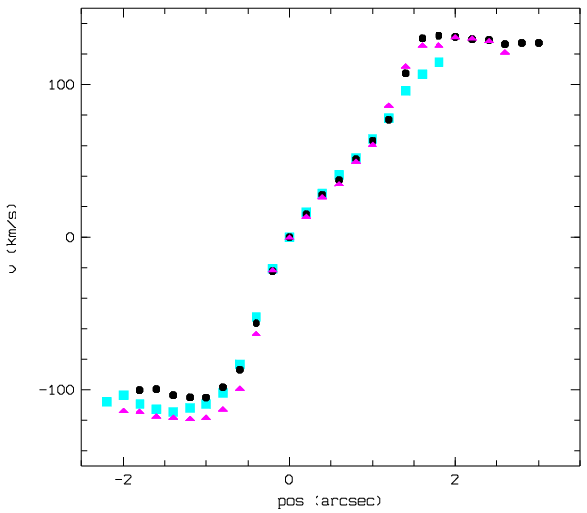}
  \caption{\small Three RCs of the object 812439 (z=0.42357) extracted from three different emission lines: H$\beta$ (cyan), [OIII](1.) (magenta) and [OIII](2.) (black).}
   \label{fig:diff_el}
\end{figure}

\begin{figure}[H]
  \centering
  \subfloat[] {\label{main:m}\includegraphics[height= 6.5 cm]{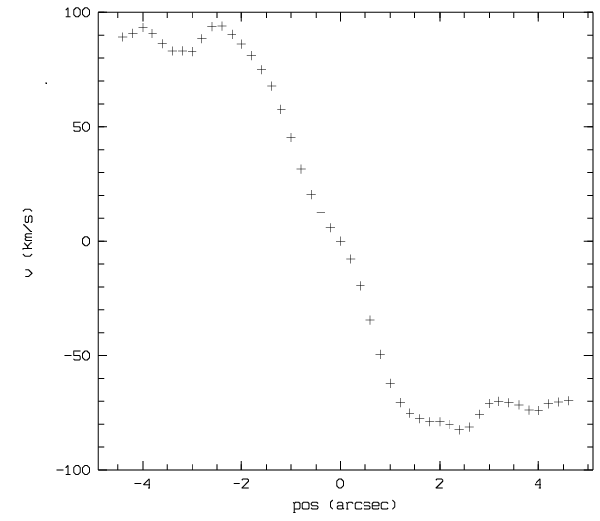}}\\
  \subfloat[] {\label{main:n}\includegraphics[width= 3.5 cm]{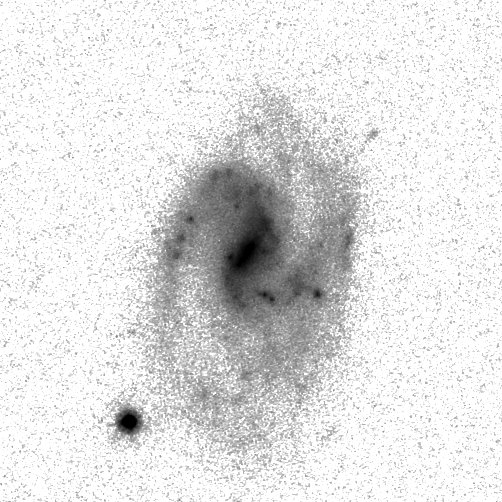}}
  \subfloat[] {\label{main:o}\includegraphics[width= 5 cm]{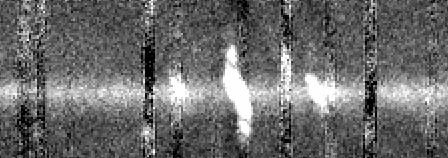}}
  \caption{\small The RC of 837366 shows rather strong bumps on both sides (\textit{a}). These features could reflect the spiral arms that can be seen on the Hubble ACS picture (\textit{b}). These "bumps" can also be seen at both ends of the H$\alpha$-line (\textit{c}).}
   \label{fig:rc_features}
\end{figure}

\newpage
\noindent
The upper two tables show only how many emission lines were used for the RC extraction, not how many lines were visible. In the majority of the slits more than one emission line were visible, but could not be used due to the reasons mentioned above.\\
Looking at the tables more closely, it can be seen that in most cases only one emission line was used for the RC extraction ($\sim 52.73\%$). In one fourth of the slits ($\sim 25.45\%$) two emission lines and in $\sim 18.18\%$ three emission lines were bright enough in order to be used for the extraction. In 3 slits the rcvimos-program was applied on four emission lines and in one slit even on five emission lines.\\
Table \ref{tab:elfreq} shows that the emission lines used most often are [OIII](2.), H$\beta$, [OII] and H$\alpha$, which are nearly always the brightest emission lines. Of course it depends on the redshift which emission lines are visible. In the lower redshift-regime up to z$\sim$0.30, H$\alpha$ is the brightest emission line, whereas between z=0.30-0.70 in the majority of cases H$\beta$ and the two [OIII]-lines can be used. For redshifts higher than z$\sim$0.70 the [OII]-line is mostly the only visible emission line. However, this subdivision is not completely strict - there is a smooth transition regarding the emission lines that are visible at a certain redshift.\\
Figure \ref{fig:diff_el} shows an example for three RCs of the object 812439 (z=0.42357), derived from different emission lines: H$\beta$ (cyan), [OIII](1.) (magenta) and [OIII](2.) (black). It can be seen that in this case the three RCs are very similar, especially in the central part. Only in the outer areas there are slight deviations, especially on the approaching side of the RC. On the receding side, the data points are very similar, particularly for the two [OIII]-lines. The H$\beta$-RC is somewhat less steep on the right end than the curves from the other two lines, and stops before getting flat. On the approaching side all three RCs have a very similar form, just with a slightly displaced flat part. Here the [OIII](2.)-line has the lowest and the [OIII](1.)-line the highest velocities. The difference is not significantly high, but at some points still around 20 km/s. This effects of course the value of the maximum velocity and thus the position of the object in the Tully-Fisher and velocity-size diagrams.\\
In some cases the observed RCs showed interesting features like bumps or dips. One example can be seen in figure \ref{fig:rc_features} which shows the RC of 837366. The RC looks symmetric and regular, but in the flat region quite strong bumps can be seen. With a redshift of z=0.1127 this object is quite near, so one possibility is that the shape of the RC reflects the spiral arms of the galaxy that can be seen very nice on the Hubble ACS picture of the object.
\newpage

\chapter{Modelling of rotation curves}
\label{simulations}
The observed RCs extracted from emission lines, as described in the previous chapter, trace only the line-of-sight rotation velocities of galaxies. This projected velocity does not correspond to the intrinsic rotation of a galaxy. This means that the maximum rotation velocity (from now on indicated as $v_{max}$) cannot be read directly from the outer parts of the extracted RC. There are several observational and geometrical effects that have to be taken into account.\\
The first important factor that has to be considered is the inclination \textit{i} of a galaxy with respect to the line-of-sight. Only in the case of a galaxy seen egde-on  the observed RC would be the same as the intrinsic RC (under the assumption that the galaxy is observed with an infinitely thin slit and at infinitely large spatial as well as spectral resolution). If $i\neq90^{\circ}$ and the inclination would not be taken into account, this would lead to an underestimation of $v_{max}$.\\
Also the position angle of the apparent major axis with regard to the slit alignment plays an important role and must be considered. Furthermore, a very crucial factor is the effect of the slit width, especially in the case of distant and thus apparently small galaxies. Assuming a typical scale length for $L^{*}$-galaxies, that is $\sim$3 kpc, and a slit width of 1", a galaxy's size at z=0.5 is almost the same as the slit width (3 kpc $\simeq 0.5$")~\cite{Boehm04}. At higher redshifts galaxies are understandably even apparently smaller than the slit. So in the case of distant galaxies, any velocity value of the observed RC is an integration perpendicular to the spatial axis, that is the slit direction, and this can lead to an effect that is similar to the "beam smearing"-effect in radio observations. This effect, as well as the seeing which is typically also in the order of 1", produce blurred RCs. If they are not taken into account, this can again lead to an underestimation of the intrinsic $v_{rot}$ values and thus to an underestimation of $v_{max}$.\\
In order to overcome this problem, a program called \textit{vel.py} was used to generate synthetic RCs by incorporating all the geometrical and observational effects mentioned above. This program will be described in more detail in the following section.

\section{Determination of the maximum rotation velocity}
\label{velpy}
\textit{vel.py} is a Python program written by Benjamin B{\"o}sch~\cite{Boesch13}. As already explained above, the observed RC of a galaxy derived from emission lines does not represent yet its intrinsic rotation. The reason are various observational and geometrical effects which have an impact on the shape of the observed RC. The purpose of \textit{vel.py} is to infer the intrinsic RC by taking into account these effects. This is done by producing a synthetic RC which is subsequently compared with the observed RC. To obtain the synthetic RC, the galaxy is modelled as an infinitely thin rotating disc with an exponential intensity profile~\cite{Boesch13}. Furthermore, the spectroscopic observation of the galaxy is simulated by taking into account the inclination of the galaxy, the slit alignment, the seeing and the slit width. By finally comparing the resulting snythetic with the observed RC, the true intrinsic RC parameters, i.e. the maximum velocity $v_{max}$ and the turnover radius $r_{t}$, are derived from the best-fitting simulated RC. In the following, the procedure of the program will be described briefly.\\
To run \textit{vel.py} and simulate a synthetic RC, the following files and parameters are required as input: a fits-file containing the table with the information obtained from the RC extraction (see Chapter \ref{extractRC}), the scale length of emitting gas $r_{d,gas}$ as an estimator for the turnover radius $r_{t}$ (in kpc), the redshift \textit{z}, the slit tilt angle $\theta$ and the inclination \textit{i} (both in degrees) as well as the full width half maximum (FWHM) of the seeing disc (in arcsec).\\
The required fits-file contains the following columns: the y-position (in pixel), the measured center of the line profile per pixel row and its error (in \AA), the measured FWHM and its error (in \AA), the measured rotation velocity \textit{v} and its error (in km/s), and the position (in arcsec). Table \ref{tab:fits_tab} shows an example of how such a table looks like:

\begin{table} [h!]
\begin{center}
\small
    \begin{tabular}{ c  c  c  c  c  c  c  c}
y [px] & center [\AA] & fwhm [\AA] & center$\_$err [\AA] & fwhm$\_$err [\AA] & v [km/s] & v$\_$err [km/s] & pos ["]\\ \hline
18 & 8570.001 & 3.873 & 0.170 & 0.400 & +65.9 & 5.9 & -1.0\\
19 & 8569.955 & 3.685 & 0.102 & 0.240 & +64.3 & 3.6 & -0.8\\
20 & 8569.749 & 4.001 & 0.066 & 0.154 & +57.1 & 2.3 & -0.6\\
21 & 8569.281 & 4.482 & 0.045 & 0.106 & +40.8 & 1.6 & -0.4\\
22 & 8568.688 & 4.651 & 0.032 & 0.075 & +20.1 & 1.1 & -0.2\\
23 & 8568.112 & 4.499 & 0.036 & 0.084 & +0.0 & 1.2 & +0.0\\
24 & 8567.521 & 4.126 & 0.048 & 0.113 & -20.7 & 1.7 & +0.2\\
25 & 8567.013 & 3.833 & 0.057 & 0.134 & -38.4 & 2.0 & +0.4\\
26 & 8566.551 & 3.520 & 0.055 & 0.129 & -54.5 & 1.9 & +0.6\\
27 & 8566.268 & 3.655 & 0.085 & 0.200 & -64.4 & 3.0 & +0.8\\
28 & 8566.173 & 4.607 & 0.202 & 0.475 & -67.7 & 7.0 & +1.0\\
    \end{tabular}
    \caption{\small Example of a table required as input for the \textit{vel.py}-program.}
    \label{tab:fits_tab}
\end{center}
\end{table}

\noindent
Furthermore, the direction of the rotation has to be indicated by specifying which side of the RC is receding and which is approaching. Also an initial value for $v_{max}$ has to be inserted. This value serves just as a rough first guess for further calculations, and does not really effect the final result. Whether the value is 1 km/s or 300 km/s, the resulting $v_{max}$-values differ only at the third or fourth decimal place. For that reason an initial value of 300 km/s was retained in the following for all objects.\\
Two further parameters that were kept fixed for all objects are the slit width (1") and the pixel scale (0.205"/px for VIMOS). The last parameter that had to be changed for every object is the boxcar. Here the same number had to be inserted as for the RC extraction.\\
The required input parameters, especially the structural parameters, were obtained mainly from the zCOSMOS-catalogue. The gas scale length $r_{d,gas}$ was calculated from the available effective radius $r_{1/2}$, taken from Sargent et al. (2007)~\cite{Sargent07}, with the following equation from B{\"o}hm et al. (2004)~\cite{Boehm04}:

\begin{equation}
r_{d,gas}=(2-\frac{z}{2})\cdot r_{d}=(2-\frac{z}{2})\cdot \frac{r_{1/2}}{1.7}
\end{equation} 

\newpage
\noindent
Here, $r_{d}$ is the scale length derived from continuum emission. Assuming an exponential profile and n=1, the relation $r_{d}\sim \frac{r_{1/2}}{1.7}$ is valid~\cite{ned}. \textit{z} is the cosmological redshift of the object.\\
The position angle of the apparent major axis and the inclination were measured from ACS-HST-images and provided by Laurence Tresse. The position angle was already given in degrees, whereas the inclination was calculated from the available ratio between the apparent minor and the apparent major axis \textit{b/a} with the following equation from Tully et al. 1998~\cite{Tully98}:

\begin{equation}
cos(i)=\sqrt{\frac{(b/a)^{2}-r_{0}^{2}}{1-r_{0}^{2}}}
\end{equation}

\noindent
Here, \textit{i} is the inclination, \textit{a} and \textit{b} are the apparent major and apparent minor axis, respectively, and $r_{0}$ is the axial ratio of a system viewed edge-on. It is assumed to be 0.2 and constant for all objects.\\
For the redshifts the same values as for the RC extraction were used, i.e. the values computed with VIPGI (see section \ref{redshift}). For the FWHM of the seeing disc the mean seeing values from table \ref{tab:seeing} were used.\\
An important point is that the spectra used in this thesis were observed with tilted slits (see Chapter \ref{data}), so it had to be taken into account that there is no deviation angle between the apparent major axis of the disk and the slit direction. For the modelling of the RCs a cosmology with $\Omega_{m}$=0.3, $\Omega_{\Lambda}$=0.7 and $H_{0}=70$ $km s^{-1} Mpc^{-1}$ is assumed.\\
The first thing that is needed in order to simulate a RC, is to assume an intrinsic rotation law $v^{int}_{rot}(r)$ which describes the shape of the RC. There are several possibilities for this. The intrinsic rotation law used in this program is defined by a linear rise of the rotation velocity at small radii, followed by a turning over into a region of constant velocity. In this flat region the dark matter halo dominates the mass distribution~\cite{Boehm04}. To describe this rotation law, the following parameterisation was used (e.g. Courteau 1997~\cite{Courteau97}):

\begin{equation}
\label{intr_rc}
v^{int}_{rot}(r)=\frac{v_{max}\cdot r}{(r^{a}+r^{a}_{t})^{1/a}}
\end{equation}

\noindent
Here, $r_{t}$ corresponds to the turnover radius and \textit{a} is a factor that defines the sharpness of the turnover of the RC at this radius. For $r\gg r_{t}$, the rotation velocity remains constant with $v^{int}_{rot}(r)=v_{max}$. As for the factor \textit{a}, it is kept fixed for all objects to minimize the number of free parameters. B{\"o}hm et al. (2004) tested different values of \textit{a} and found that for distant galaxies the shape of the turnover region is best reproduced by a value of $a=5$~\cite{Boehm04}. Hence, this value is applied also for the modelling of RCs in this work. An example for RCs with different \textit{a}-values can be seen in figure \ref{fig:diff_a}~\cite{Asmusthesis}:

\newpage
\begin{figure}[H]
  \centering
  \includegraphics[height= 8.5 cm]{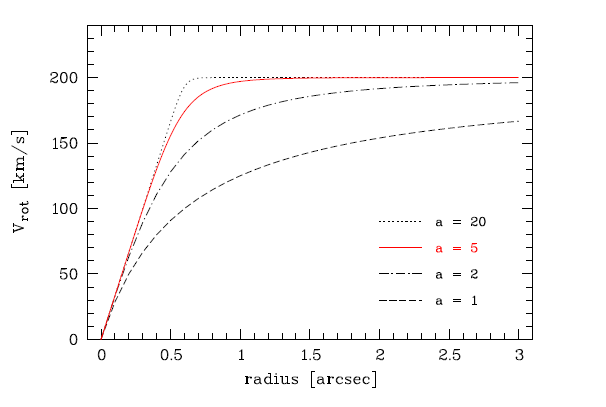}
  \caption{\small Intrinsic RCs with a rise-turnover-flat shape and different values for the parameter \textit{a} in equation \ref{intr_rc}~\cite{Asmusthesis}.}
   \label{fig:diff_a}
\end{figure}

\begin{figure}[H]
  \centering
  \subfloat[] {\label{main:p}\includegraphics[height= 6 cm]{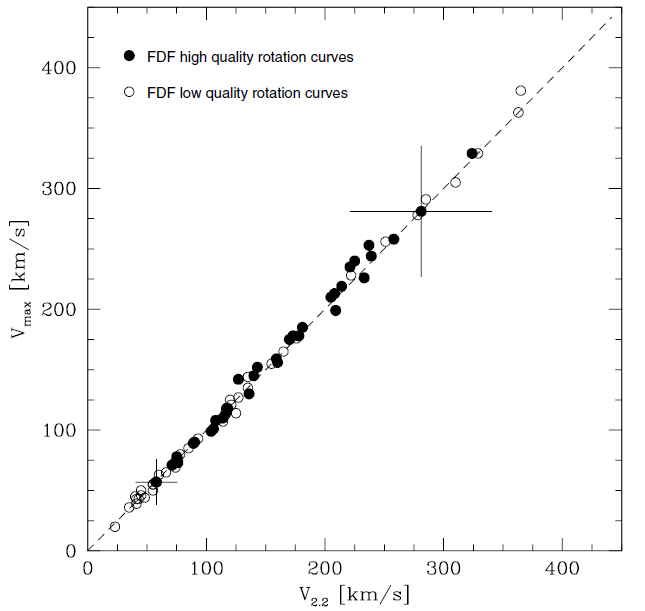}}
  \subfloat[] {\label{main:q}\includegraphics[height= 6 cm]{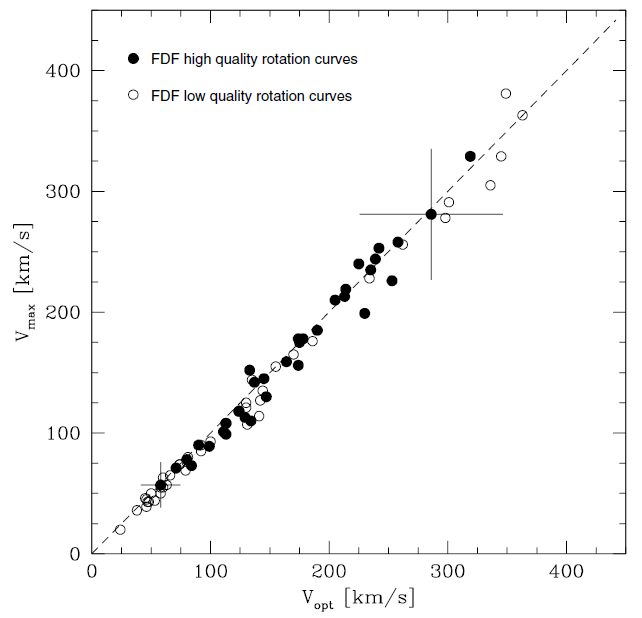}}
  \caption{\small Comparison between RC fitting with a rise-turnover-flat shape (y axis) with the URC91 shape (\textit{a}) and the URC96 shape (\textit{b}), respectively. In both figures, typical error bars are shown for two objects~\cite{Boehm04}.}
   \label{fig:vmax_comp}
\end{figure}

\noindent
Two other possibilities for the intrinsic rotation law should be explained briefly. Both are from the Universal Rotation Curve (URC) framework, the first of Persic \& Salucci (1991) (~\cite{Persic91}, URC91) and Persic et al. (1996) (~\cite{Persic96}, URC96). The difference between these two intrinsic RC shapes and the RC described by eq. \ref{intr_rc} is that the first two incorporate the dependence of the RC morphology on luminosity~\cite{Boehm04}. Regarding the URC shapes, not all spiral galaxies have constant rotation velocities at large radii. Only in the case of $L^{*}$-galaxies the RC remains flat in the outer regions. However, for sub-$L^{*}$ spiral galaxies the RCs rise even beyond a certain characteristic radius, whereas in the case of objects with luminosities higher than $L^{*}$, the rotation velocity decreases again at large radii. This is valid for the URC91 as well as the URC96 form. The only difference between them is the definition of the characteristic radius that defines the characteristic rotation velocity that can later on be used for the scaling relations. For the URC91 form, the characteristic radius is defined to be 2.2 scale lengths ($r_{2.2}$), whereas in the case of the URC96 form it is 3.2 scale lengths ($r_{3.2}$).\\
In order to know how much the resulting rotation velocity value of our method depends on the used intrinsic RC shape, B{\"o}hm et al. (2004) used a sample of 77 FORS Deep Field (FDF) galaxies~\cite{Boehm04}. For each galaxy they simulated the synthetic RCs three times, first using the rise-turnover-flat shape and then the two URC-shapes. Then they compared the resulting characteristic rotation velocity values. This can be seen in figure \ref{fig:vmax_comp}. The velocities derived from the URC forms are denoted as $v_{2.2}$ (URC91) and $v_{opt}$ (URC96), respectively. $v_{max}$ is the asymptotic velocity from the rise-turnover-flat shape. Also, B{\"o}hm et al. (2004) distinguish between high quality (solid symbols) and low quality RCs (open symbols). The difference is that high quality objects cover the region of constant rotation velocity at large radii.
\newpage

\noindent
As can be seen, the different rotation velocity values are in quite good agreement. B{\"o}hm et al. (2004) state that the values of $v_{max}$ and $v_{2.2}$ are in agreement to within 5\% for 79\% of the FDF sample and to within 10\% for 95\% of the objects. There is no dependence on the velocity regime nor the quality of the RC. In contrast, the values of the UCR96 $v_{3.2}$ are on average larger by about 7\% than $v_{max}$. Also, the differences increase slightly towards the slow rotator regime. However, they conclude that the impact on the TFR is small, even for the slow rotators. So in principle all three intrinsic RC shapes are suitable for the derivation of the synthetic RCs and furthermore for the Tully-Fisher analysis. For example Pelliccia et al. (2017), who reduced all 28 masks of the data set used in this thesis and studied the kinematics of star-forming galaxies at $z\sim 0.9$, use $v_{2.2}$ and $r_{2.2}$ for their analysis~\cite{Pelliccia17}.\\
However, in this work henceforward only the rise-turnover-flat shape is used as the intrinsic RC shape.\\
The first thing the program does is to generate a 2D exponential intensity profile on a pixel grid~\cite{Boesch13}:

\begin{equation}
I(x,y)=I_{0}\cdot e^{-\frac{r}{r_{d}}}
\end{equation}

\begin{figure}[H]
  \centering
  \includegraphics[height= 6 cm]{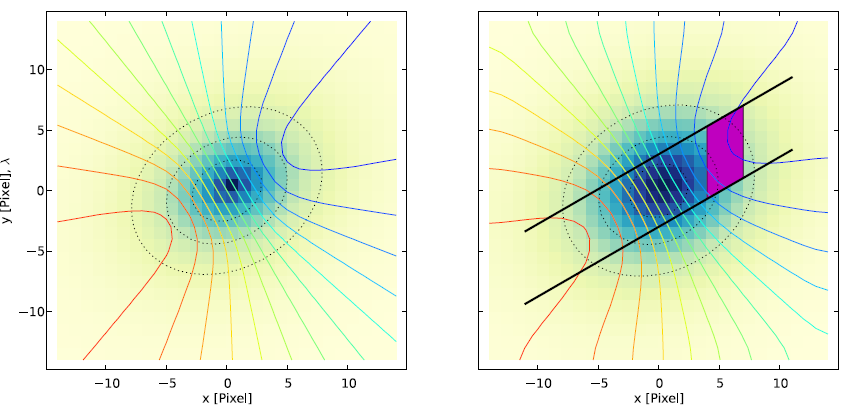}
  \caption{\small Example of a simulated 2D intensity and velocity field of a galaxy with $i=40^{\circ}$ and $\theta=30^{\circ}$. On the right panel the intensity and velocity field are convolved with a Gaussian point spread function. The placing of the simulated slit along the galaxy's apparent major axis is indicated. The magenta area indicates the region where an intensity-weighted average of the velocity values is calculated~\cite{Boesch13}.}
   \label{fig:velpy_exp}
\end{figure}

\noindent
Here, $I_{0}$ is the intensity at the galactic centre (thus at the pixel position (x,y)=(0,0)), and r is defined as $r=\sqrt{x^{2}+y^{2}}$. On this intensity profile a 2D velocity field is superimposed, using the chosen rotational law for the intrinsic RC shape. This procedure produces a rotating disc that is seen face-on. As this is mostly not the case, the next step is to take into account the geometrical factors. This means that the simulated galaxy is rotated according to the inclination angle and the position angle. If the geometrical projection would not be considered, the velocity values would be underestimated, as already discussed above. Furthermore, the blurring effect due to the seeing is taken into account. This is done by convolving the intensity and velocity field with a Gaussian point spread function. Next, an observation of the galaxy is simulated by placing a "slit" on the galaxy and extracting this strip from the velocity field. As in the case of the objects used in this thesis tilted slits were used, the slit is placed along the apparent major axis. Then the program performs a slit integration. At this step the velocity field is weighted by the normalised surface brightness profile. This is necessary because velocities originating from brighter regions contribute stronger to the signal in a spectrum and thus to the velocity shift at a certain radius. The last computation step is that for each pixel position along the spatial axis the velocity values within the simulated slit are integrated along the direction of dispersion. As the slits are tilted, this integration is not perpendicular to the slit direction. As B{\"o}sch et al. (2013) describe it, "an intensity-weighted average of the velocity values inside the slit at each spatial position"~\cite{Boesch13} is calculated. Here it is important that the same boxcar filter is used that was also used for the extraction of the RC. This final step is basically the projection of the extracted strip of the velocity field onto the spatial axis and results in a simulated position-velocity information, the synthetic RC.\\
Figure \ref{fig:velpy_exp} shows an example for a simulated 2D intensity and velocity field from B{\"o}sch et al. (2013)~\cite{Boesch13}. The fields on this figure are projected according to the inclination and the position angle (in this case $i=40^{\circ}$ and $\theta=30^{\circ}$). The spatial axis is along the x-axis and the direction of dispersion is along the y-axis. The exponential intensity profile is shown in a blue colour scale and the black dotted ellipses are the isophotes, corresponding to 50\%, 30\% and 15\% of the central intensity. The coloured curves are the iso-velocity lines (the so-called spider diagram) and range in this case from -60\% $v_{max}$ (red) to 60\% $v_{max}$ (blue). The difference between the two panels is that on the right panel the intensity and velocity field are already convolved with a Gaussian point spread function to account for the blurring effect of the seeing (here FWHM=0.7"). Futhermore, it is indicated how the simulated slit is placed along the galaxy's apparent major axis and how at a certain spatial position, an intensity-weighted average of the velocity values inside the slit borders is calculated (magenta area). In this case a boxcar of 3 pixels was used. It can also be seen that the slit integration is not perpendicular to the slit direction.

\noindent
This program was now applied on the extracted RCs. As mentioned in chapter \ref{extractRC}, emission lines from a total of 110 spectra (out of 160) were used for the RC extraction. However, the fact that it was possible to extract RCs for 110 objects does not mean that these RCs are all good and can be used for further analysis. So before applying the python-program in order to model the RCs and compute the $v_{max}$-values, the extracted RCs were all visually examined. RCs where it was clear that it would not be possible to simulate them - because of e.g. strange and very asymmetric shapes - were excluded already at this stage. Two examples for such asymmetric RCs are shown in figure \ref{fig:badobj}.\\
20 objects were rejected for further analysis. The majority of these objects was rejected because of their non-symmetric and peculiar shaped RCs. In three cases the RCs looked  rather okay, but for one of these objects no inclination and no position angle were given (819178) and for the other two objects no scale length was available (1482306 and 1482419).\\
After sorting out of bad RCs, a sample of a total of 90 objects was left over. The RCs of these objects were subsequently simulated with the \textit{vel.py}-program.\\
Figure \ref{fig:rc_example} shows an example of a simulated RC and the corresponding simulated 2D velocity and intensity field of the object 839451 (z=0.3054):

\begin{figure}[H]
  \centering
    \subfloat[] {\label{main:r}\includegraphics[height= 7 cm]{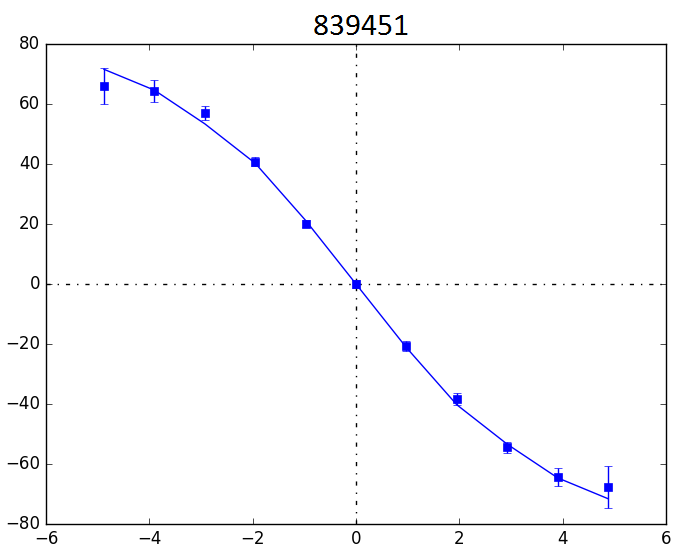}}
    \subfloat[] {\label{main:s}\includegraphics[height= 7 cm]{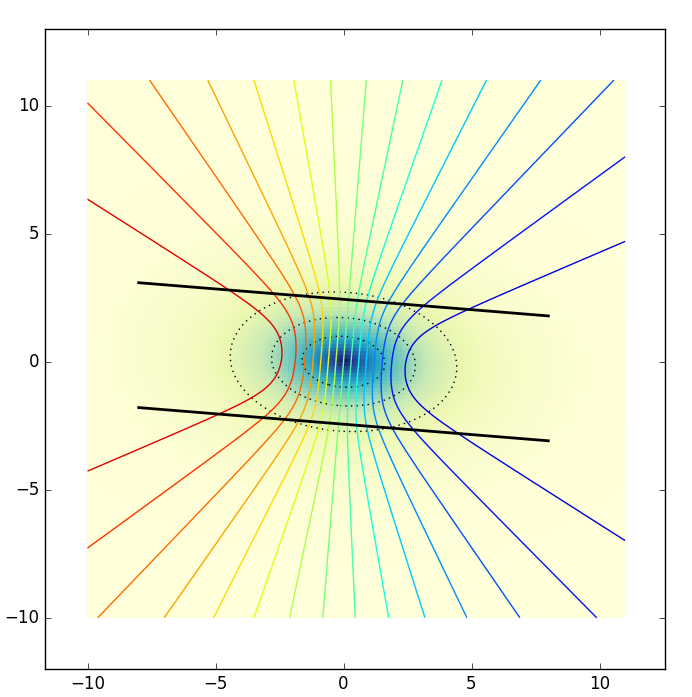}}
  \caption{\small Example of RC fitting for the object 839451 (z=0.3054). Panel \textit{a} shows the observed RC (blue squares) and the best fitting simulated RC (blue fit). The calculated $v_{max}$-value for this example is 104.03 km/s. Panel \textit{b} shows the corresponding 2D velocity and intensity field.}
   \label{fig:rc_example}
\end{figure}

\begin{figure}[H]
  \centering
  \subfloat[] {\label{main:ai}\includegraphics[width= 7.5 cm]{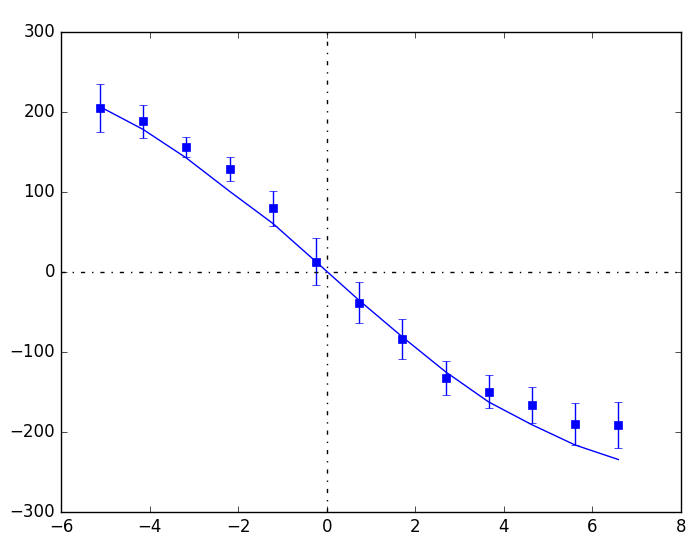}}
  \subfloat[] {\label{main:aj}\includegraphics[width= 7.5 cm]{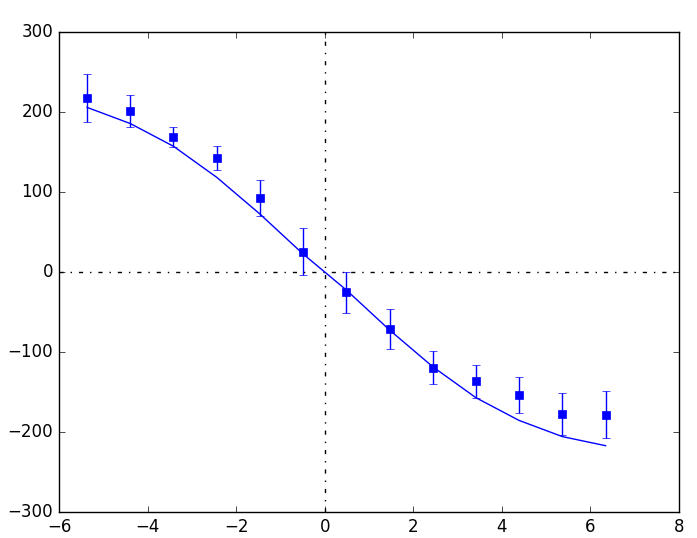}}
  \caption{\small Two RCs of the object 837433 (z=0.796). In the left panel $r_{t}$ is held fixed, in the right panel it is used as a free parameter.}
   \label{fig:rdfitflag}
\end{figure}

\noindent
The simulation program was not run only once per object, but multiple times in order to find the best possible results. The main reasons for this are explained in the following.\\
First of all, for almost every object the extraction of the RCs from emission lines (as described in chapter \ref{extractRC}) was performed using various boxcar filters, mostly 3 and 5 pixels. Therefore, RCs extracted with different boxcars were simulated in order to see if the resulting $v_{max}$-values are similar or differ rather strongly.\\
Another important point is that in some cases of the RCs the kinematic centre did not fit so well, but looked rather offset. A possible reason for this is that the optical and the kinematic centre do not coincide. As has been explained before, the program assumes that both centres are the same. For most regular rotating galaxies this is the case, but in some cases the gas disk of a galaxy can be somewhat offset from the stellar disk which results in shifted centres. So in cases where it seemed that the centre was not completely accurate, the centre was moved manually to make the RC more symmetrical. Then the RCs with the "new" centre were simulated again to see if the outcoming results were better than before. In most cases it was enough to move the centre between a half and one pixel, but sometimes larger shifts were required. It should be mentioned here, that the centre should not be moved by more than 2-3 kpc, but a shift of the centre below this value is tolerable~\cite{Asmuspersonal}. With the help of the cosmological calculator by Ned Wrigth~\cite{cosmcal} the pixel shifts were converted into kpc according to the redshift (with 1 pixel corresponding to $\sim$ 0.2"). This showed that in the majority of the cases the centre was shifted by less than 2 kpc, mostly even less than 1 kpc. Hence, the performed centre shifts were reasonable. Only in three cases a shift of more than 2 kpc was necessary to get a symmetric RC. If the centre of a RC looked offset, in most cases several "new" centres were tried out to find the most suitable.\\
The third point is that it is possible to define whether the turnover radius $r_{t}$ is held fixed, thus making the maximum rotation velocity $v_{max}$ the only free parameter, or if it is a free parameter as well and calculated by the program. In the first case, $r_{t}$ is assumed to be the gas scale length $r_{d}$. In general it is recommended to keep this parameter fixed, especially for RCs that do not probe very deeply into the flat regime. Because if the radial extent of a RC is not large enough, fitting $v_{max}$ and $r_{t}$ simultaneously could lead to strong degeneracies~\cite{Asmuspersonal}. Still, every single RC was simulated twice, once with $r_{t}$ as a free parameter and once held fixed. It should be mentioned here that in the majority of the cases the resulting $v_{max}$-values did not really depend on whether $r_{t}$ was held fixed or not, but were very similar for both options. In almost all cases $v_{max}$ was smaller, if $r_{t}$ was a free parameter as well, but the difference was mostly less than 5 km/s. However, in some cases the differences were rather large, up to 30 km/s. In these cases the individual RCs had to be examined more precisely in order to see in which case the modelled RC fits better to the observed one.\\
Figure \ref{fig:rdfitflag} shows an example for a simulated RC of the object 837433 (z=0.796), once with $r_{t}$ held fixed (panel \textit{a}), and once used as a free parameter (panel \textit{b}). In both cases, the modelled RC fits the observed data points very well, however the difference between the resulting $v_{max}$-values is about $\sim 30$ km/s. If $r_{t}$ is held fixed, $v_{max}$ is $285.14\pm 10.81$ km/s, whereas in the case that $r_{t}$ is used as a free parameter, the galaxy rotates slower ($254.04\pm 36.05$ km/s). Of course, the error on $v_{max}$ is larger in the second case, and if it is taken into account, the $v_{max}$-values are in agreement.

\noindent
As has been mentioned in the previous chapter, the spectra of most objects contained more than one emission line. If it was possible to extract RCs from different emission lines, the different RCs were all simulated and compared with one another. The RCs were examined regarding their spatial extent and their resulting $v_{max}$-values (see also figure \ref{fig:diff_el}).\\
As an example for simulated RCs from different emission lines, figure \ref{fig:emlines} shows two RCs of the object 830319 (z=0.5772):

\begin{figure}[H]
  \centering
  \subfloat[] {\label{main:ak}\includegraphics[width=7 cm]{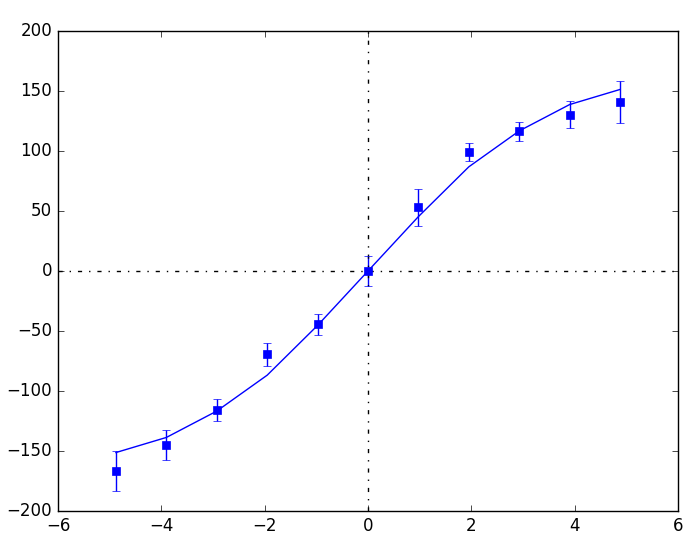}}
  \subfloat[] {\label{main:al}\includegraphics[width= 7 cm]{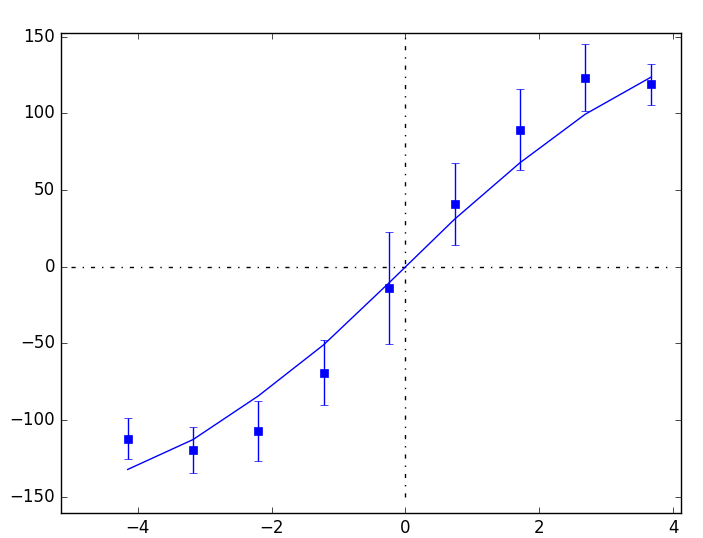}}\\
   \subfloat[] {\label{main:am}\includegraphics[width= 7 cm]{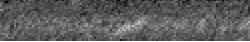}}
  \subfloat[] {\label{main:an}\includegraphics[width= 7 cm]{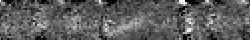}}
  \caption{\small Two RCs of the object 830319 (z=0.5772), extracted from the H$\beta$-line (\textit{a}) and the [OIII](2.)-line (\textit{b}). Panels \textit{c} and \textit{d} show the corresponding emmision lines.}
   \label{fig:emlines}
\end{figure}

\noindent
The RC in figure \ref{main:ak} was extracted from the H$\beta$-line, whereas the RC in figure \ref{main:al} was extracted from the [OIII](2.)-line. Below the RCs the corresponding emission lines are shown. The resulting $v_{max}$-values are $171.93\pm 4.87$ km/s for the H$\beta$-RC and $163.16\pm 8.72$ km/s for the [OIII](2.)-RC, respectively. Thus the RC from the slightly brighter H$\beta$-line, which also extends further than the RC from the [OIII](2.)-line, yields a slightly higher $v_{max}$. However, the difference is not very large and the two $v_{max}$-values are in agreement within the errors. It should be also noted that the [OIII](2.)-line is truncated by a sky line on one side.\\

\noindent
Due to the points explained above, every object was simulated multiple times. At the end, all simulated RCs were inspected with their corresponding $v_{max}$-values and for every object the best result was chosen to be used for the scaling relation analysis.\\
As a summary it can be said that if there were more emission lines in a spectrum, in most cases RCs from the brightest emission line and with the highest velocities were chosen because, naturally enough, they usually probe further into the flat regime than RCs from weaker emission lines. In star-forming galaxies Balmer lines are mostly the most prominent emission lines. So if a spectrum contained a H$\alpha$ or H$\beta$-line, in most cases the RCs from these lines were selected. Only in a few cases the [OIII](2.)-line was brighter or yielded a more extended RC. Conversely, RCs from H$\gamma$ or [SII]-lines were never selected because these lines were always considerably weaker than other emission lines. Concerning objects with $z> 0.6$, almost always RCs from the [OII]-doublet were chosen.\\
If possible RCs extracted with a boxcar of 3 pixels were selected. RCs extracted with a boxcar of 5 pixels were only chosen if they probed further into the flat regime. Furthermore, mostly RCs were picked where $r_{t}$ was held fixed (see table \ref{tab:table_1}). Only in some cases the simulated RC fitted better if $r_{t}$ was a free parameter.

\chapter{Results}
\label{results}
At the end of the previous chapter I explained on what criteria the simulated RCs and the corresponding $v_{max}$-values were chosen for each object. The next step was to examine the 90 simulated RCs and classify them regarding their quality. The objects were subdivided into three classes: high, low and bad quality. The first class contains objects that have very nice and symmetric RCs that mostly probe quite deeply into the flat regime. Furthermore, for these objects the fit of the simulated RCs fits very well to the observed RC. The low quality-class is composed of objects that have RCs that do not look completely symmetric, but are still rather regular in order to be used for further analysis. The RCs of the objects in this class have for example some irregular features. In the case of some low quality-objects, there are also some more or less strong deviations between the simulated fit and the observed data points, especially in the outer parts of the RC. The third class consists of the bad objects, i.e. (i) objects that have RCs that look very asymmetric, (ii) objects where the simulated fit does not match the observed RC at all, and (iii) objects where the RCs do not reach the flat regime, but have a completely linear shape.\\
It should be mentioned here that the classification of the objects into these three classes was based primarly on the shape of the RCs. In some publications, the inclination \textit{i} and the slit tilt angle $\theta$ are also criteria whether an object is used for further analysis or not. For example, B{\"o}hm and Ziegler (2016) use only galaxies with $i>30^{\circ}$ (to avoid face-on disks and ensure sufficient rotation along line-of-sight) and with $\theta < \pm 45^{\circ}$~\cite{Boehm16}. In the case of the 90 objects that were analysed in the course of this thesis, only one object had an inclination smaller than $30^{\circ}$, but this object was characterised as a bad object due to its RC's low velocity extent and linear shape. So the criterion regarding the inclination was fulfilled also for the objects used in this work. In contrast, 13 out of the 90 objects have position angles of the apparent major axis and thus slit tilt angles larger than $\pm 45^{\circ}$, and only one of them falls into the category "bad objects" due to its asymmetric shaped RC. (For the other 12 objects, see table \ref{tab:table_1}.) Although the requirement $\theta < \pm 45^{\circ}$ would ensure a robust sky subtraction and wavelength calibration, the other 12 objects were nonetheless kept in the high- and the low quality-classes.\\
After examining the RCs, 57 objects were characterised as high quality and 19 as low quality-objects. Both classes, that is in total 76 objects, will be used for the scaling relation analysis. The remaining 14 "bad" objects were dismissed. Figure \ref{fig:quality} shows one example for a high quality, a low quality and a bad object, respectively:

\begin{figure}[H]
  \centering
    \subfloat[] {\label{main:t}\includegraphics[height= 4.5 cm]{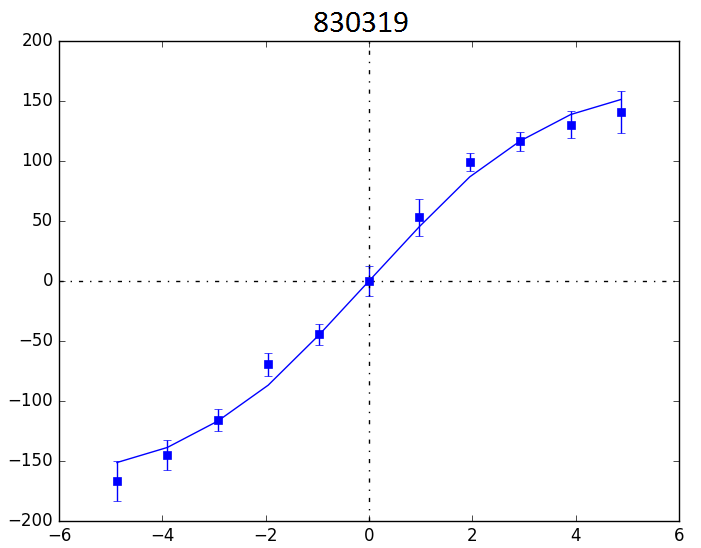}}
    \subfloat[] {\label{main:u}\includegraphics[height= 4.5 cm]{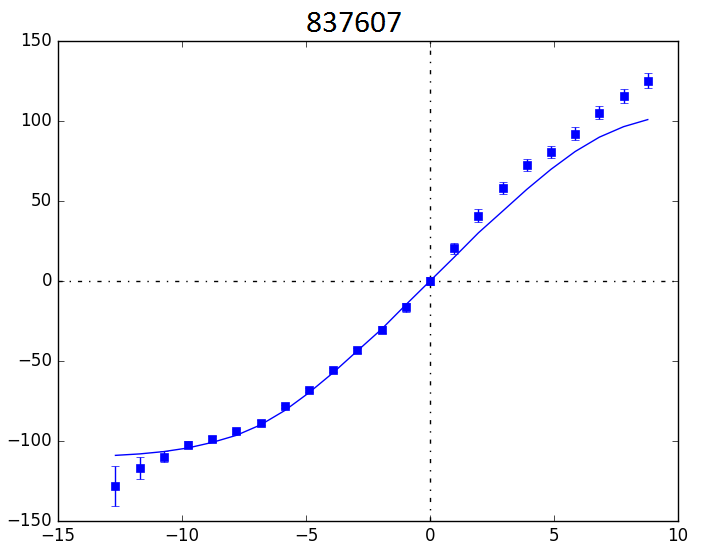}}
    \subfloat[] {\label{main:v}\includegraphics[height= 4.5 cm]{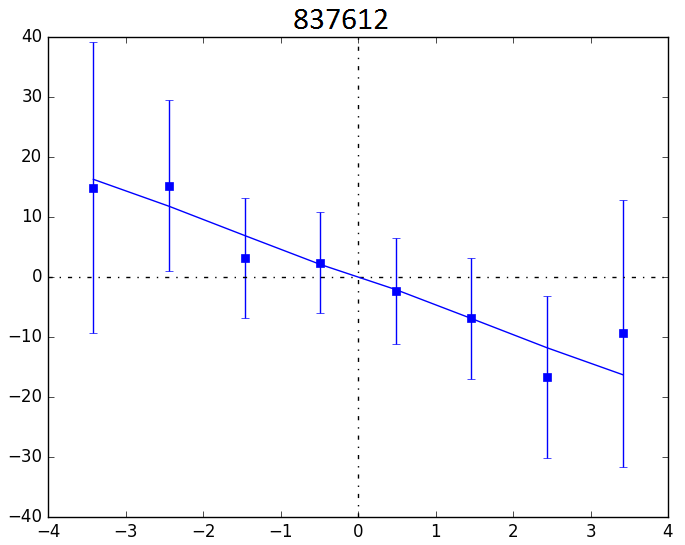}}
  \caption{\small Examples for a high quality object (830319; z=0.5772 (\textit{a})), a low quality object (837607; z=0.193 (\textit{b})) and a "bad" object (837612; z=7485 (\textit{c})), respectively.}
   \label{fig:quality}
\end{figure}

\noindent
The most important parameters of the 76 objects used for the kinematic analysis are shown in tables \ref{tab:table_1} and \ref{tab:table_2}.\\
Panel \textit{a} from figure \ref{fig:z_distr} shows the redshift distribution of the 76 objects that are subsequently used for the scaling relation analysis. The objects are subdivided into high quality (blue) and low quality (red) objects. The total distribution is indicated by the black line. For comparison, panel \textit{b} shows the redshift distribution of the zCOSMOS-bright sample from Lilly et al. (2009)~\cite{Lilly09}:

\begin{figure}[H]
  \centering
  \subfloat[] {\label{main:ax}\includegraphics[height= 7 cm]{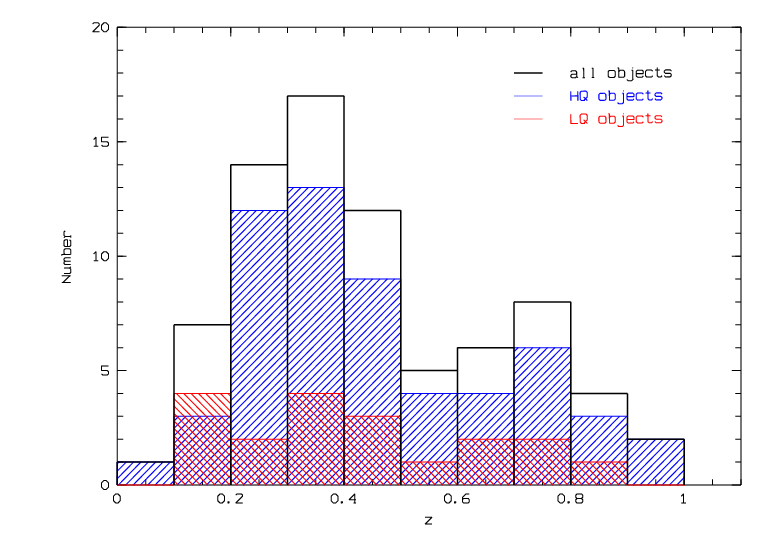}}
  \subfloat[] {\label{main:ay}\includegraphics[height= 6.5 cm]{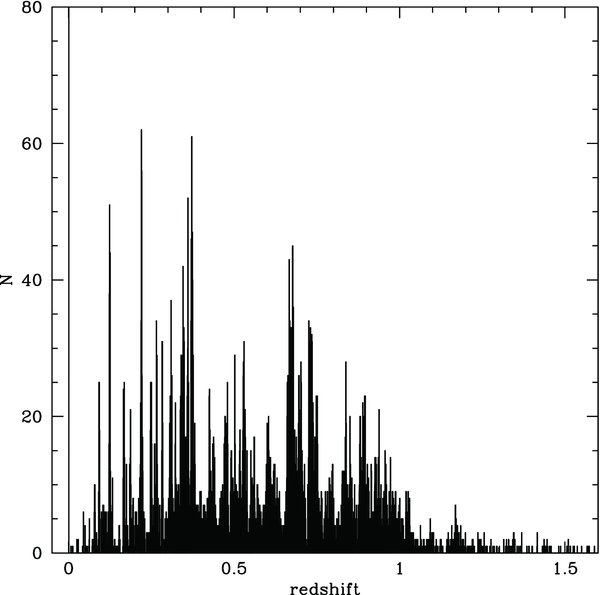}}
  \caption{\small Panel \textit{a} shows the redshift distribution of the 57 high quality (blue) and the 19 low quality (red) objects. The black line marks the overall distribution. Panel \textit{b} shows the redshift distribution of the zCOSMOS-bright sample from Lilly et al. (2009)~\cite{Lilly09}.}
   \label{fig:z_distr}
\end{figure}

\begin{figure}[H]
  \centering
  \includegraphics[height= 6 cm]{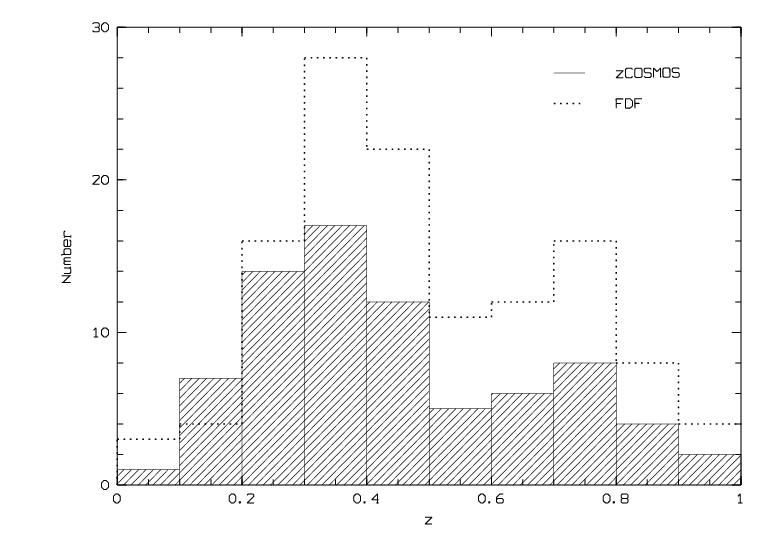}
  \caption{\small The redshift distribution of the 76 zCOSMOS-galaxies used in this work (hashed histogram) compared to the distribution of 124 FDF-galaxies used by BZ16 (dotted line).}
   \label{fig:Asmus_z_distr}
\end{figure}

\noindent
In the case of all three groups in figure \ref{main:ax} - high and low quality separated and combined - a similar distribution can be seen. In the total distribution two peaks are clearly visible: the first between $0.2\leq z\leq 0.45$ and the second between $0.7\leq z\leq 0.8$. These peaks are also visible in the distributions of the high and the low quality objects, although in the latter case they are less pronounced. The redshift distribution of the total zCOSMOS-bright sample in figure \ref{main:ay} is binned in smaller intervals ($\Delta z=0.001$). Still also in this histogram peaks between $0.2\leq z\leq 0.4$ and at $z\sim 0.7$ can be seen, implying that the \textit{z}-distribution of the reduced sample used in this thesis is representative of the whole zCOSMOS-bright sample.\\
Figure \ref{fig:Asmus_z_distr} shows once more the overall redshift distribution of the 76 objects, compared to the redshift distribution of 124 FDF (FORS Deep Field) galaxies used by B{\"o}hm and Ziegler (2016) (hereinafter referred to BZ16) to investigate disk galaxy scaling relations at intermediate redshifts~\cite{Boehm16}. It can be seen that apart from the larger number of FDF galaxies, the redshift distributions of both samples are very similar. Also in the case of FDF galaxies two peaks at $z\sim 0.3$ and $z\sim 0.7$ can be seen. The corresponding mean redshifts are $\langle z\rangle _{zCOSMOS}=0.4452$ and $\langle z\rangle _{FDF}=0.4929$.

\noindent
As has been mentioned in section \ref{scalrel}, parameters describing the energy characteristic, the kinematic characteristic and the spatial scale of a galaxy are needed for the construction of the scaling relations. In this work, these three characteristics are represented mainly by the absolute magnitude in the B-band $M_{B}$, the maximum velocity $v_{max}$ and the scale length derived from the continuum emission $r_{d}$, as these are also most commonly used for the TFR and the VSR. For $v_{max}$ the values from the RC modelling were used, whereas the $M_{B}$ and the $r_{d}$ values were obtained from the zCOSMOS-catalogue. The scale length $r_{d}$ was calculated from the effective radius $r_{1/2}$ with the following equation (see chapter \ref{simulations}):

\begin{equation}
\label{rd_eq}
r_{d}=\frac{r_{1/2}}{1.7}
\end{equation}

\noindent
The effective radii $r_{1/2}$ used for the computation of $r_{d}$ were taken from Sargent et al. (2007) who derived them from the GIM2D (Galaxy IMage 2D) Sersic fits, using HST images taken in the ACS I(814) filter~\cite{Sargent07}.\\
The $M_{B}$-values for the objects were taken from Zucca et al. (2009). In order to derive the absolute magnitudes for the zCOSMOS 10k bright sample, they computed the K-correction using a set of templates and all available photometric information~\cite{Zucca09}. Then, to derive the rest-frame absolute magnitude in each band, they used the apparent magnitude from the closest observed band, shifted to the redshift of the galaxy. That way the template dependency was reduced and the applied K-correction was as small as possible.\\
Before using the $M_{B}$-values from Zucca et al. (2009), additional corrections for the intrinsic absorption $A_{B}^{i}$ due to dust extinction had to be applied. The reason for this is the choice of the form of the local TFR as the reference. In this work the local TFR as given by Tully et al. (1998) is used for the comparison between local and distant galaxies~\cite{Tully98}:

\begin{equation}
\label{localTFR}
M_{B}=-7.79\cdot log(v_{max})-2.91
\end{equation} 

\noindent
For a correct comparison, the luminosities of the distant galaxies had to be corrected for intrinsic dust absorption in the same way as the galaxies of the local sample used to construct the local TFR. Hence, the B-band magnitudes provided by the zCOSMOS-catalogue were corrected for intrinsic dust absorption following the approach of Tully et al. (1998):

\begin{equation}
\label{absorption}
A_{B}^{i}(a/b, v_{max})=log(a/b)[-4.48+2.75\cdot log(v_{max})]
\end{equation}

\begin{figure}[H]

    \subfloat[] {\label{main:w}\includegraphics[height= 6 cm]{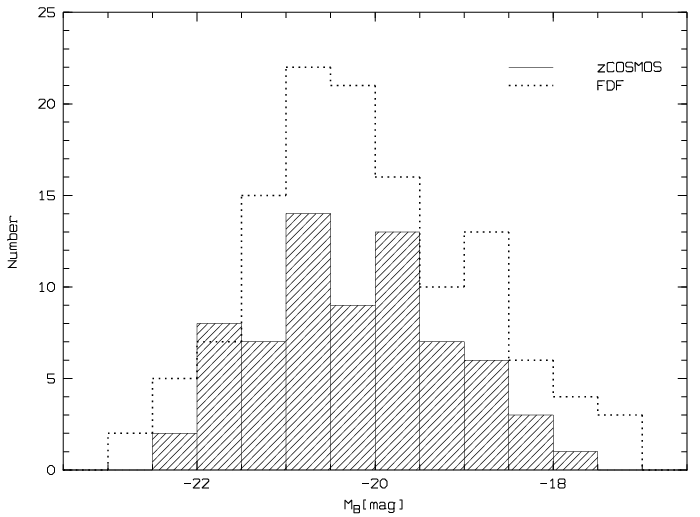}}
    \subfloat[] {\label{main:x}\includegraphics[height= 6 cm]{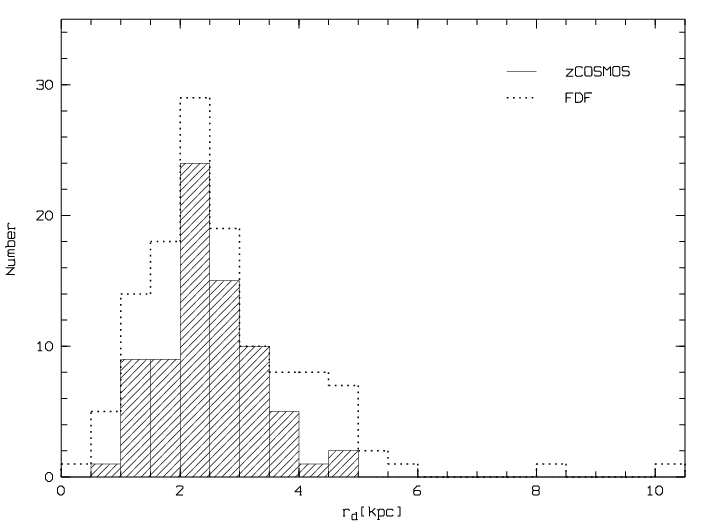}}\\
    \subfloat[] {\label{main:y}\includegraphics[height= 6 cm]{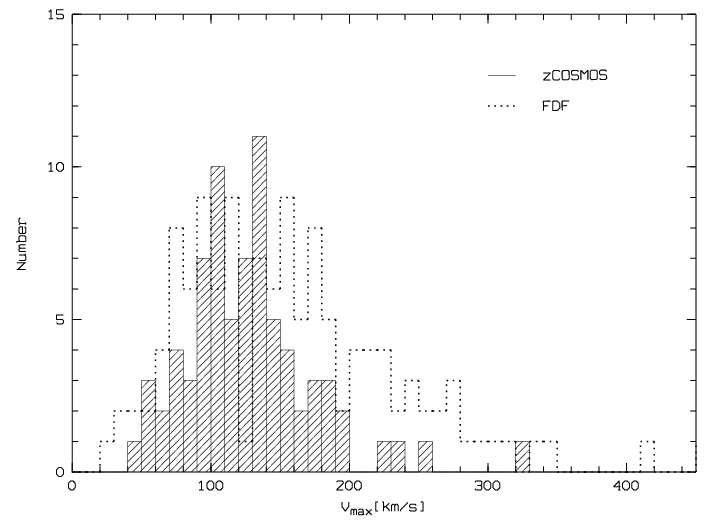}}
     \subfloat[] {\label{main:az}\includegraphics[height= 6 cm]{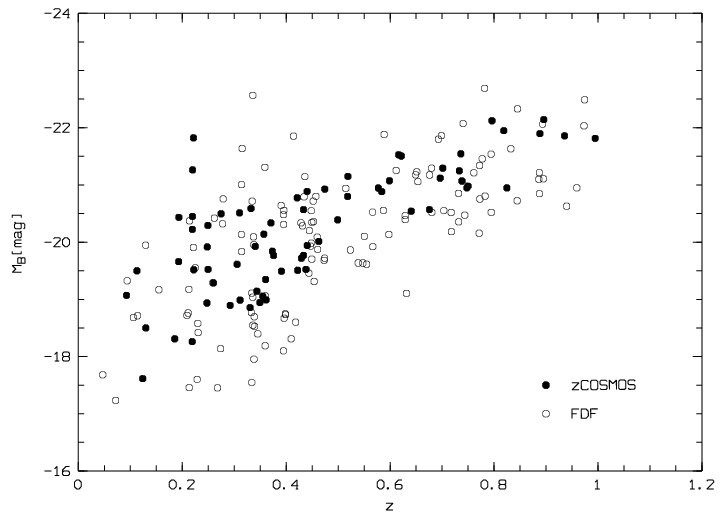}}
  \caption{\small $M_{B}$-histogram (\textit{a}), $r_{d}$-histogram (\textit{b}), $v_{max}$-histogram (\textit{c}) and $M_{B}$ vs. \textit{z} plot (\textit{d}) of the 76 zCOSMOS-galaxies used in this work (hashed area and filled circles, respectively). For comparison, the equivalent distributions of 124 FDF-galaxies used by BZ16 are indicated with the dotted line in the histograms and with open circles in the $M_{B}$-\textit{z} plot, respectively.}
   \label{fig:distributions}
\end{figure}

\noindent
The intrinsic dust absorption in a galaxy depends primarily on its inclination \textit{i} which is expressed in the above equation by means of the axis ratio \textit{a/b} of the disk. This means that the correction is minimum for a face-on and maximum for an edge-on galaxy. In addition to this, Tully et al. (1998) also added a correction factor correlated with the maximum rotation velocity of an object. This is based on the findings of Giovanelli et al. (1995) that extinction is stronger in more massive galaxies, and thus in galaxies with higher $v_{max}$~\cite{Giovanelli95}.\\
To give an overview of what range the objects used in this thesis cover in luminosity, size and rotation velocity, figure \ref{fig:distributions} shows three histograms of the $M_{B}$-, $r_{d}$- and $v_{max}$-distributions of the 76 objects. Furthermore, the distribution of $M_{B}$ over redshift is shown. No distinction is made between high and low quality objects. In order to compare the distributions of the parameters of this sample to other samples, the equivalent distributions of the 124 FDF galaxies from BZ16 are indicated as well~\cite{Boehm16}.\\
The $M_{B}$-values of the zCOSMOS galaxies in figure \ref{fig:distributions} are already corrected for the intrinsic dust absorption. BZ16 used the same approach by Tully et al. (1998), therefore their magnitudes did not have to be corrected anymore with equation \ref{absorption}. However, the different photometric systems had to be taken into account. BZ16 use magnitudes in the Vega-system, whereas the magnitudes from the zCOSMOS-objects are in the AB-system. The Vega-system uses the spectrum of the star Vega as the reference, defining that Vega's magnitudes are 0.0 in all bands. The AB-systems however does not depend on a star spectrum, but assumes that an object with a flat energy distribution, thus $F_{\nu}=const.$, has the same magnitude at all frequencies~\cite[45]{Weigert}. The two magnitude systems are defined in such a way that the AB- and the Vega-magnitude are nearly the same in the V-band. In the B-band, the difference can be calculated with the following equation from Blanton and Roweis (2007)~\cite{Blanton07}:

\begin{equation}
m_{AB}-m_{Vega}=-0.09
\end{equation}

\noindent
So in order to compare the magnitude distribution of both object samples, a value of 0.09 (which is simply the AB magnitude of the star Vega) was subtracted from the $M_{B}$-values of the FDF galaxies.

\newpage

\noindent
Looking at figure \ref{main:w}, one can see that the magnitude distribution is very similar for both samples. The mean magnitudes are $\langle M_{B}\rangle _{zCOSMOS}=-20.25$ mag and $\langle M_{B}\rangle _{FDF}=-20.09$ mag, respectively. The only difference is that in the case of the FDF galaxies the distribution is slightly broader, having a few brighter and also a few less luminous objects. The distribution of the FDF galaxies peaks between $-21.5<M_{B}<-19.5$ mag, and also in the case of the zCOSMOS galaxies there is a peak between $-21<M_{B}<-19.5$ mag. There is a difference regarding the objects with magnitudes between -21 mag and -21.5 mag, with the FDF objects having considerably more objects in this range. But besides that, one can say that both samples have similar magnitude distributions.\\
Regarding the scale length $r_{d}$, the two samples have also quite similar distributions, with most objects having scale lengths between $1<r_{d}<3.5$ kpc  (\ref{main:x}). Also the mean scale lengths are very similar: $\langle r_{d}\rangle _{zCOSMOS}=2.45$ kpc and $\langle r_{d}\rangle _{FDF}=2.69$ kpc. The difference is that in the case of the zCOSMOS galaxies the distribution flattens rather quickly after 3.5 kpc, having only a few objects with larger scale lengths. In contrast, the FDF sample contains quite many galaxies with scale lengths between 3.5 kpc and 5 kpc, and five objects with $r_{d}>5$ kpc. Two of these five objects have even scale lengths around 8 and 10 kpc, respectively.\\
Concerning the maximum rotation velocity, the situation appears to be a little bit different  (\ref{main:y}). The distributions of both samples are quite similar up to 150 km/s. However, there are big differences at higher velocities. The velocity distributions of both samples become flatter after 150 km/s, but the FDF galaxies extend much further into the high velocity regime. There are many FDF objects that have very high rotation velocities, including one object with even $v_{max}>400$ km/s (object 4730). In contrast, the zCOSMOS sample contains only 4 objects with velocities higher than 200 km/s, and the highest value is around 330 km/s. The different distributions are mirrored in the mean maximum velocities, with $\langle v_{max}\rangle _{FDF}=154.10$ km/s being considerably higher than $\langle v_{max}\rangle _{zCOSMOS}=127.97$ km/s. It is somewhat curious that the $v_{max}$-distributions of the two samples differ so much and that there are not more objects with higher velocities in the zCOSMOS sample. However, one reason for the higher $v_{max}$-values of the FDF sample are the differences between the $M_{B}$-distributions and the $r_{d}$-distributions of the two samples, especially the larger number of FDF galaxies with $M_{B}$'s between -21 mag and -21.5 mag and the lack of zCOSMOS galaxies with $M_{B}<-22$ mag and $r_{d}>4$ kpc. From the 14 FDF objects with $M_{B}<-21.5$ mag, 11 have $v_{max}$ larger than 200 km/s and 9 have $r_{d}$'s larger than 3.5 kpc. The velocity distribution will also play a role for the scaling relations.\\
In figure \ref{main:az} the $M_{B}$-values of both samples are plotted against the redshift. Both distributions look very similar, covering a larger luminosity range at lower redshifts than at higher redshifts. However, the $M_{B}$-range of the FDF galaxies is almost in all redshift bins slightly broader, covering lower, but often also larger luminosities than the zCOSMOS galaxies.

\section{The B-band Tully-Fisher relation}
\label{TFRMB}
\begin{figure}[H]
  \centering
  \includegraphics[height= 8 cm]{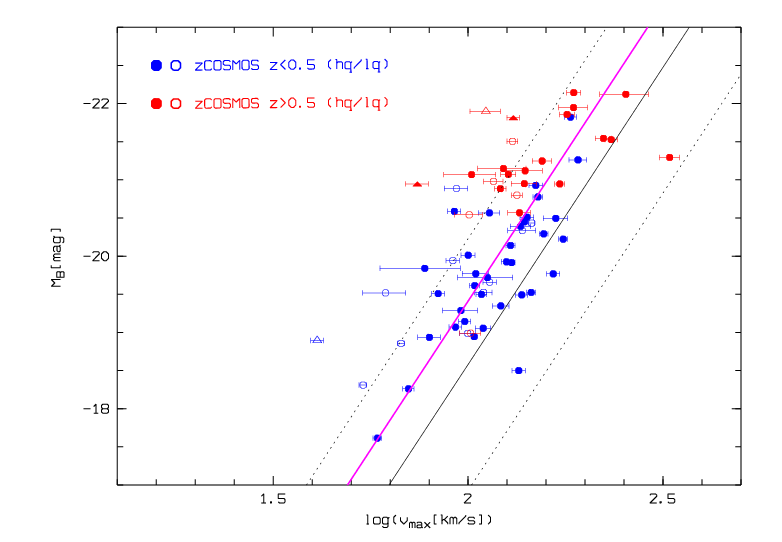}
  \caption{\small The B-band Tully-Fisher diagram of the 70 zCOSMOS galaxies with available $M_{B}$-values. The objects with redshifts $z<0.5$ and $z>0.5$ are represented by blue and red symbols, respectively. The solid circles are the high quality, and the open circles are the low quality objects. The black solid line indicates the local Tully-Fisher relation and the dotted black lines its $3\sigma$ scatter. The magenta solid line shows the fit to the zCOSMOS sample with the slope fixed to the local value. It is offset from the local relation by $\Delta M_{B}=-0.83\pm 0.09$ mag.}
   \label{fig:TFR}
\end{figure}

There are various variants of the TFR, and the most famous is probably the optical TFR, often also called the classical TFR~\cite{Boehm16}. Several years later it was discovered that also e.g. the total baryonic mass~\cite{McGaugh00} as well as the stellar mass $M_{*}$~\cite{Bell01} of a galaxy correlate with its maximum rotation velocity. This means that there are thus several variants of the same relation between a spiral galaxy's dark matter and its baryonic matter content.\\
In the case of the optical variant, different bands can be used in order to get luminosities and construct the TFR, for example the K-band and the B-band. The B-band TFR is probably the most popular TFR and thus this section will focus on this variant of the TFR.\\
Figure \ref{fig:TFR} shows the B-band TFR for the objects of the zCOSMOS sample that were selected for the kinematic analysis. It should be mentioned that the $M_{B}$-values were available only for 70 from the 76 objects, reducing the sample for the Tully-Fisher analysis even more. The $M_{B}$-values are already corrected for the intrinsic dust absorption following equation \ref{absorption}. The $v_{max}$-values are the results from the RC modelling from chapter \ref{simulations} and the errors on $v_{max}$ were computed with the following equation:

\begin{equation}
\label{error}
\sigma^{2}_{vmax}=\sigma^{2}_{\chi}+v_{max}^{2}\cdot tan(i)^{-2}\cdot \sigma^{2}_{i}+v_{max}^{2}\cdot tan(\delta)^{2}\cdot \sigma^{2}_{\delta}
\end{equation}

\noindent
Here, $\sigma^{2}_{\chi}$ is the error from the $\chi^{2}$-fits of the synthetic to the observed RC. This error was automatically computed by the program \textit{vel.py} in addition to the maximum rotation velocity. $i$ and $\delta$ are the inclination and the misalignment angle between the slit position and the apparent major axis, respectively, and $\sigma^{2}_{i}$ and $\sigma^{2}_{\delta}$ are the corresponding errors. As in the case of the objects used in this thesis the slits were aligned along the apparent major axis, the misalignment angle $\delta$ is assumed to be 0, and thus also the last part of equation \ref{error} ($v_{max}^{2}\cdot tan(\delta)^{2}\cdot \sigma^{2}_{\delta}$) is 0. For the inclinations no errors were available, therefore for $i>30^{\circ}$ an error of $\sigma^{2}_{i}=2^{\circ}$ and for $i<30^{\circ}$ an error of $\sigma^{2}_{i}=5^{\circ}$ was assumed because the smaller the inclination is, the larger its contribution to the error of $v_{max}$ is. However, all 76 objects used for the kinematic analysis have inclinations larger than $30^{\circ}$, and thus the assumed inclination error is in all cases $2^{\circ}$. Taking this into account, it can be seen that the first term of \ref{error}, $\sigma^{2}_{\chi}$, has the largest influence on the total error $\sigma_{vmax}$. It should also be noted that in most cases the errors on $v_{max}$ are very small, often below 10 km/s. The expected error would be about $\sim 10\%$ of the $v_{max}$-value (except for very low-mass galaxies where it should be even larger), but for the majority of the 76 zCOSMOS objects this is not the case. It is probable that the errors should actually be larger and that the python module that is used by the program \textit{vel.py} to calculate the errors from the $\chi^{2}$-fits of the synthetic to the observed RC, computes too small values for the errors~\cite{Asmuspersonal}.\\

\noindent
The objects in figure \ref{fig:TFR} are subdivided according to their quality, with filled circles being high quality and open circles being low quality objects, but also according to their redshifts. Blue symbols have lower redshifts up to $z\sim 0.5$ ("low-z"), whereas objects with redshifts between z=0.5 and z=1 ("high-z") have red symbols. In total there are 23 high-z objects (18 of them being high quality objects) and 47 low-z objects (35 high quality).\\
In order to compare the distribution of the zCOSMOS sample in the $M_{B}-v_{max}$ space with the distribution of local galaxies, the local TFR relation from Tully et al. (1998) (equation \ref{localTFR}) is used~\cite{Tully98}. However, as has been mentioned above, the magnitudes used in this thesis are in the AB-system, whereas the local TFR is expressed in the Vega-system. To adjust it for the AB-system, the value 0.09 is subtracted from the zeropoint, leading to the following equation for the local relation:

\begin{equation}
\label{localTFR_AB}
M_{B}=-7.79\cdot log(v_{max})-3.0
\end{equation}

\begin{figure}[H]
  \centering
  \includegraphics[height= 7.5 cm]{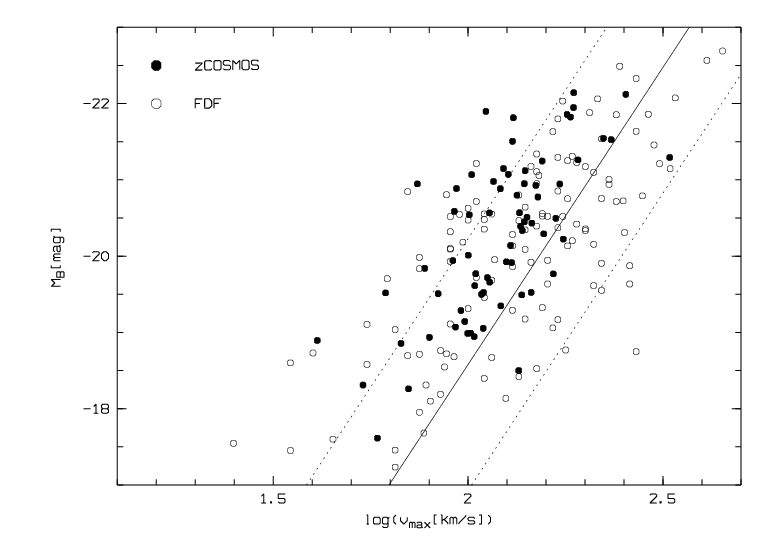}
  \caption{\small The Tully-Fisher diagram of the 70 zCOSMOS galaxies (solid circles) compared to the 124 FDF galaxies from BZ16 (open circles). The local Tully-Fisher relation and its $3\sigma$ scatter are indicated by the solid line and the two dotted lines, respectively.}
   \label{fig:TFR_comp}
\end{figure}

\noindent
The local TFR is plotted as the continuous black line in figure \ref{fig:TFR}. The dotted black lines indicate the $3\sigma$ scatter of the local B-band relation ($\sigma_{obs} = 0.55$ mag).\\
In order to investigate the evolution of the TFR in luminosity, a line with the slope fixed to the local value (a=-7.79) was fitted through the distant sample. This fit is indicated by the solid magenta line in the above diagram, and is offset from the local relation toward higher luminosities by $\Delta M_{B}=-0.83\pm 0.09$ mag. The scatter of this sample is $\sigma_{obs}\sim 0.95$ mag, which is about 1.8 x $\sigma_{obs}$ of the local sample from Tully et al. (1998). By looking at the diagram above one can see that the objects are concentrated primarily to the "left" of the local TFR, that is towards higher luminosities. On the "right" side there are only six objects, five with redshifts $z<0.5$ and one with $z>0.5$. However, the majority of the objects are within the area of the $3\sigma$ scatter of the local relation. Only 14 objects lie outside this area. From these 14 objects the majority are low quality objects.\\
Another thing that is noticeable is that in comparison to the objects with lower redshifts, the objects with $z>0.5$ are concentrated in the brighter magnitudes regime, and thus also in the high-velocity regime. Whereas the low-z objects have B-band magnitudes between -18.31 mag and -21.8 mag, the lowest magnitude found in the high-z sample is -20.5 mag. The reason for this is the so-called Malmquist bias, a selection effect due to a brightness limit that was first described by the Swedish astronomer Karl Gunnar Malmquist in 1922~\cite[111]{Binney2}. It says that in the case of brightness-limited observations, objects with magnitudes below a certain limiting magnitude are not detected. As the apparent magnitude depends on the distance, and thus on the redshift, this can lead to an overrepresentation of luminous objects, because objects with lower luminosities are not included. The zCOSMOS survey is, as has been mentioned in section \ref{zCOSMOS}, a magnitude limited survey. So the different distribution of low-z and high-z objects is probably mainly due to this selection effect, which can also be seen very well in figure \ref{main:az}.\\
As has been mentioned before, it is expected that spiral galaxies at given $v_{max}$ were more luminous at higher redshifts than in the local universe. However, a mean brightening by $\sim \Delta M_{B}=-0.8$ mag is rather large. For example, by fixing the slope to the local value, BZ16 get a mean offset of $\langle\Delta M_{B}\rangle=-0.47\pm 0.16$ mag and a scatter of $\sigma_{obs}=1.28$ mag~\cite{Boehm16}. Whereas the scatter is $\sim 1.3$ times larger than in the case of the zCOSMOS galaxies, the mean offset is only almost half as large. This is certainly a considerable difference. For illustration, figure \ref{fig:TFR_comp} shows the Tully-Fisher diagram of the 70 zCOSMOS galaxies compared to the 124 FDF galaxies from BZ16. It can be seen that the FDF galaxies are distributed more uniformly around the local TFR. Compared to the zCOSMOS galaxies there are considerably more objects on the right side of the local TFR, that is towards lower luminosities. It should be noted that BZ16 used almost twice as many objects for their analysis. Still the different distributions are noticeable.

\newpage

\begin{figure}[H]
  \centering
    \subfloat[] {\label{main:z}\includegraphics[width= 7 cm]{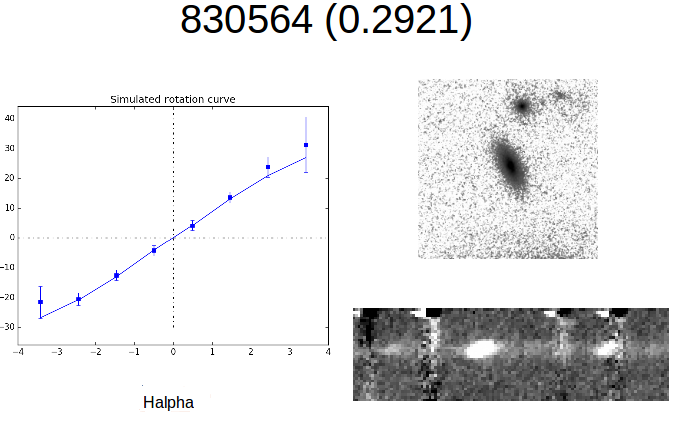}}
    \subfloat[] {\label{main:aa}\includegraphics[width= 7 cm]{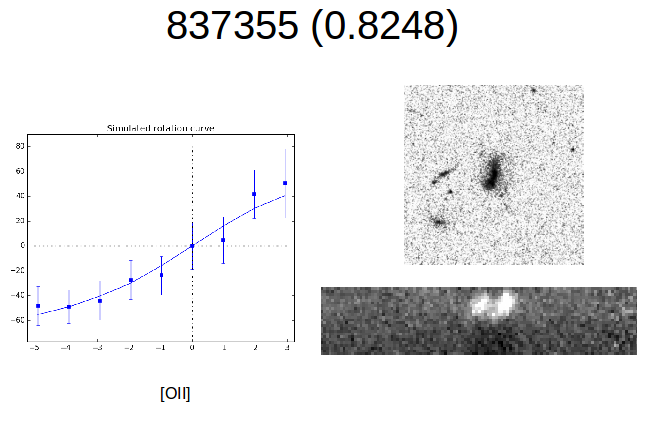}}\\
    \subfloat[] {\label{main:ab}\includegraphics[width= 7 cm]{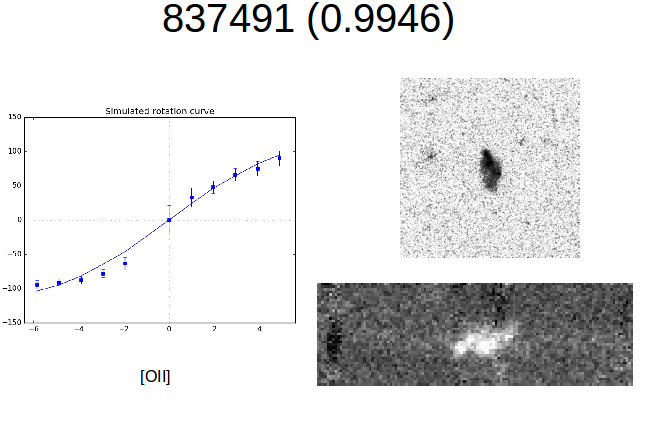}}
    \subfloat[] {\label{main:ac}\includegraphics[width= 7 cm]{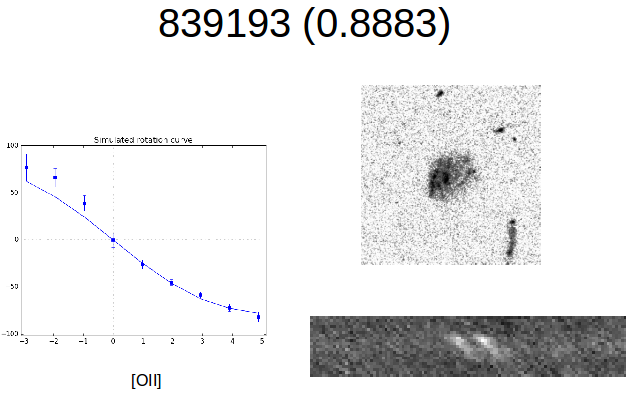}}
  \caption{\small The RCs, emission lines used for the RC extraction and HST-images of the two high quality galaxies (\textit{b}, \textit{c}) and the two low quality galaxies (\textit{a}, \textit{d}) from table \ref{tab:offsets_exp}.}
   \label{fig:offset_exp}
\end{figure}

\noindent
BZ16 write in their paper that interestingly almost all FDF galaxies with $log(v_{max})<2$ are located on the high-luminosity side of the local relation. Eight of these galaxies, which have $log(v_{max})<1.8$, lie even above the $3\sigma$-limit~\cite{Boehm16}. In the case of the zCOSMOS galaxies the problem is that there are, as has been mentioned, only six objects on the low-luminosity side of the local TFR. These six objects all have velocities $log(v_{max})>2$. This means that the 18 objects of this sample that have $log(v_{max})<2$ are all on the high-luminosity side. Still, due to the small number of objects on the "right" side of the local relation, the statement for this sample is not as clear as for the FDF sample. The same applies to the objects with velocities $log(v_{max})<1.8$. In the zCOSMOS sample there are only four objects with such low velocities, and only one of them lies within the $3\sigma$-limit. But as four objects is not a very high number, it is difficult to make a clear statistical statement.\\
Still three possible explanations for this, mentioned by BZ16, should be addressed briefly. The first possible reason is that $v_{max}$ is underestimated for low-mass objects~\cite{Boehm16}. This is likely as local low-mass spirals often have RCs with a positive rotation velocity gradient even at large radii. So it is possible that the spatial extent of the RCs at higher redshifts is not sufficient to constrain a potential velocity gradient. Secondly, a mass-dependent evolution in luminosity that is larger for low-mass than for high-mass objects could be an explanation. And thirdly, the apparent mass-dependency could be mimicked by the magnitude limit. For example, Willick (1994) showed that toward lower velocities, mostly objects on the high-luminosity side enter a sample~\cite{Willick94}. BZ16 also state that overestimated luminosities from galaxies with low $v_{max}$-values are very unlikely. Also an evolution in redshift cannot be the reason as the three outliers with $log(v_{max})<1.8$ have a mean redshift of $\sim 0.23$, which is considerably lower than the mean redshift of all 76 objects, $\langle z\rangle \sim 0.45$.\\
As all three objects above the $3\sigma$-limit are low quality objects, the most plausible explanation is that $v_{max}$ is underestimated due to a insufficient extent of the RC.\\
In the following, four objects that lie considerably outside the $3\sigma$-area, will be examined in more detail (in figure \ref{fig:TFR} indicated by triangles): three high-z objects (two high and one low quality) and one low-z object (low quality). The RCs of these four objects as well as the emission lines that were used for the RC extraction and the HST-images are shown in figure \ref{fig:offset_exp}. Table \ref{tab:offsets_exp} lists the redshifts, maximum velocities, B-band magnitudes and quality category of these objects. 

\begin{table}[H]
\begin{center}
\small
  \begin{tabular}{ l | c  c  c  c  c }  
    ID & z & $v_{max}$ [km/s] & $v_{err}$ [km/s] & $M_{B}$[mag] & Quality \\ \hline
    830564 & 0.2921 & 41.00 & 1.54 & -18.90 & L\\ 
    837355 & 0.8248 & 74.09 & 4.93 & -20.95 & H\\ 
    837491 & 0.9946 & 130.66 & 4.83 & -21.81 & H\\ 
    839193 & 0.8883 & 111.10 & 10.04 & -21.90 & L\\ 
  \end{tabular}
  \caption{\small Redshifts, maximum rotation velocities + errors, B-band magnitudes and quality categories of four objects with large offsets from the local TFR.}
  \label{tab:offsets_exp}
\end{center}
\end{table}

\noindent
In the case of the low-z object 830564 the $H_{\alpha}$-line was used for the RC extraction. The HST-image shows a symmetric galaxy. The emission line is very bright and clearly visible, and also the RC looks very regular and symmetric. However, the flat part is not really distinguishable. It is thus possible that the rotation velocity is underestimated in this case. Considering the low $v_{max}$ of 41 km/s this is probable. Also the very small velocity error is unusual. Understandably an error of only $\sim 1.5$ km/s is too low, but as has been mentioned before in this chapter, it is very probable that the errors in $v_{max}$ are generally too small. Possible reasons for this RC are an inclination problem or solid body rotation. However with an inclination of $i=64.74^{\circ}$ the first possibility can be ruled out.\\
The RCs of the three high-z objects were all extracted from the [OII]-line doublet which can be seen very well on the cutouts of the spectra. All three objects also have rather high redshifts. Compared to 830564, the HST-images of these objects show galaxies that do not have a very symmetric and regular shape, but look rather clumpy. Still it was possible to obtain nice-looking RCs. In the case of 837355 the emission lines are bright, but also rather compact. Unfortunately, the spectrum is cut in a way that the emission lines are located at the upper edge. This could explain the seemingly missing points on the right approaching side of the RC compared to the left receding side.\\
The [OII]-emission lines of 837491 are somewhat more diffuse. The gap between the two lines is not as clearly visible as in the other two cases, and there are two sky lines at the same position than the [OII]-lines. Still, the resulting RC looks very good and symmetric. There is one data point missing in the centre (due to the sky line), but the simulated RC fits very well. From all four objects, this galaxy has the smallest offset to the local TFR.\\
The emission lines of the fourth object, 839193, are the weakest. It seems that the lines are sharper on the upper side and more diffuse at the bottom. Looking at the RC it can be seen that it is not completely symmetric. On the receding side the simulated RC fits very well to the observed data points, but on the approaching side there is a slight difference. On the HST-image it seems that the galaxy extends a bit further on one side than on the other. This is probably reflected by the slightly higher velocities on the approaching side. Although the simulated RC fits the observed mostly well, it is probable that the $v_{max}$ for this object is also underestimated.\\

\begin{figure}[H]
  \centering
  \includegraphics[height= 8 cm]{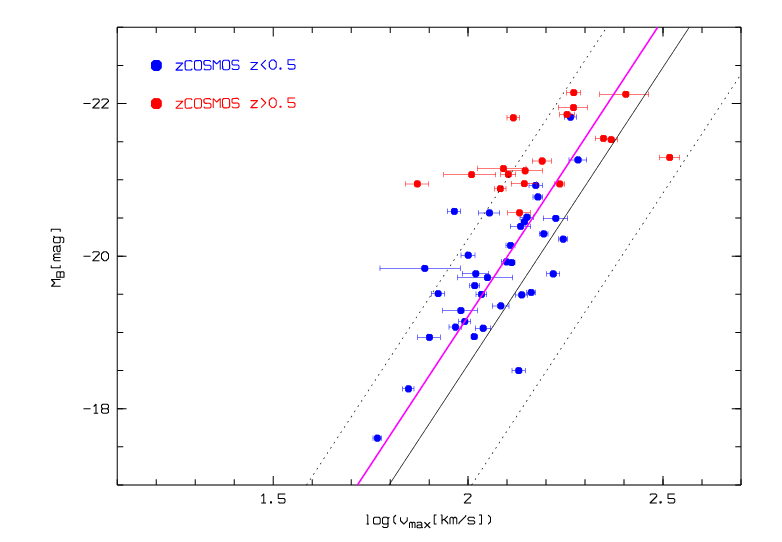}
  \caption{\small The B-band Tully-Fisher diagram of the 53 zCOSMOS galaxies with high quality RCs. The objects with redshifts $z<0.5$ and $z>0.5$ are represented by blue and red symbols, respectively. The black solid line indicates the local Tully-Fisher relation and the dotted black lines its $3\sigma$ scatter. The magenta solid line shows the fit to this sample with the slope fixed to the local value. It is offset from the local relation by $\Delta M_{B}=-0.63\pm 0.10$ mag.}
   \label{fig:TFR_hq}
\end{figure}

\noindent
Figure \ref{fig:TFR_hq} shows again the B-band Tully-Fisher diagram for the zCOSMOS galaxies, but this time taking into account only the high quality objects. The objects are again subdivided according to their redshifts. Again a line with the slope fixed to the local value is fitted through the data (magenta line). Including only the high quality objects, the fit is also offset from the local value toward higher luminosities. But with a mean value of $\Delta M_{B}=-0.63\pm 0.10$ mag, the offset here is considerably smaller. However, it is still larger than the value found by BZ16. Also the scatter is slightly smaller than if all objects are included ($\sigma_{obs}\sim 0.88$ mag).

\newpage

\noindent
Beside the fact that high-z objects are distributed mostly in the high luminosity-regime, another thing stands out in figures \ref{fig:TFR} and \ref{fig:TFR_hq}: the distribution in velocities of the high-z objects is also wider, that is the scatter is considerably larger. The reason for this are probably selection effects. Therefore it is interesting to look by how much the mean offset changes if only low-z objects ($z<0.5$) are taken into account. Fitting a line with the slope fixed again to -7.79 yields a mean offset of $\Delta M_{B}=-0.74\pm 0.11$ mag (only high quality objects: $\Delta M_{B}=-0.53\pm 0.10$). This is still larger than the value BZ16 get by taking into account galaxies up to $z\sim 1$, however the brightening is by $\sim 0.1$ mag weaker than if high-z objects are included as well. So it can be seen that although there are not so many high-z galaxies, they still strongly impact the mean offset due to their large scatter.\\
As a summary, figure \ref{tab:deltaMB_offsets} shows the mean TF offsets $\Delta M_{B}$ for the different zCOSMOS samples and for the FDF sample from BZ16:

\begin{table}[H]
\begin{center}
\small
  \begin{tabular}{ l | l  c | l  c  | c}  
      & All &    & Low-z ($z<0.5$) &  & \\ 
     & HQ \& LQ & HQ  & HQ \& LQ &  HQ & BZ16\\ \hline
    $\Delta M_{B}$ [mag] & $-0.83\pm 0.09$ & $-0.63\pm 0.10$ & $-0.74\pm 0.11$ & $-0.53\pm 0.10$ & $-0.47\pm 0.16$\\ 
  \end{tabular}
  \caption{\small The mean offsets from the local TFR $\Delta M_{B}$ for the different zCOSMOS samples and for the FDF sample from BZ16.}
  \label{tab:deltaMB_offsets}
\end{center}
\end{table}

\noindent
The large difference between the offsets of the zCOSMOS galaxies and the FDF galaxies is on one hand resulting from the fact that there are considerably less objects on the low-luminosity side of the local TFR. But also the lack of objects with very low luminosities (only 1 object with $M_{B}>-18$) does surely play a role.\\

\newpage
\noindent
In order to investigate an evolution with look-back time, a possible correlation of the TF offsets $\Delta M_{B}$ with redshift will be tested. The $\Delta M_{B}$-values for each object are calculated by subtracting the magnitude which would be expected from the local TFR from the observed magnitude:

\begin{equation}
\Delta M_{B}=M_{B}-[-7.79\cdot log(v_{max})-3.0]
\end{equation}

\noindent
These offsets are plotted against $log(1+z)$ which can be seen in the following figure:

\begin{figure}[H]
  \centering
  \includegraphics[height= 8 cm]{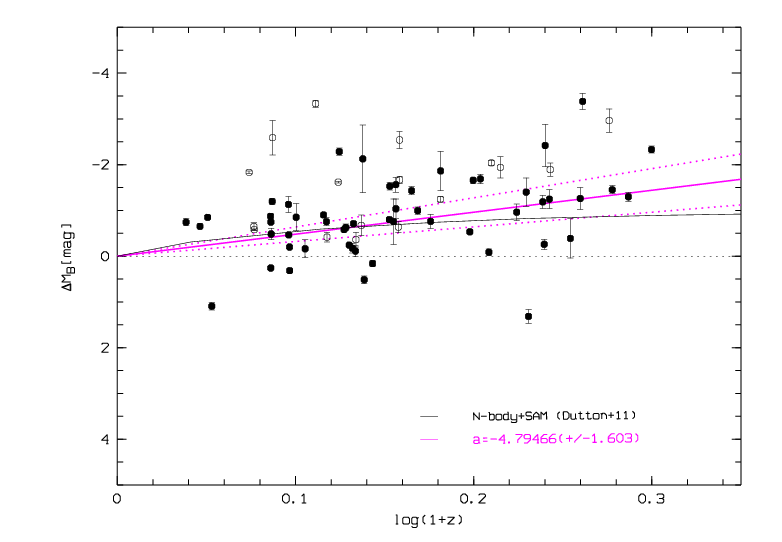}
  \caption{\small TF offsets $\Delta M_{B}$ plotted as a function of look-back time. With increasing redshift, the galaxies show increasing overluminosities. Plotting a linear fit (solid magenta line), we get $\Delta M_{B}=(-4.79\pm 1.6)\cdot log(1+z)$ which corresponds to a brightening of $\Delta M_{B}=-1.44\pm 0.5$ mag at z=1. The dotted black line indicates no evolution in luminosity, and the solid black line reflects the predictions from numerical simulations by Dutton et al. (2011).}
   \label{fig:deltaMB_z}
\end{figure}

\noindent
The errors on $\Delta M_{B}$ have been calculated through error propagation from the errors on $v_{max}$ and $M_{B}$. The dotted black line indicates no evolution in luminosity, which would mean that all objects lie on the local TFR. The solid black line reflects the predictions from numerical simulations by Dutton et al. (2011), who found an average brightening by $\Delta M_{B}=-0.9$ mag at z=1. A linear fit (solid magenta line) through the data yields: $\Delta M_{B}=(-4.79\pm 1.6)\cdot log(1+z)-(0.29\pm 0.27)$, which also shows increasing overluminosities toward higher redshifts, and corresponds to a mean brightening by $\Delta M_{B}=-1.44$ at z=1. However, compared to the results from BZ16, who get a brightening by $\Delta M_{B}=-1.2\pm 0.5$ mag at z=1, and the results from Dutton et al. (2011) ($\Delta M_{B}=-0.9$ mag), the resulting evolution in luminosity from the zCOSMOS sample is considerably larger. The reason for this strong evolution is probably again the lack of objects with positive TF offsets.\\

\begin{figure}[H]
  \centering
  \includegraphics[height= 8 cm]{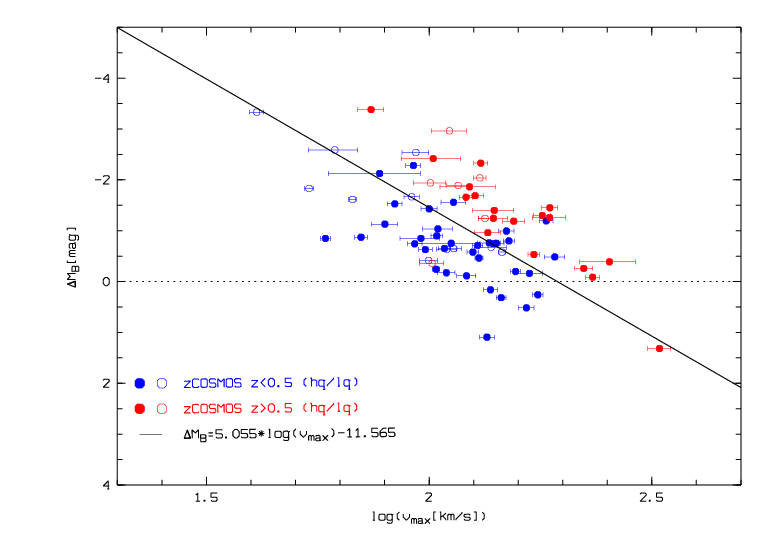}
  \caption{\small TF offsets $\Delta M_{B}$ plotted as a function of maximum rotation velocity. In this plot objects with slower rotation (lower masses) have larger overluminosities, whereas fast-rotating galaxies show more similar luminosities than their local counterparts. The dotted line corresponds to no evolution in luminosity, and the solid line indicates the linear fit through the data.}
   \label{fig:deltaMB_v}
\end{figure}

\noindent
In figure \ref{fig:deltaMB_v} the TF offsets $\Delta M_{B}$ are plotted against the maximum rotation velocity $v_{max}$. A linear fit to the sample yields: $\Delta M_{B}=5.055\cdot log(v_{max})-11.565$ (solid black line), whereas the dotted black line indicates no evolution in luminosity. There is a clear distinction between low-z and high-z objects. The high-z galaxies at given $v_{max}$ are almost always brighter than low-z galaxies with the same rotation velocity. Furthermore, it appears in this diagram that there is a correlation between $\Delta M_{B}$ and $v_{max}$. Objects with slower rotation velocities have almost all larger overluminosities (up to $\Delta M_{B}\sim 3$ mag), whereas the fast-rotating objects have more similar luminosities than their local counterparts. This could mean that the TFR slope does not remain constant, but changes with redshift which would imply a mass-dependent luminosity evolution of the TFR. However, as the zCOSMOS sample is not debiased and corrected for the incompleteness effect, the previously mentioned Malmquist bias could as well cause the effect seen in the figure below.

\begin{figure}[H]
  \centering
  \includegraphics[height= 8 cm]{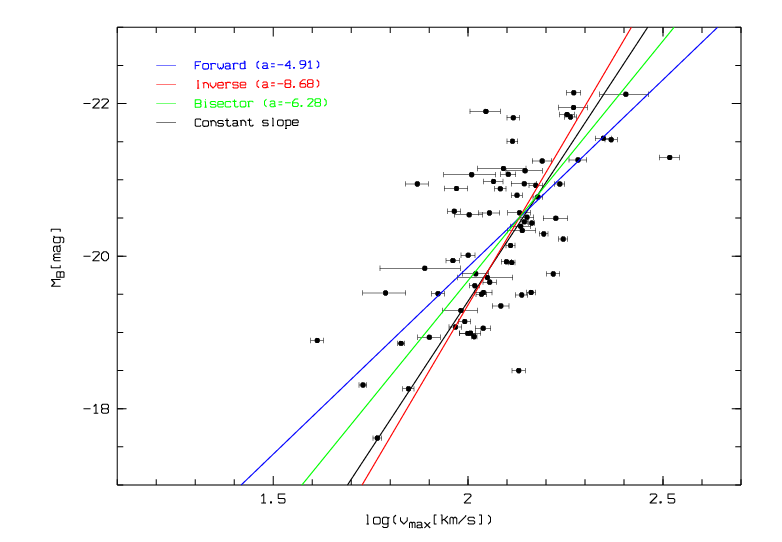}
  \caption{\small The Tully-Fisher diagram of the 70 zCOSMOS galaxies. The blue, green and red solid lines show the forward, bisector and inverse fit to this sample, respectively. The black solid line indicates the fit with the slope fixed to the local value (eq.\ref{localTFR_AB}).}
   \label{fig:fitmethods}
\end{figure}

\noindent
So far it was assumed that the slope of the TFR remains constant with redshift. In the following, the subject of slope evolution should be briefly addressed.\\
As has been already mentioned in section \ref{scalrel}, ever since the TFR of spiral galaxies was first observed, there was a debate about the origin of its slope and scatter, as well as about a possible evolution with redshift. A change of the magnitude zero-point implies an evolution in luminosity, whereas a change of the slope would furthermore mean an evolution depending on $v_{max}$, and thus a mass-dependent evolution. Through the years, observational studies of the evolution of the TFR, yielded different and also discrepant results. For example, B{\"o}hm et al. (2004) found that low-mass spirals at $z\sim 0.5$ were considerably brighter than local counterparts, whereas high-mass spirals evolved very weakly, if at all, in luminosity at given $v_{max}$~\cite{Boehm04}. Weiner et al. (2006) observed the opposite and found a stronger brightening of the high-mass than of the low-mass galaxies toward higher redshifts~\cite{Weiner06}.\\
Theoretical studies led as well to different results. In some N-body simulations, e.g. in Portinari and Sommer-Larsen (2007), only the zero-point is predicted to change, whereas the slope remains constant with cosmic look-back time~\cite{Portinari07}. In contrast, other studies, for example Ferreras and Silk (2001), who modelled the mass-dependent chemical enrichment history of disk galaxies using the local TFR as a constraint, found that the TF slope increases with redshift~\cite{Ferreras01}.\\
In their paper of 2004, in which they found a shallower slope of the distant TFR using a bisector fit, and thus a potential mass-dependent luminosity evolution, B{\"o}hm et al. (2004) also tested and ruled out some systematic errors that could bias the observed distant TF slope~\cite{Boehm04}. Amongst others, a tidally induced star formation in close galaxy pairs, different intrinsic absorption corrections and the RC quality could be excluded as the origin of the flatter slope. They also tested whether an incompleteness effect due to the apparent magnitude limit in the target selection could mimick a slope change (see figure \ref{main:az}). Such a limit leads to higher luminosities, thus higher masses, at higher redshifts, because a fraction of the low-luminosity (=low-mass) objects is missed. The low-mass galaxies that are still selected, might rather lie on the high-luminosity side of the TFR. However, no clear indications for this were found, hence they concluded that the tilt of the distant slope is flatter and that this could be partly attributed to a population of small, star-forming galaxies which in the local universe contribute less to the luminosity density. In 2007, B{\"o}hm and Ziegler tested this effect on a considerably larger sample and found that if the magnitude limit is taken into account, the slope of the distant TF would be consistent with the local slope if the TFR scatter decreased by more than a factor of 3 between $z\sim 0.5$ and 0~\cite{Boehm07}. With other words, a strong increase of the TFR scatter with redshift, e.g. due to a broader distribution in star formation rates, could mimick an apparent slope evolution.\\
Another important thing that must be mentioned in this context is that if the slope is not kept fixed, its value strongly depends on which fitting method is used. BZ16 performed a correction of the magnitude bias and derived the forward, inverse and bisector slopes for the uncorrected as well as for the corrected sample: $a_{f}=-3.71\pm 0.35$, $a_{b}=5.02\pm 0.47$, $a_{i}=-7.62^{+0.63}_{-0.78}$ and $a_{f}=-5.82\pm 0.35$, $a_{b}=-6.86\pm 0.41$, $a_{i}=-8.33^{+0.47}_{-0.53}$, respectively~\cite{Boehm16}. It can be seen that there are large differences between the slopes of the three different fit types. Furthermore, the forward as well as the bisector fit both yield considerably different slope values for the uncorrected and the de-biased sample, respectively. The slopes of the inverse fits though agree within the errors. This shows that the inverse fit is the least sensitive to the effect of the magnitude bias, whereas the other two methods are both rather strongly affected by it. BZ16 thus state that the inverse fit should be used in the case of the TFR. Additionally, the inverse fit slopes they find agree very well with the local slope a=-7.79 of Tully et al. (1998) who also used the inverse fitting method~\cite{Tully98}. From this one can conclude that it is justified to assume that the slope of the TFR remains constant with redshift, as has been done so far in this section.\\
In order to illustrate the differences between the three fitting methods, a forward, inverse and bisector fit were calculated for the 70 zCOSMOS galaxies. The three fits are plotted in the Tully-Fisher diagram seen in figure \ref{fig:fitmethods}. The resulting slopes for this sample are $a=-4.91$ (forward - blue line), $a=-6.28$ (bisector - green line) and $a=-8.68$ (inverse - red line). Although these values are all larger than the values found by BZ16 for their uncorrected sample, also in this case the large differences between the different fitting methods can be seen. The inverse slope is also larger than the slope from the local TFR, but taking into account its errors ($a=-8.68^{+0.81}_{-1.00}$), the two slope values are consistent. This can also be seen in figure \ref{fig:fitmethods}. The black solid line shows the fit to the zCOSMOS sample with the slope fixed to the local value. One can see that there is a difference between the inverse fit and the fixed-slope fit, however this difference is rather small, especially compared to the forward and the inverse fit.

\section{The velocity-size relation}
\label{VSR}
\begin{figure}[H]
  \centering
  \includegraphics[height= 6 cm]{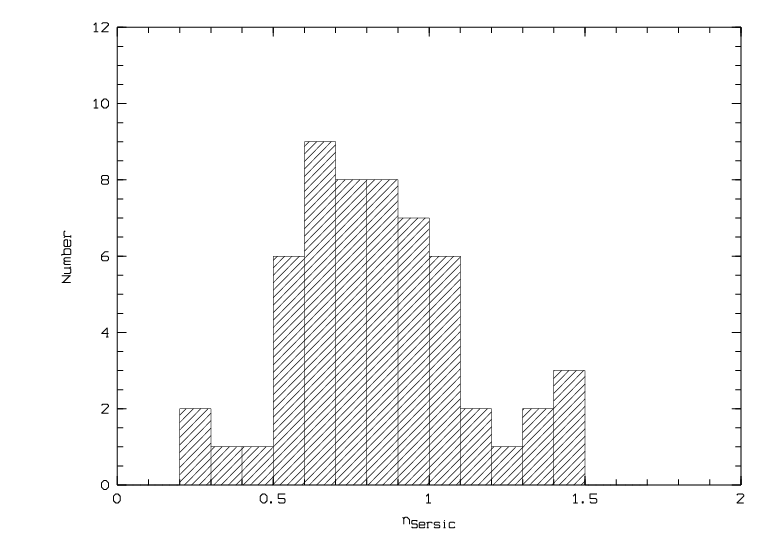}
  \caption{\small Distribution of the Sersic indices of 56 zCOSMOS galaxies.}
   \label{fig:Sersic}
\end{figure}

As has been mentioned in section \ref{scalrel}, the velocity-size relation (VSR) is another important scaling relation of disk galaxies, and as the TFR a projection of the so-called fundamental plane for spiral galaxies. It correlates the kinematic characteristics of spirals (mostly $v_{max}$) to a typical spatial scale (mostly the disk scale length $r_{d}$), showing that fast rotating galaxies are also larger than slower rotating galaxies. Just like the TFR, the VSR can also be used to analyse a possible evolution of the disk sizes with look-back time. As has also been explained before, it is believed that at given $v_{max}$, spiral galaxies at higher redshifts are on average smaller than local spiral galaxies (hierarchical merging scenario).\\
Before constructing the VSR with the zCOSMOS galaxies used in this thesis, two problems relating to the disk scale length $r_{d}$ should be considered: first, as mentioned earlier, the disk scale length was calculated from the effective radius $r_{1/2}$ with equation \ref{rd_eq}. However, as has been explained in section \ref{velpy}, this equation is valid for an exponential profile with the Sersic index n=1. If the Sersic index is $n>1$, $r_{d}$ becomes larger, and if $n<1$, $r_{d}$ becomes smaller. The Sersic index was available only for 56 of the 76 zCOSMOS galaxies, and the Sersic indices that were available, had large error bars. Therefore it was assumed that all galaxies have $n=1$. Figure \ref{fig:Sersic} shows a histogram of the available 56 Sersic indices. It can be seen that the assumption of $n=1$ for all galaxies is a good approximation as the majority of the values scatters around $n=1$.\\
The second point is that the effective radii $r_{1/2}$, as well as the Sersic indices, were all measured in the 814 nm-band. However, the restframe wavelength at which $r_{1/2}$ was actually measured, changes rather strongly with redshift, decreasing with $(1+z)^{-1}$. For example, the restframe wavelengths for objects at z=0.2, z=0.5 and z=0.8 would be 678 nm, 543 nm and 452 nm, respectively. The different restframe wavelengths affect the size of the effective radii which is smaller in redder filters. This effect was however not taken into account in this work and the $r_{1/2}$ values were not corrected for the wavelength dependence.

\begin{figure}[H]
  \centering
  \includegraphics[height= 8 cm]{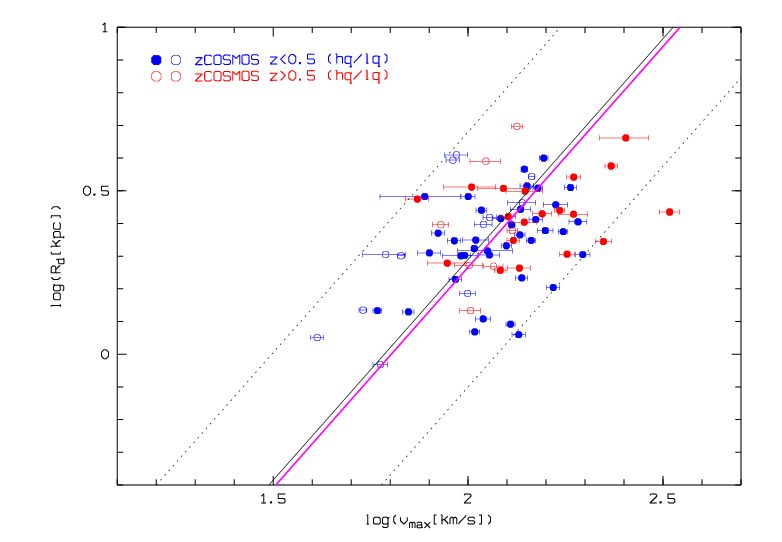}
  \caption{\small The velocity-size diagram of the 76 zCOSMOS galaxies. The objects with redshifts $z<0.5$ and $z>0.5$ are represented by blue and red symbols, respectively. The solid circles are the high quality, and the open circles are the low quality objects. The black solid line indicates the local velocity-size relation for the sample of Haynes et al. (1999) and the dotted black lines its $3\sigma$ scatter. The magenta solid line shows the fit to the zCOSMOS sample with the slope fixed to the local value. It is offset from the local relation on average by $\Delta log(r_{d})=-0.02\pm 0.02$ dex.}
   \label{fig:VSR}
\end{figure}

\noindent
In the following the VSR is constructed using $r_{d}$-values calculated with equation \ref{rd_eq}. One should keep in mind that due to the two facts mentioned above, in most cases the $r_{d}$'s should be somewhat larger or smaller. However, the calculated $r_{d}$ values still serve as a good approximation in order to analyse the evolution of the VSR.\\
Figure \ref{fig:VSR} shows the VSR-diagramm for the 76 zCOSMOS galaxies. The objects are again subdivided according to their quality and according to their redshifts. There are 25 high-z objects (19 of them being high quality objects) and 51 low-z objects (38 high quality). As a reference for the comparison of the distribution of the zCOSMOS galaxies in the $r_{d}-v_{max}$ space with the local distribution, the sample of $\sim 1.100$ local galaxies from Haynes et al. (1999) is used~\cite{Haynes99}. BZ16 calculated the bisector fit to these data which is in the following used as the local VSR~\cite{Boehm16}:

\begin{equation}
log(r_{d})=1.35\cdot log(v_{max})-2.41
\end{equation}

\noindent
In the figure below the local relation is indicated by the solid black line, and its $3\sigma$ scatter ($\sigma_{obs}=0.13$ dex) by the dotted black lines. As for the TFR, again a line with the slope fixed to the local value (=1.35) is fitted through the distant zCOSMOS sample. This fit is indicated by the solid magenta line and is offset from the local relation toward smaller sizes by $\Delta log(r_{d})=-0.02\pm 0.02$ dex. The scatter of the distant sample is $\sigma_{obs}\sim 0.2$ dex, which is about 1.5 x $\sigma_{obs}$ of the local sample from Haynes et al. (1999).
\newpage

\noindent
A mean evolution in disk sizes of $\Delta log(r_{d})=-0.02\pm 0.02$ dex is very small or one could say even almost no evolution at all. For comparison, BZ16 find a median offset of $\langle\Delta log(r_{d})\rangle =-0.10\pm 0.05$, which is significantly larger~\cite{Boehm16}. By looking at the diagram above one can see that the objects are distributed very uniformly around the local relation, and that almost all of them lie within the local $3\sigma$ scatter. Only three objects (one low-z and two high-z) lie outside this area, and all three on the smaller size-side of the local relation. However, from these three objects only one is considerably offset, the other two lie very close to the $3\sigma$-line. The object with the largest offset is the galaxy 830569 that has a redshift of $z=0.7014$. Interstingly it is also the object with the largest offset to the local TFR towards smaller luminosities. This means that on one hand it has a considerably smaller disk scale length than would be expected from the local VSR, but on the other hand its magnitude in the B-band is also smaller than expected from the local TFR. A possible explanation for the low luminosity would be an interaction process that caused quenching.\\

\begin{figure}[H]
  \centering
  \includegraphics[height= 6 cm]{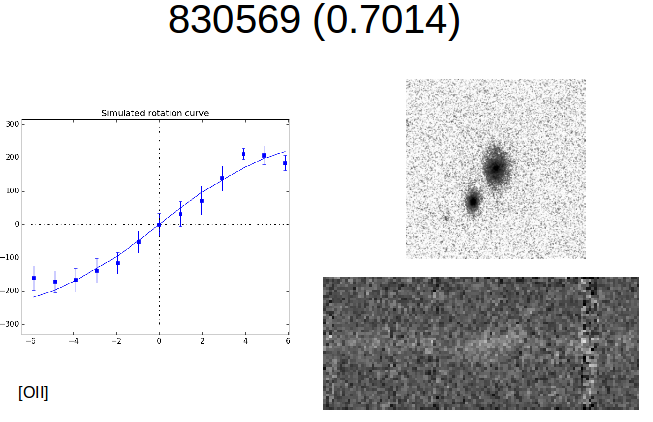}
  \caption{\small The RC, [OII]-emission line used for the RC extraction and HST-image of galaxy 830569 (z=0.7014).}
   \label{fig:830569}
\end{figure}

\noindent
Figure \ref{fig:830569} shows the RC, emission line and HST-image of 830569. For the RC extraction the [OII] emission line was used. On the cutout of the spectrum it can be seen that the emission line is very weak. The line as well as the continumm are not clearly distinguishable from the background. Also, the double structure of the emission line cannot be seen, it looks more like a single line. Still, the resulting RC looks very nice and symmetric, and also the HST-image shows a symmetric and undisturbed looking disk.
\newpage

\noindent
Another thing that can be see in figure \ref{fig:VSR} is that the objects with $z>0.5$ are located mostly on the "right" side of the local relation, meaning that they have smaller sizes than local galaxies with the same $v_{max}$. Only 8 out of 25 objects have larger sizes. The low-z galaxies however are distributed very uniformly on both sides of the local relation.\\
The majority of the low quality-objects has larger sizes than local galaxies, especially in the case of the low-z galaxies. Therefore, as in the case of the TFR, the VS-offset of just the high quality-objects should be analysed as well.

\begin{figure}[H]
  \centering
  \includegraphics[height= 8 cm]{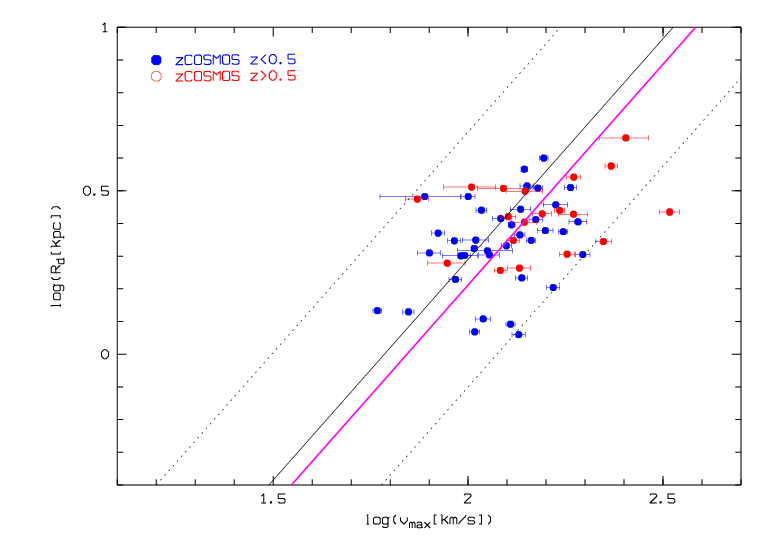}
  \caption{\small The velocity-size diagram of the 57 zCOSMOS galaxies with high quality RCs. The objects with redshifts $z<0.5$ and $z>0.5$ are represented by blue and red symbols, respectively. The black solid line indicates the local velocity-size relation and the dotted black lines its $3\sigma$ scatter. The magenta solid line shows the fit to this sample with the slope fixed to the local value. It is offset from the local relation on average by $\Delta log(r_{d})=-0.08\pm 0.02$ dex.}
   \label{fig:VSR_hq}
\end{figure}

\noindent
Figure \ref{fig:VSR_hq} shows the VSR for the 57 high quality galaxies. The local relation and its $3\sigma$ scatter are again indicated by the solid black and the dotted black lines, respectively. The solid magenta line shows the fit through the data with the slope fixed to the local value.\\
It can be seen that if only high quality objects are taken into account, the offset to the local VSR is much larger. The fit with the slope fixed to the local value yields a mean decreasing in size by $\Delta log(r_{d})=-0.08\pm 0.02$ dex, which is - taking into account the error bars - in agreement with the value found by BZ16 ($\Delta log(r_{d})=-0.10\pm 0.05$).\\

\begin{figure}[H]
  \centering
  \includegraphics[height= 8 cm]{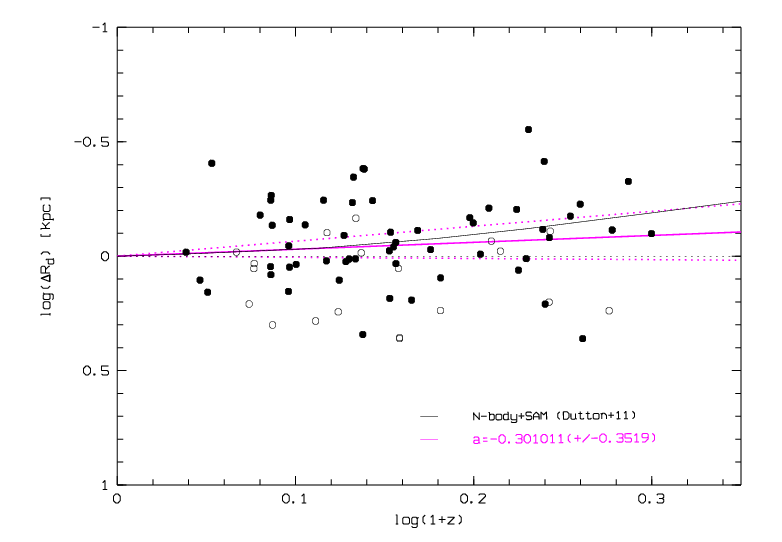}
  \caption{\small VSR offsets $\Delta log(r_{d})$ plotted as a function of look-back time. With increasing redshift, the size of the galaxies decreases. Plotting a linear fit (solid magenta line), we get $\Delta log(r_{d})=(-0.3\pm 0.35)\cdot log(1+z)$ which corresponds to a decrease in size by a factor $\sim 1.23$ at z=1. The dotted black line indicates no evolution in size, and the solid black line reflects the predictions from numerical simulations by Dutton et al. (2011).}
   \label{fig:Deltard_z}
\end{figure}

\noindent
In order to get information about a possible evolution with redshift, again the offsets from the local VSR $\Delta log(r_{d})$ are plotted against $log(1+z)$ which can be seen in figure \ref{fig:Deltard_z}. The VS offsets are calculated by subtracting the expected disk scale length from the observed disk scale length:

\begin{equation}
\Delta log(r_{d})=log(r_{d})-[1.35\cdot log(v_{max})-2.41]
\end{equation}

\noindent
The dotted black line in figure \ref{fig:Deltard_z} indicates no evolution in size, which would mean that all objects lie on the local VSR. The solid black line reflects the predictions from numerical simulations by Dutton et al. (2011), who found that disk galaxies at z=1 are on average smaller by $\sim -0.19$ dex in logarithmic disk scale length than their local counterparts~\cite{Dutton11}. A linear fit (solid magenta line) through the data yields: $\Delta log(r_{d})=(-0.30\pm 0.35)\cdot log(1+z)+(0.02\pm 0.06)$. According to this fit, galaxies at z=1 were smaller by $\sim 0.09\pm 0.09$ dex, or a factor of $1.23^{+0.34}_{-0.26}$ than local galaxies with the same $v_{max}$. This is a rather small evolution which can also be seen in figure \ref{fig:Deltard_z}, and it is also smaller than the evolution predicted by Dutton et al. (2011) or by BZ16 who find a decrease of $\Delta log(r_{d}) = -0.16$ dex at z=1~\cite{Boehm16}. Vergani et al. (2012) also found a slightly weaker evolution and a decrease in size by $\Delta log(r_{d})=-0.12$ at z=1.2~\cite{Vergani12}. The linear fit through the data in figure below would yield an average decrease of -0.1 dex at this redshift. This implies an evolution smaller than predicted by Vergani et al. (2012), but the difference is not as large as compared to the predictions by Dutton et al. (2011) and by BZ16.\\

\noindent
Of course one has to keep in mind that the statements and predictions from the plots in this section are not completely correct because, as has been referred to at the beginning of this section, the scale disk lengths have been calculated based on the assumption that the Sersic index is $n=1$ for all objects and that the rest frame wavelength is always 814 nm.\\
In the case of the FDF sample, BZ16 applied the correction for the wavelength dependence of $r_{d}$. But they state that if this correction would have been omitted, the resulting size evolution would be weaker and would correspond to $\Delta log(r_{d})=-0.11$ dex at z=1, which is more similar to the value found in this work~\cite{Boehm16}. They also write that the effect of the wavelength dependence of $r_{d}$ is very weak in the F814W filter for redshifts up to $z\sim 1$, which is also the case for the zCOSMOS sample. In the case of their FDF sample the maximum overestimate of $r_{d}$ is 11\% at z=0.97 which is not very large.\\
So despite the fact that the values of the disk scale lengths of the zCOSMOS galaxies are not entirely accurate, it can be said that they are still representative and that the plots \ref{fig:VSR} and \ref{fig:Deltard_z} reflect the ongoing disk growth with cosmic time which is expected in the hierarchical growth scenario (e.g. Mao et al (1998)~\cite{Mao98}).

\section{$\Delta M_{B}$ vs. $\Delta logr_{d}$}
\begin{figure}[H]
  \centering
    \subfloat[] {\label{main:ad}\includegraphics[width= 8.5 cm]{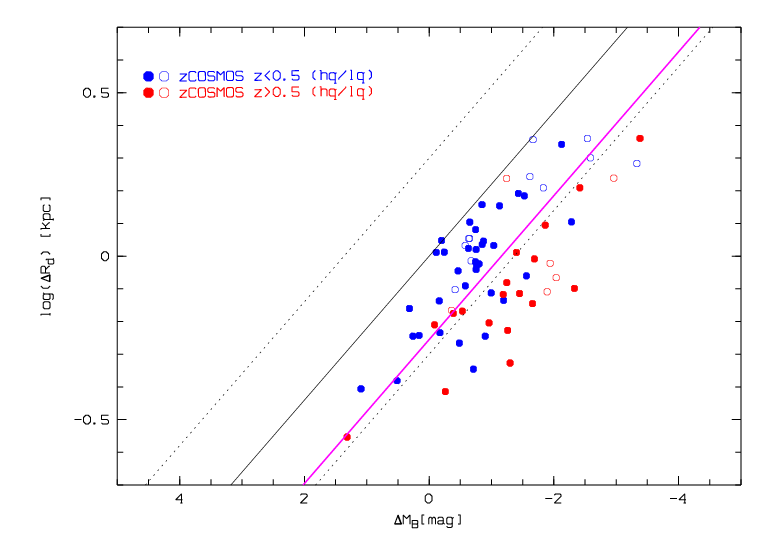}}\\
    \subfloat[] {\label{main:ae}\includegraphics[width= 8.5 cm]{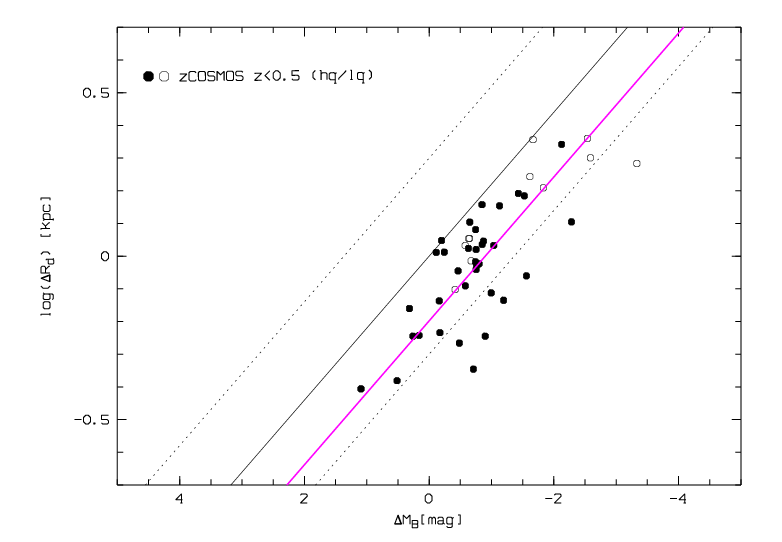}}
    \subfloat[] {\label{main:af}\includegraphics[width= 8.5 cm]{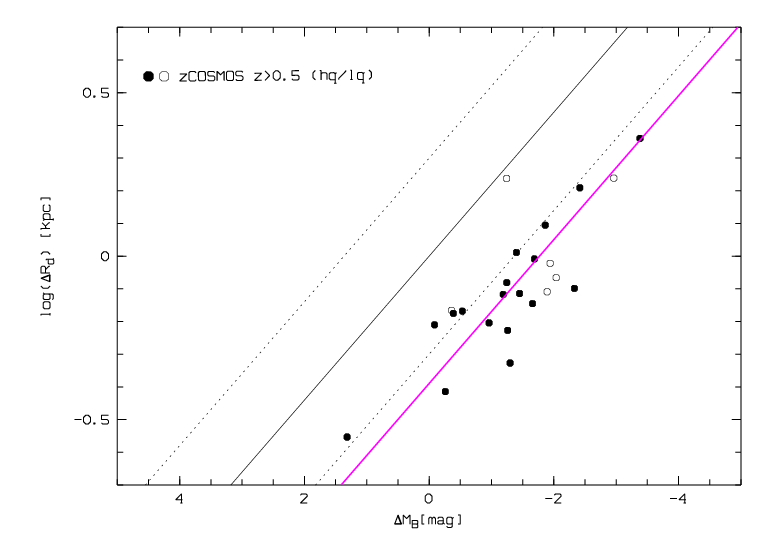}}
  \caption{\small Correlation between the offsets $\Delta M_{B}$ from the TFR and $\Delta log(r_{d})$ from the VSR. The solid and dotted black lines indicate the correlation of these parameters for the local sample from Haynes et al. (1999) and its $3\sigma$-scatter. Panels \textit{a}, \textit{b} and \textit{c} show the distribution of all 70 zCOSMOS objects, of the low-z and of the high-z objects, respectively. A fixed-slope fit to the data is indicated by a solid magenta line.}
   \label{fig:deltaMB_deltard}
\end{figure}

The plots and results from the previous two sections have shown that disk galaxies obviously evolve with time in luminosity as well as in size. But they do not say anything about how these evolutions are linked, or if they are linked at all. Do the galaxies with the largest TF offsets also have the largest VS offsets? In order to investigate this question in more detail, the $\Delta M_{B}$ offsets are compared to the $\Delta log(r_{d})$ offsets. This can be seen in figure \ref{fig:deltaMB_deltard}. The sample from Haynes et al. (1999) is again used as the local reference~\cite{Haynes99}. BZ16 calculated the TF offsets $\Delta M_{B}$ and the VS offsets $\Delta log(r_{d})$ from the local relations for these galaxies~\cite{Boehm16}. The relation was constructed in such a way that the median of both parameters is zero. In figure \ref{fig:deltaMB_deltard} it is depicted with the black solid line. The dotted black lines show the $3\sigma$-local scatter which in terms of $\Delta log(r_{d})$ is $\sigma =0.09$ dex. The fact that the offsets $\Delta M_{B}$ and $\Delta log(r_{d})$ are correlated can be explained by means of the fundamental plane of spiral galaxies that has been described in chapter \ref{scalrel} (e.g. Koda et al. (2000)~\cite{Koda00a}). The TFR as well as the VSR are both projections of this three-dimensional plane. This means that e.g. a deviation of a galaxy in the local universe from the local TFR is related to its position in the luminosity-size space, and thus to its size. As BZ16 describe it, a $\Delta M_{B}$-$\Delta log(r_{d})$ diagram "is equivalent to an edge-on view on the fundamental plane of disk galaxies in the direction of, but not parallel to, the $v_{max}$ axis"~\cite{Boehm16}.\\

\noindent
Figure \ref{fig:deltaMB_deltard} shows three panels. In all subfigures the local relation and its $3\sigma$-scatter are plotted for comparison. In \ref{main:ad} all 70 objects with available $\Delta M_{B}$ and $\Delta log(r_{d})$ values are shown. The objects are again subdivided according to their quality category and their redshift. In order to investigate how these two parameters are correlated in the case of distant galaxies, a line with the slope fixed to the local value was plotted to the data (solid magenta line). It can be seen already in \ref{main:ad} that objects with $z<0.5$ are distributed differently than objects with $z>0.5$. Therefore a line with the slope fixed was plotted to the low-z as well as the high-z objects separately which can be seen in the subfigures \ref{main:ae} and \ref{main:af}.

\noindent
It can be seen that in all three cases the fixed-slope fits are shifted away from the local $\Delta M_{B}-\Delta log(r_{d})$ relation. If all objects are taken into account, the offset in terms of $\Delta log(r_{d})$ is $-0.26\pm 0.02$ dex. If however low-z and high-z objects are used seperately, the fixed-slope fits imply an evolution of $-0.20\pm 0.02$ dex and $-0.39\pm 0.03$ dex, respectively. Regarding the plots above it is important to bear in mind that any evolution in luminosity and size is imprinted on the correlation between the parameters $\Delta M_{B}$ and $\Delta log(r_{d})$.

\noindent
The plots in figure \ref{fig:deltaMB_deltard} state that this correlation between $\Delta M_{B}$ and $\Delta log(r_{d})$ holds at least up to $z\sim 1$.
There are only half as many high-z objects than low-z objects, but still it can be seen how the evolution gets stronger with increasing redshift. It also seems that the shape of the distribution of distant galaxies does not change compared to the local distribution, but that it is just shifted towards smaller sizes and higher luminosities. Another thing that can be seen is that the galaxies with the strongest evolution in luminosity do not automatically have also the strongest evolution in size. The same applies to galaxies with the weakest evolution in luminosity. As there are not so many objects, especially with $z>0.5$, it is not easy to make precise statements. In the low-z bin the galaxies with the largest $\Delta M_{B}$ have mostly $\Delta log(r_{d})>0$ which means that they are at given $v_{max}$ a little bit larger than local galaxies. In contrast, the galaxies that have the strongest decrease in size seem to lie in the region around $\Delta M_{B}\sim 0$, ranging from $\Delta M_{B}\sim -1$ to 1. For the high-z objects it is more difficult to make such statements due to the even smaller number of objects. But also here the three galaxies with the largest $\Delta M_{B}$'s have larger disks than expected from the local VSR. And galaxies with the the strongest decrease in size have smaller luminosity offsets than galaxies that lie around $\Delta log(r_{d})\sim 0$.

\section{The Stellar Tully-Fisher relation}
\label{STFR}
Now I analyse and examine in more detail the stellar TFR. As the B-band TFR, also the stellar TFR correlates the visible matter to the total matter of a galaxy. But instead of the B-band magnitude, now the relation between the stellar mass $M_{*}$ of a galaxy and its maximum rotation velocity $v_{max}$ is investigated.\\
The stellar masses were computed as described in Maier et al. (2015), using the software HyperzMass for the SED fitting analysis and the estimation of the galaxy stellar masses based on the optical to IRAC photometric data from COSMOS~\cite{Maier15}. The software HyperzMass fits photometric data points with synthetic stellar population models and selects the best-fit parameters. The stellar mass of a galaxy is then obtained by integrating the star formation rate (SFR) over the age of the galaxy and correcting for the mass loss in the course of stellar evolution.\\
 
\noindent
The SFR of a galaxy can be measured by means of emission line luminosities. Here, the $H_{\alpha}$-line is one of the most reliable and well calibrated SFR indicators~\cite{Maier09}. However, in the case of optical surveys, this line mostly cannot be observed for objects with redshifts $z>0.5$ because it is redshifted out of the observed range. Instead of $H_{\alpha}$, the [OII]-emission line is often used as a SFR tracer for objects with $z>0.5$. This was also the case for the zCOSMOS galaxies. The SFR of the low-z galaxies were measured by means of the $H_{\alpha}$-line, and the SFR of the high-z galaxies by means of the [OII]-line. The equations used for the conversion of emission line luminosities to SFR were as follows:

\begin{equation}
SFR (M_{\odot}yr^{-1})=7.9\cdot 10^{-42}L(H_{\alpha}) ergs/s
\end{equation}

\noindent
and

\begin{equation}
log[SFR (M_{\odot}yr^{-1})] = log[L(OII)(ergs/s)]-41-0.195\cdot M_{B}-3.434
\end{equation}

\noindent
The equation for the conversion of the $H_{\alpha}$-luminosity is taken from Kennicutt (1998)~\cite{Kennicutt98} and the equation for the conversion of the [OII]-luminosity from Maier et al. (2009)~\cite{Maier09}.\\

\noindent
The information about the SFR and the stellar mass was not available for all 76 objects, but only for 51 (32 low-z and 19 high-z galaxies). In order to decide which of the 51 objects should be used for the stellar TFR, the so-called main sequence of star-forming galaxies was used. The main sequence is a relation that correlates the stellar mass of spiral galaxies to their specific star formation rate (sSFR). The sSFR is defined as the SFR per mass and could therefore be calculated from the available SFR and stellar masses~\cite{Peng10}.\\
The SFR is not constant, but depends on the redshift. This means that also the main sequence changes with cosmic time. For a certain redshift, the corresponding main sequence can be calculated with the following equation from Peng et al. (2010)~\cite{Peng10}:

\begin{equation}
\label{sSFR_eq}
sSFR(m,t)=2.5\left( \frac{m}{10^{10}M_{\odot}}\right) ^{\beta}\left( \frac{t}{3.5 Gyr}\right) ^{-2.2}Gyr^{-1}
\end{equation}

\noindent
Here, \textit{m} is the stellar mass and \textit{t} is the age of the universe at the selected redshift which depends on the selected cosmology. For the slope $\beta$ a value of -0.24 is assumed~\cite{Renzini15}.\\
By comparing galaxies of a certain redshift to the corresponding main sequence, it is possible to see which galaxies are "typical" star-forming galaxies. Galaxies that lie far from the main sequence should be excluded from further analysis like the stellar TFR as their SFR are either far too high than expected (e.g. starburst galaxies) or far too low.\\
As the 51 zCOSMOS galaxies with available SFR and stellar masses data span a wide redshift range, two main sequences were computed: for the low-z objects the relation was calculated for $z=0.3$ and for the high-z objects for $z=0.7$. The expressions for the two main sequences are:

\begin{equation}
\label{mainseq}
\begin{aligned}
log(sSFR(z=0.3))=-0.24\cdot log(M_{*})+1.79\\
log(sSFR(z=0.7))=-0.24\cdot log(M_{*})+2.10
\end{aligned}
\end{equation}

\noindent
The sSFR-$M_{*}$ plots for the low-z as well as the high-z objects can be seen in figure \ref{fig:main_seq}. The main sequences from equation \ref{mainseq} are indicated by the solid magenta line. For both relations a scatter in terms of log(sSFR) of $\pm 0.3$ dex was assumed. The criterion to be included into the stellar TFR-sample was that the galaxies are not more than $\pm 0.3$ dex away from the scatter area of the main sequence. This is because the sSFR are assumed to have an error of a factor two which in logarithmic space corresponds to 0.3 dex~\cite{Christianpersonal}. Based on these conditions, nine objects from the low-z sample and one object from the high-z sample are excluded. These objects are shown as filled red dots in the following plots:
\newpage

\begin{figure}[H]
  \centering
    \subfloat[] {\label{main:ag}\includegraphics[height= 8 cm]{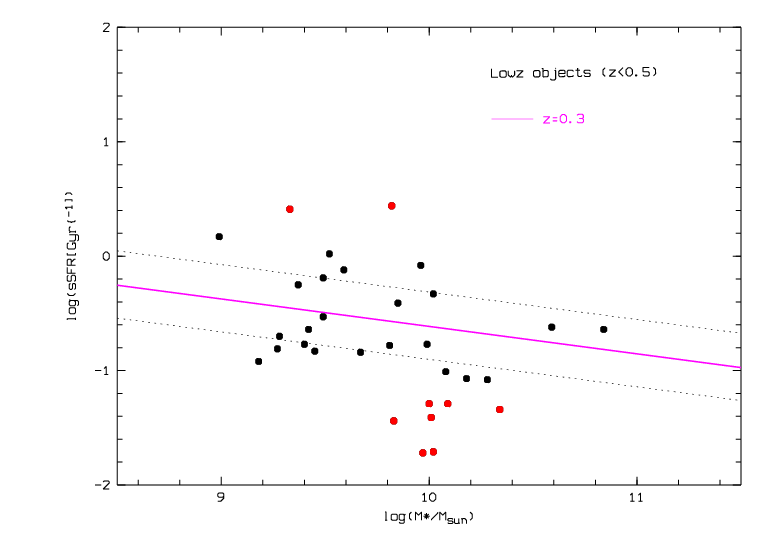}}\\
    \subfloat[] {\label{main:ah}\includegraphics[height= 8 cm]{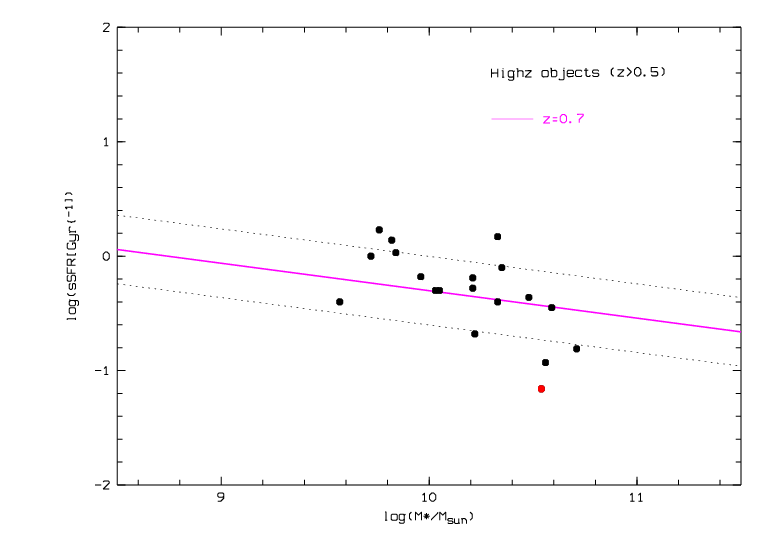}}
  \caption{\small sSFR vs. stellar mass for objects with redshifts $z<0.5$ (\textit{a}) and with redshifts $z>0.5$ (\textit{b}). The main sequences for z=0.3 and z=0.7 are indicated by the solid magenta line, and the scatter of $\pm 0.3$ dex with the black dotted lines. The objects that lie more than $\pm 0.3$ away from the scatter area are shown with filled red dots.}
   \label{fig:main_seq}
\end{figure}

\newpage
\noindent
The ID's and redshifts of the ten objects excluded from the stellar TFR analysis are listed in table \ref{tab:mainseq}:

\begin{table} [h!]
\begin{center}
  \begin{tabular}{ c  c | c  c }    
   ID & \textit{z} & ID & \textit{z} \\ \hline
810934 & 0.4401 & 837485 & 0.4334\\
811102 & 0.4404 & 837603 & 0.2492\\
812329 & 0.2195 & 837768 & 0.3404\\
812516 & 0.3706 & 837846 & 0.4212\\
830598 & 0.5187 & 837865 & 0.4221\\
  \end{tabular}
    \caption{\small The ten objects (red dots in figure \ref{fig:main_seq}) excluded from the stellar TFR-sample.}
  \label{tab:mainseq}
\end{center}
\end{table}

\noindent
Figure \ref{fig:TFRMstar} shows the stellar Tully-Fisher diagram for the remaining 41 objects. The low-z objects (23) are shown with blue symbols and the high-z objects (18) are shown with red symbols. A distinction between high quality (filled circles) and low quality objects (open circles) is also made. As the local reference the relation from Reyes et al. (2011) is used who constructed different versions of the local TFR by means of a sample of 189 disc galaxies at redshifts $z<0.1$~\cite{Reyes11}. For the stellar masses they used two different photometric stellar mass estimates, on one hand estimates from Bell et al. (2003) (~\cite{Bell03}) and on the other hand estimates based on fits to the \textit{u, g, r, i, z} spectral energy distribution (SED) from the MPA/JHU group (~\cite{mpa}). In this thesis the local relation using the stellar masses from Bell et al. (2003), $M_{*,Bell}$, is used. However, this relation was computed based on the Kroupa initial mass function (IMF), whereas the $M_{*}$-values from the zCOSMOS galaxies are based on the Salpeter IMF. Stellar masses from the Salpeter IMF are on average by a factor of $\sim 1.5$ larger than masses from the Kroupa IMF, assuming the same amount of ionizing radiation~\cite{Brinchmann04}. So in order to convert masses from a Kroupa IMF to a Salpeter IMF, 0.18 dex have to be added to the zeropoint, which corresponds to a factor of 1.5. This value of 0.18 dex is, as described by Pozzetti et al. (2007), a "systematic median offset [...] in the masses derived with the two different IMFs", which means that individual offsets can of course deviate from this value, but it still serves as a good estimator regarding the conversion of Kroupa stellar masses into Salpeter stellar masses~\cite{Pozzetti07}.\\
In addition, Reyes et al. (2011) used $v_{80}$ as the measure for the rotation velocity which is defined as the rotation velocity "at a radius equal to the major axis of an ellipse containing 80\% of the total integrated $H_{\alpha}$ flux, $r_{80}$"~\cite{Tiley16}. Reyes et al. (2011) write that in most cases $v_{80}$ samples the flat region of a RC and that for most galaxies it is within 10\% of $v_{2.2}$ (rotation velocity at 2.2 $r_{d}$) and on average only about 5\% larger than the latter~\cite{Reyes11}. So although it is not the same as the maximum rotation velocity, it still serves as a good estimate for $v_{max}$. In figure \ref{fig:TFRMstar}, the resulting fit of the local relation, given by $log(M_{*})=3.64\cdot log(v_{80})+2.49$, is indicated by the solid black line:
\newpage

\begin{figure}[H]
  \centering
  \includegraphics[height= 8 cm]{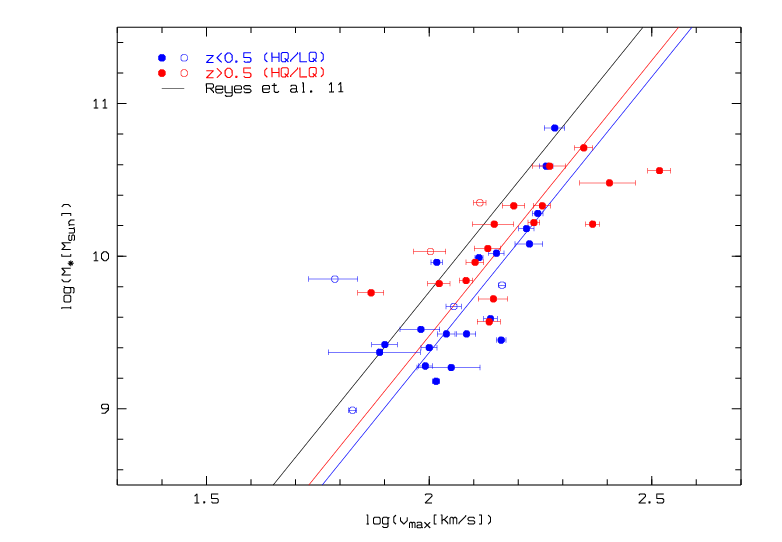}
  \caption{\small The stellar Tully-Fisher diagram of the 41 zCOSMOS galaxies with available SFR and $M_{*}$ information. The objects with redshifts $z<0.5$ and $z>0.5$ are represented by blue and red symbols, respectively. The black solid line indicates the local Tully-Fisher relation from Reyes et al. (2011). The blue and the red solid lines show the fixed-slope fit to the low-z and the high-z sample, respectively.}
   \label{fig:TFRMstar}
\end{figure}

\noindent
In order to investigate an evolution of the stellar TFR with redshift, a line with the slope fixed to the local value was fitted to the low-z and the high-z data. In figure \ref{fig:TFRMstar} these fits are shown with the blue and the red solid line, respectively. As $v_{max}$ represents the total mass of a galaxy (dark matter, gas and stars) and it is assumed that the total mass of a galaxy does not really change much with time (except in the case of minor mergers), also $v_{max}$ should stay more or less the same since $z\sim 1$. The stellar mass however should change because over time more and more gas is converted into stars. The consequence of this assumption is thus that distant galaxies at given $v_{max}$ should have lower $M_{*}$'s than local galaxies. This is definitely the case for the zCOSMOS galaxies seen in the diagram above. The fits through the data are shifted to the right relative to the local relation which implies smaller stellar masses. However, the blue fit is shifted somewhat more strongly to the right than the red fit, which means that according to the fits the low-z galaxies have on average smaller stellar masses than the high-z galaxies. The mean offset of the low-z galaxies is $\Delta M_{*}=-0.397\pm 0.058$ dex, whereas the stellar masses of the high-z galaxies are on average smaller by $\Delta M_{*}=-0.295\pm 0.089$ dex. If only the high quality objects are taken into account, the offsets get a little bit larger, but with $\Delta M_{*}=-0.416\pm 0.067$ dex the low-z galaxies still have on average smaller stellar masses than the high-z galaxies with $\Delta M_{*}=-0.380\pm 0.084$ dex.\\
However it can be seen that the low-z and high-z fits are consistent within scatter. One should also bear in mind that with only 41 galaxies the number of objects used for this analysis is rather small. There are also less high-z galaxies than low-z galaxies. It can also be seen that below $log(M_{*}/M_{\odot})$ $\sim 9.5$ there are only low-z galaxies which can probably also be ascribed to selection effects. Enlarging the sample would allow to make more reliable statements about the behaviour of low-z and high-z galaxies.\\
Another possible explanation could be again that some $v_{max}$ are underestimated. There are three high-z objects (one high and two low quality) on the left side of the local relation. The high quality object that has a rather large offset to the local relation, is the galaxy 837355 that also had a very strong offset to the local B-band TFR. Its RC, emission line and HST-image are shown in figure \ref{fig:offset_exp}. As has been explained in section \ref{TFRMB}, it is quite possible that $v_{max}$ is underestimated for this object. The same applies to the two low quality objects.\\
It is interesting to compare the positions of the galaxies with the highest $v_{max}$ in the stellar and the B-band TF diagram. In the stellar TF diagram it can be seen that three of the four fastest rotating high-z galaxies are clearly offset from the local relation, having much smaller stellar masses than local galaxies with the same $v_{max}$. The offset of the galaxy which has the fourth-highest rotation velocity is considerably smaller. In the case of the B-band TFR (figure \ref{fig:TFR}) the situation is different. One can see that three of these four galaxies lie very near or rather on the local B-band TFR. The galaxy with the highest $v_{max}$ (830569) however has a quite large offset. But it is not, as would be expected, brighter than local galaxies, but fainter. It is interesting that the offset of this galaxy in the stellar TF diagram meets the expectations, but the offset in the B-band TF diagram is just the opposite as expected for high-z galaxies.\\
A fit with the slope fixed to the local value through all 41 objects yields a mean offset of $\Delta M_{*}=-0.372\pm 0.047$ dex. This is in quite good agreement with the results from Puech et al. (2008) and Tiley et al. (2016). Tiley et al. (2016) constructed the stellar TFR for a subsample of 56 galaxies from the KMOS Redshift One Spectroscopic Survey (KROSS) with redshifts between $z=0.8-1.0$ and found an evolution of the zero-point of $\Delta M_{*}=-0.41\pm 0.08$ dex~\cite{Tiley16}. Puech et al. (2008) constructed the stellar TFR only for galaxies up to $z\sim 0.6$ and also found a significant evolution of $\Delta M_{*}=-0.36$ dex~\cite{Puech08}.\\

\begin{figure}[H]
  \centering
  \includegraphics[height= 7 cm]{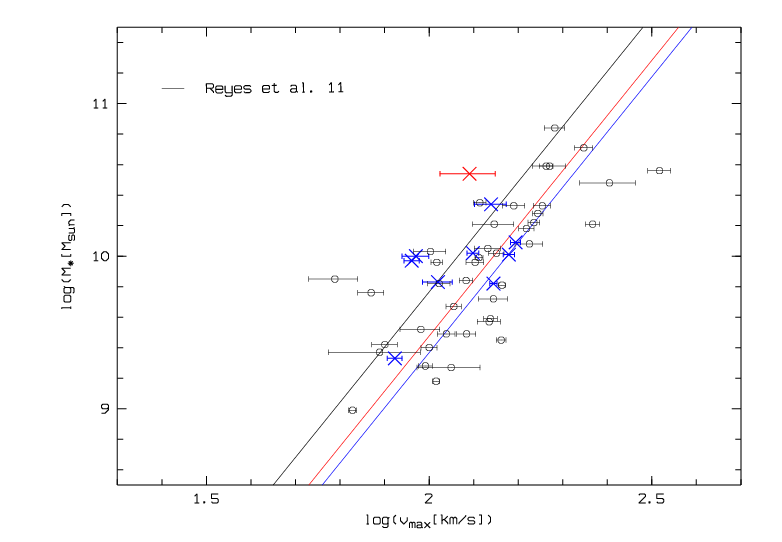}
  \caption{\small The stellar Tully-Fisher diagram of the 51 zCOSMOS galaxies with available SFR and $M_{*}$ information. The 41 objects from figure \ref{fig:TFRMstar} are represented by black open circles, whereas the 10 excluded objects from table \ref{tab:mainseq} are represented by blue (low-z) and red (high-z) crosses. The black solid line indicates the local Tully-Fisher relation from Reyes et al. (2011). The blue and the red solid lines show the fixed-slope fit to the low-z and the high-z sample from figure \ref{fig:TFRMstar}.}
   \label{fig:TFRMstar_bad}
\end{figure}

\noindent
In order to examine how objects with too high or too low sSFR fit into the stellar Tully-Fisher diagram, figure \ref{fig:TFRMstar_bad} shows the same plot as figure \ref{fig:TFRMstar}, but now including the ten objects from table \ref{tab:mainseq}. These ten objects are represented by blue (low-z) and red (high-z) crosses. The 41 objects from figure \ref{fig:TFRMstar} are represented by black open circles.\\
It can be seen that the majority of the low-z objects (seven objects if the error bars are taken into account) lie on the right side of or on the local relation. Three of these objects are located very close to the fixed-slope fit of the low-z sample from figure \ref{fig:TFRMstar}. So although these seven objects were excluded due to their strong deviation from the main sequence, they fit very good into the stellar TF diagram. Five of these seven objects, as well as the two objects on the left of the local relation (810934 and 811102), have sSFR smaller than expected from the main sequence. In contrast, the remaining two low-z objects (837865 and 812329) have considerably higher sSFR.\\
The only excluded high-z object (830598) lies on the left side of the local relation, i.e. towards higher stellar masses, but has a smaller sSFR than "typical" star-forming galaxies, as can be seen in figure \ref{fig:main_seq}.

\newpage

\section{The Stellar mass-Halo mass relation}
With stellar masses available for a large part of the 76 zCOSMOS galaxies that were used for the kinematic analysis, it is interesting to study also the relation between $M_{*}$ and $M_{h}$, that is the mass of the host dark matter halos of the disk galaxies. It is assumed that galaxy formation is driven by the growth of the underlying large-scale structure and formation of dark matter halos~\cite{Moster13}. With other words: the cooling and condensation of gas at the centres of the potential wells of these dark matter halos leads to the formation of galaxies. This however means that there should be a tight correlation between the physical properties of galaxies, e.g. their SFR or their stellar masses, and the masses of the halos surrounding them. A popular approach to study and understand this correlation are numerical simulations. In this section the $M_{*}$-$M_{h}$ relation will be constructed for the 51 zCOSMOS galaxies and compared with the predictions of N-body simulations performed by Moster et al. (2010).\\
Based on Newton's law of gravity, the mass interior to a certain orbit can be calculated if the radius of the orbit as well as the rotation speed are known. For the halo mass $M_{h}$ this can be expressed with the following equation~\cite{Cattaneo14}~\cite{Kravtsov13}:

\begin{equation}
\label{Mhalo}
M_{h}=\dfrac{v_{circ}^{2}\cdot r_{vir}}{G}
\end{equation}
 
\noindent
Here, the virial radius of the dark matter halo $r_{vir}$ is used as an approximation for the size of the halo, $v_{circ}$ is the circular velocity of the halo and G is Newton's gravitational constant ($6.674\times 10^{-11} N\cdot m^{2}/kg^{2}$). However, $v_{circ}$ and $r_{vir}$ are not available for the galaxies used in this thesis, as they are quantities that are not easily measured. This means that $M_{h}$ cannot be calculated directly from $v_{circ}$ and $r_{vir}$. What can be calculated, is the dynamical mass of a galaxy. The "dynamical" mass $M_{dyn}$ is defined as the total mass (that is stars, gas and dark matter) of the galaxy within a certain radius. It can be estimated with the equation:

\begin{equation}
\label{Mdyn}
M_{dyn}=\dfrac{v_{max}^{2}\cdot r_{1/2}}{G},
\end{equation}

\noindent
which is the same as for $M_{h}$, just that now the half-light radius $r_{1/2}$ is used as the size (scale) of the galaxy and $v_{max}$ as the rotation velocity at this radius. Using the half-light radius instead of the disk scale length $r_{d}$ in this equation has the advantage that the influence of the Sersic indices on $r_{d}$ and thus the problem of possibly incorrect disk scale lengths (as mentioned in the sections above, e.g. \ref{VSR}) does not play a role. It should be noted that the equation above assumes that galaxies are supported mainly by rotation, and is thus not valid for dispersion-dominated galaxies. For simplicity, it is assumed here that all zCOSMOS galaxies used for this analysis are completely rotation-supported.\\
If the dynamical mass of a galaxy is known, the mass of its dark matter halo can be estimated by means of it. A link between $M_{dyn}$ and $M_{h}$ can be established roughly by finding conversions of $r_{vir}$ and $v_{circ}$ into $r_{1/2}$ and $v_{max}$. This can be achieved by using three relations. First of all, Cattaneo et al. (2014) find the following correlation between the maximum rotation velocity $v_{max}$ of a galaxy and the circular velocity $v_{circ}$ of the dark matter halo~\cite{Cattaneo14}:

\begin{equation}
v_{circ}=\dfrac{v_{max}}{1.3}
\end{equation}

\noindent
Furthermore, with the following equation from Kravtsov (2013) the virial radius can be expressed by means of the half-mass radius $r_{1/2_{m}}$~\cite{Kravtsov13}:

\begin{equation}
\label{Kravtsov}
r_{vir}=\frac{r_{1/2_{m}}}{0.015}
\end{equation}

\noindent
The half-mass radius is not known, but it is linked to the half-light radius $r_{1/2}$ by the following relation from Szomoru at al. (2013)~\cite{Szomoru13}:

\begin{equation}
r_{1/2}=1.33\cdot r_{1/2_{m}},
\end{equation}

\noindent
which inserted in equation \ref{Kravtsov} yields:

\begin{equation}
r_{vir}=\frac{r_{1/2}}{0.02}=50\cdot r_{1/2}
\end{equation}

\noindent
If all these relations are now inserted in equation \ref{Mhalo}, we obtain a correlation between $M_{dyn}$ and $M_{h}$:

\begin{equation}
\label{Mh_Mdyn}
M_{h}\sim 30\cdot M_{dyn}
\end{equation}

\begin{figure}[H]
  \centering
  \includegraphics[height= 8 cm]{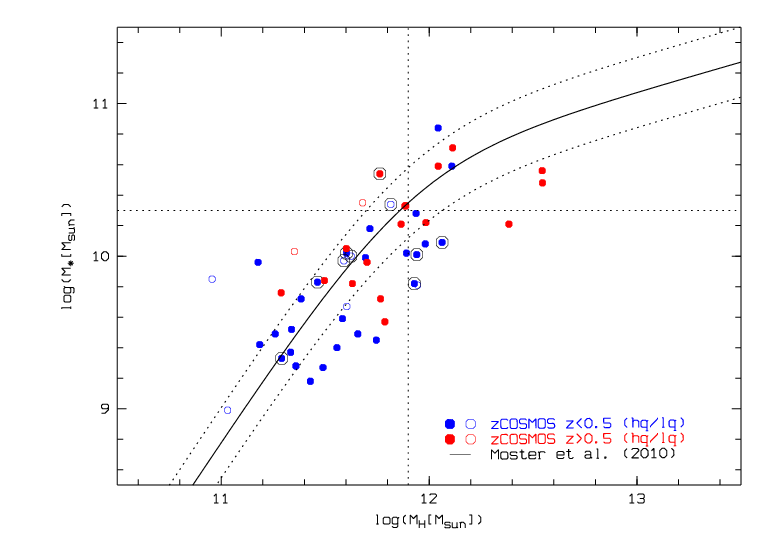}
  \caption{\small Stellar mass plotted against halo mass of 51 zCOSMOS galaxies. The objects with redshifts $z<0.5$ and $z>0.5$ are represented by blue and red symbols, respectively. The solid circles are the high quality, and the open circles are the low quality objects. The black solid line indicates the fit from Moster et al. (2010), and the black dotted lines the scatter of $\sigma_{m}=\pm 0.15$ dex. The black circles mark the objects from table \ref{tab:mainseq}.}
   \label{fig:MsMh}
\end{figure}

\noindent
It should be noted that the transformations above are all only rough estimates and only for given assumptions. The factor 30 in equation \ref{Mh_Mdyn} is thus just an average factor for the ratio $\frac{M_{h}}{M_{dyn}}$. Normally this ratio should be calculated for each galaxy separately, so one should keep in mind that there are probably slight deviations for individual galaxies. However, equation \ref{Mh_Mdyn} serves as an approximation in order to study the behaviour of the $M_{*}$-$M_{h}$ relation. Therefore the "dynamical mass" $M_{dyn}$ (eq. \ref{Mdyn}) was calculated for the 51 zCOSMOS galaxies with available stellar masses, and subsequently multiplied by 30. The resulting halo masses are plotted in figure \ref{fig:MsMh} against the stellar masses.\\

\noindent
In order to compare the correlation of $M_{*}$ and $M_{h}$ of the zCOSMOS galaxies with predictions from simulations, the following equation for the stellar-to-halo mass (SHM) ratio of the local universe from Moster et al. (2010) is used~\cite{Moster10}:

\begin{equation}
\label{Moster10}
\frac{M_{*}}{M_{h}}=2\left( \frac{M_{*}}{M_{h}}\right)_{0}\left[ \left( \frac{M_{h}}{M_{1}}\right)^{-\beta}+ \left( \frac{M_{h}}{M_{1}}\right)^{\gamma}\right] ^{-1}
\end{equation}

\begin{figure}[H]
  \centering
  \includegraphics[height= 6.5 cm]{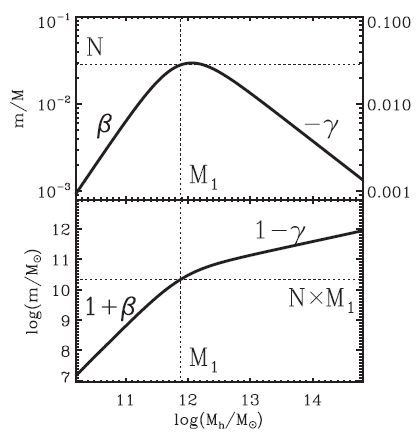}
  \caption{\small Upper panel: sketch of the SHM ratio $M_{*}/M_{h}$ as a function of $M_{h}$ peaking around the characteristic mass $M_{1}$ where it has the normalization $N=(M_{*}/M_{h})_{0}$. $\beta$ and $-\gamma$ are the low-mass and the high-mass slope, respectively. Lower panel: sketch of the SHM relation as a function of $M_{h}$. The low-mass and the high-mass log-slopes are $1+\beta$ and $1-\gamma$, respectively~\cite{Moster13}.}
   \label{fig:Moster13}
\end{figure}

\noindent
There are four free parameters in this equation. $(M_{*}/M_{h})_{0}$ is the normalization of the SHM ratio, $M_{1}$ ist the characteristic mass where the SHM ratio is equal to $(M_{*}/M_{h})_{0}$, and $\beta$ and $\gamma$ are two slopes that indicate the behavior of $M_{*}/M_{h}$ at the low-mass and high-mass ends, respectively.\\
Moster et al. (2010) compared the halo mass function $n(M_{h})$ to the galaxy mass function $\varphi (M_{*})$ in order to constrain the stellar-to-halo mass (SHM) relation~\cite{Moster10}. First they derived the galaxy stellar-mass function (SMF) trivially from the halo mass function (HMF), assuming that every system has exactly the same SHM ratio. So the resulting galaxy SMF has understandably the same features as the HMF. Then they compared the galaxy SMF derived for $M_{*}/M_{h} = 0.05$ to the observed Sloan Digital Sky Survey (SDSS) galaxy SMF and found that the observed SMF is steeper for high masses and shallower for low masses than the one derived from the HMF. This implies that the actual SHM ratio $M_{*}/M_{h}$ is not constant, but that baryons are converted into stars with different efficiencies in halos of different mass~\cite{Moster13}. From the comparison of the observed SMF to the one derived from the HMF, Moster et al. (2010) adopted the parametrization shown in equation \ref{Moster10} which implies that the SHM ratio first increases with increasing mass, but after reaching a maximum at the characteristic mass $M_{1}$ it decreases again~\cite{Moster10}. A sketch of the resulting SHM relation can be seen in figure \ref{fig:Moster13}.

\noindent
In this sketch, \textit{N} is the normalization $(M_{*}/M_{h})_{0}$ of the SHM ratio, $M_{1}$ is the characteristic mass with $(M_{*}/M_{h})=(M_{*}/M_{h})_{0}$, and $\beta$ and $-\gamma$ (or ($1+\beta$) and ($1-\gamma$)) are the slopes of the low- and the high-mass ends, respectively.\\
Assuming a scatter of $\sigma_{m}=0.15$ dex in stellar mass, Moster et al. (2010) found the following best fit values for the free parameters~\cite{Moster10}:

\begin{table} [h!]
\begin{center}
  \begin{tabular}{ l  c  c  c }    
   log$(M_{1}/M_{\odot})$ & $(M_{*}/M_{h})_{0}$ & $\beta$ & $\gamma$ \\ \hline
   11.899 & 0.02817 & 1.068 & 0.611 \\
  \end{tabular}
    \caption{\small Fitting results for stellar-to-halo mass relationship including a scatter of $\sigma_{m}=0.15$ dex~\cite{Moster10}.}
  \label{tab:Moster_par}
\end{center}
\end{table}

\noindent
Inserting these values in equation \ref{Moster10} yields the fit shown in figure \ref{fig:MsMh} which shows the stellar masses plotted against the halo masses. The fit from Moster et al. (2010) is indicated by the black solid line. The dotted black lines show the scatter of $\sigma_{m}=\pm 0.15$ dex. For visualization, also the characteristic mass $log(M_{1}/M_{\odot})=11.899$ which separates the low-mass and the high-mass regime is indicated by a vertical dotted line. The horizontal dotted line marks the "characteristic stellar mass", that is $N\times M_{1}$, or rather $(M_{*}/M_{h})_{0}\times M_{1}$.\\
Just like in previous plots, the objects in figure \ref{fig:MsMh} are devided into low-z (blue symbols) and high-z (red symbols) galaxies, as well as high quality (filled symbols) and low quality (open symbols) galaxies.

\noindent
It can be seen that the majority of the objects have halo masses below the characteristic mass $M_{1}$. Only 14 objects have masses $log(M_{h}/M_{\odot})>11.899$. The objects in the low-mass regime ($log(M_{h}/M_{\odot})<11.899$) seem to be in quite good agreement with the fit by Moster et al. (2010). Almost 60\% of these galaxies lie within the $1\sigma$-regime. As for the high-mass objects, it seems that there is a change in slope, but it is not really possible to make precise statements because of the lack of objects in this regime. The greater part of these objects has halo masses $log(M_{h}/M_{\odot})<12.2$, only three objects have larger masses. Furthermore, in the low-mass as well as the high-mass regime the objects outside the $1\sigma$-area lie for the most part at smaller stellar masses, and thus smaller SHM ratios. Only eight objects have stellar masses larger than expected. These eight objects are listed in the following table:

\begin{table} [h!]
\begin{center}
  \begin{tabular}{ c  c  c  c | c  c  c  c}    
   ID & \textit{z} & log$(M_{*}/M_{\odot})$ & log$(M_{h}/M_{\odot})$ & ID & \textit{z} & log$(M_{*}/M_{\odot})$ & log$(M_{h}/M_{\odot})$ \\ \hline
   811183 & 0.2221 & 9.85 & 10.96 & 830598 & 0.5187 & 10.54 & 11.76\\
   812388 & 0.6215 & 10.35 & 11.68 & 837355 & 0.8248 & 9.76 & 11.29\\
   817366 & 0.6406 & 10.03 & 11.35 & 837461 & 0.2199 & 10.84 & 12.04  \\
   830408 & 0.2475 & 9.42 & 11.19 &839451 & 0.3054 & 9.96 & 11.18 \\   
  \end{tabular}
    \caption{\small ID's, redshifts, stellar masses and halo masses of the eight objects that lie outside the $1\sigma$-area of the fit from Moster et al. (2010) (eq. \ref{Moster10}) and have larger stellar masses, and thus larger SHM ratios, than expected.}
  \label{tab:large_stellmass}
\end{center}
\end{table}

\noindent
The smaller SHM ratios are in agreement with the assumption mentioned in section \ref{STFR} that distant galaxies should have lower stellar masses than local galaxies, and thus lower SHM ratios. However, there is no clear difference between the distributions of low-z and high-z galaxies visible. The scatter in $M_{*}$ of the objects in this plot gives us information about how much gas can be converted into stars at a given $M_{h}$.\\
For the diagrams of this section also the objects from table \ref{tab:mainseq}, which were excluded from the stellar TFR due to their strong deviation from the main sequence of star-forming galaxies, were included. These ten objects are marked in the above plot and the following plots with black circles. It can be seen that they fit well into the diagram and that their distribution does not differ from the distribution of the other 41 objects. Six of these objects lie within the $1\sigma$-area. Three objects, all with redshifts $z<0.5$ (837603, 837846 and 812329), have stellar masses smaller than expected from the fit. In contrast, the only high-z object (830598) lies above the upper $1\sigma$-line and thus has a higher $M_{*}$ than expected. Also in figure \ref{fig:TFRMstar_bad} this object lies above the local stellar TFR, i.e. towards higher stellar masses.\\

\begin{figure}[H]
  \centering
  \includegraphics[height= 8 cm]{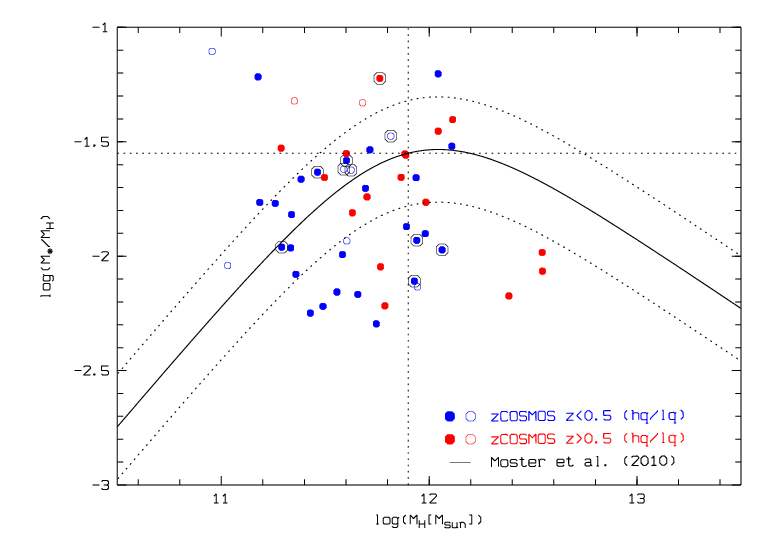}
  \caption{\small SHM ratio plotted against halo mass of 51 zCOSMOS galaxies. The objects with redshifts $z<0.5$ and $z>0.5$ are represented by blue and red symbols, respectively. The solid circles are the high quality, and the open circles are the low quality objects. The black solid line indicates the fit from Moster et al. (2010), and the black dotted lines the scatter of $\sigma_{m}=\pm 0.15$ dex. The black circles mark the objects from table \ref{tab:mainseq}.}
   \label{fig:MsMh_ratio1}
\end{figure}

\noindent
Figure \ref{fig:MsMh_ratio1} is in principle the same as figure \ref{fig:MsMh}, just that now the SHM ratio of the 51 zCOSMOS galaxies is plotted against the halo mass. The fit from Moster et al. (2010) is again indicated by the black solid line, and the $\sigma_{m}=0.15$ dex-scatter with the black dotted lines. As in the previous plot, the objects are devided according to their redshift and quality. Also the characteristic mass $log(M_{1}/M_{\odot})=11.899$ and the normalization $(M_{*}/M_{h})_{0}$ of the SHM ratio are indicated by a vertical and a horizontal dotted line, respectively. The objects from table \ref{tab:mainseq} are again marked with black circles.\\
In this plot it is more difficult to see a clear trend because of the large scatter regarding the SHM ratio of the objects. Still a slight trend is visible, especially in the low-mass regime. In the high-mass regime it is again not really possible (as also in figure \ref{fig:MsMh}) to make clear statements about the behaviour of the objects in this area, because first, there are not so many objects and second, there are no objects with really high masses. As already mentioned, the three objects with the highest masses have halo masses between $log(M_{h}/M_{\odot})\simeq 12.4-12.6$, but there are no objects with even higher masses. Furthermore, there is also a "gap" between these three and the other objects. It looks like the three massive galaxies would follow the fit, but with lower $M_{*}/M_{h}$. But because of the lack of objects with halo masses between $log(M_{h}/M_{\odot})\simeq 12.2-12.4$, it is hard to say how strictly the objects in this regime follow the expected trend.\\
Again, no real differences are visible between low-z and high-z objects, except that at the low-mass end of the distribution there are for the most part objects with $z<0.5$, and at the high-mass end objects with $z>0.5$. This was also already visible in figure \ref{fig:MsMh}. Also by looking at how many objects lie below and how many above the fit, no distinctive trend can be seen. From the low-z objects 15 from 33 lie above the fit, which corresponds to $\sim 45.5\%$. In the high-z regime 8 from 18 objects have larger SHM ratios than expected ($\sim 44.4\%$).

\begin{figure}[H]
  \centering
  \includegraphics[height= 8 cm]{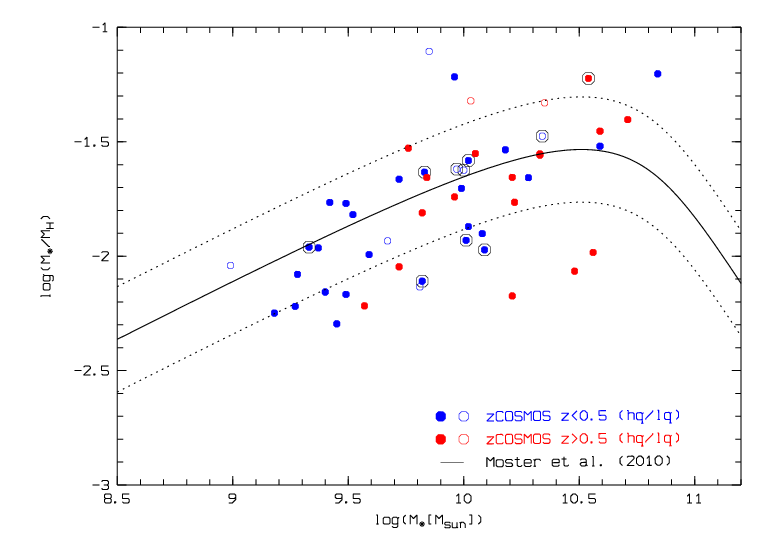}
  \caption{\small SHM ratio plotted against stellar mass of 51 zCOSMOS galaxies. The objects with redshifts $z<0.5$ and $z>0.5$ are represented by blue and red symbols, respectively. The solid circles are the high quality, and the open circles are the low quality objects. The black solid line indicates the fit from Moster et al. (2010), and the black dotted lines the scatter of $\sigma_{m}=\pm 0.15$ dex. The black circles mark the objects from table \ref{tab:mainseq}.}
   \label{fig:MsMh_ratio2}
\end{figure}

\noindent
The same can now be done for the stellar mass, by plotting the SHM ratio against $M_{*}$. This can be seen in figure \ref{fig:MsMh_ratio2}. Although in principle only the x-axis changed compared to figure \ref{fig:MsMh_ratio1}, the plot looks quite different now, because of the "broader" spectrum of stellar masses in log-space populated by the objects. Whereas the halo masses $M_{h}$ in the plot above lie mostly between $10^{11}$ and $10^{12.2} M_{\odot}$, the stellar masses $M_{*}$ in figure \ref{fig:MsMh_ratio2} range from $10^{9}$ to $10^{10.8} M_{\odot}$.\\
As in the previous plots, the objects seem to follow the fit quite well until the peak at about $log(M_{*}/M_{\odot})\sim 10.5$, but after the peak there are only about 6 objects. And these objects are quite widely dispersed which makes it even more difficult to say if they match the expected behaviour. Only three of these objects are within the $1\sigma$-area, the others are rather distant to the fit. Again, no clear difference is visible between low-z and high-z objects.

\newpage

\noindent
In order to analyse further, how well the zCOSMOS objects follow the expected SHM ratio, a line is fitted to the objects in figure \ref{fig:MsMh}. The purpose of this is to test how well the resulting slope matches the slope from Moster et al. (2010). As there are very few objects with halo masses larger than $log(M_{h}/M_{\odot})>11.899$, only the slopes in the lower-mass regime are compared, that is the slopes $\beta$ and $1+\beta$ (see figure \ref{fig:Moster13}). The three rightmost high-z objects are excluded from this analysis because they are clearly in the regime where the $\gamma$-slope dominates the SHM ratio.\\
An inverse fit through the remaining 48 objects yields a slope of $a=2.0995$. As the slope in figure \ref{fig:MsMh} is denoted with $1+\beta$, $\beta$ is 1.0995. If only the high quality objects are used for the fit, the resulting slopes are $1+\beta=2.094$ and $\beta=1.094$, respectively. As can be seen in table \ref{tab:Moster_par}, Moster et al. (2010) use $\beta=1.068$~\cite{Moster10}. This value is very similar to the values found for the zCOSMOS sample, indicating that the zCOSMOS galaxies follow the expected local SHM ratio rather good.\\
If the slope is calculated for the low-z and high-z objects separately, slight differences are found. The respective slopes of the low-z and the high-z subsample are $1+\beta=1.984$ ($\beta=0.984$) and $1+\beta=2.321$ ($\beta=1.321$), respectively. From the steeper high-z slope it could be inferred that the star formation efficiency at $z>0.5$ increased stronger with halo mass than at $z<0.5$.\\
In their paper, Moster et al. (2010) also study the evolution of the parameters in eq. \ref{Moster10} with redshift. They find that the normalization of the SHM ratio $(M_{*}/M_{h})_{0}$ decreases with increasing redshift, while the characteristic mass $M_{1}$ increases~\cite{Moster10}. This means that at higher redshifts there is less stellar content in a halo of a given mass. The values of the high-mass slope $\gamma$ also decrease with increasing redshift, but have very large error bars. Finally, the low-mass slope $\beta$ increases until $z\sim 2$ and then seems to drop to a low value.\\
The results found in this thesis agree with the findings of Moster et al. (2010) that the low-mass slope increases with increasing redshift. The difference between the local value of $\beta$ and the value for the zCOSMOS galaxies is not very large, but still it implies an increase with look-back time. If only low-z galaxies are taken into account, the resulting $\beta$ is slightly smaller than the local value. However, the different low-mass slopes of the low-z and the high-z sample also imply an increase of $\beta$ with increasing redshift.

\chapter{Discussion}
\label{discussion}
The purpose of this master thesis was to study the kinematics of spiral galaxies up to redshifts of $z\sim 1$ and draw conclusions from this about their evolution in the last eight billion years. In order to do that, scaling relations were constructed for a sample of zCOSMOS galaxies with redshifts up to $z\sim 1$ and compared with local relations. Scaling relations link fundamental characteristics of galaxies to one another, and are thus a very powerful tool to analyse the nature and evolution of galaxies. With regard to spiral galaxies, the Tully-Fisher relation, which links the rotation velocity (e.g. $v_{max}$) of spiral galaxies to their luminosity, and the velocity-size relation are of particular importance.\\
For the kinematic analysis, spectra taken with the "Visible Multi-Object Spectrograph" (VIMOS) were used. Five masks, each containing spectra of $\sim 30$ zCOSMOS galaxies, were reduced with the software package VIPGI ("VIMOS Interactive Pipeline and Graphical Interface"). After the data reduction, rotation curves (RCs) were extracted from emission lines that were sufficiently extended and bright. Subsequently, the maximum rotation velocity $v_{max}$ was derived by modelling the observed RCs, considering several observational and geometrical effects, like e.g. inclination and seeing. $v_{max}$ is one of the parameters required for the construction of the scaling relations. The other parameters - the absolute B-band magnitudes $M_{B}$, the disk scale lengths $r_{d}$ and the stellar masses $M_{*}$ - were obtained from the zCOSMOS catalogues.\\
The modelled RCs were visually inspected and selected according to their RC quality and symmetry. From a total of 110 spectra with clearly recognizable emission lines, 76 objects were selected for the kinematic analysis. These 76 objects were furthermore subdivided into high and low quality RCs.\\
The main findings from the kinematic analysis are summarized and discussed below:\\

\noindent
\textbf{The B-band Tully-Fisher relation:} The distant B-band TFR, which links the absolute restframe B-band magnitudes $M_{B}$ to $v_{max}$ and investigates the ongoing and recent star formation, was constructed using 70 zCOSMOS galaxies (figure \ref{fig:TFR}). As it is assumed that only the zero-point of the TFR changes with time, but not its slope, a line with the slope fixed to the local value was fitted to the data points and compared to the local TFR. The comparison between the two relations reveals an average brightening of $\langle\Delta M_{B}\rangle=-0.83\pm 0.09$ mag for the distant galaxies. This mean offset is larger than the offset found by other studies. E.g. BZ16 find a mean brightening of $\langle\Delta M_{B}\rangle = -0.47\pm 0.16$ mag for a sample of FDF galaxies~\cite{Boehm16}. If only high quality objects of the zCOSMOS sample are taken into account, the mean offset is $\langle\Delta M_{B}\rangle=-0.63\pm 0.10$ mag and thus smaller than if all objects are included. This value is still larger than the value found by BZ16, but it coincides within the error margins.\\
In order to investigate an evolution with look-back time, the TF offsets were plotted against the redshift (figure \ref{fig:deltaMB_z}). A linear fit through the data implies that galaxies at z=1 are on average by about $\Delta M_{B}=-1.44\pm 0.5$ mag brighter than local galaxies with the same $v_{max}$. The evolution implied by this is also stronger than the evolution found by other studies. BZ16 find a brightening of $\Delta M_{B}=-1.2\pm 0.5$ at z=1~\cite{Boehm16}. Numerical simulations also find a smaller brightening at z=1, e.g. Dutton et al. (2011) ($\Delta M_{B}=-0.9$ mag) or Portinari and Sommer-Larson (2007) ($\Delta M_{B}=-0.85$ mag)~\cite{Dutton11}~\cite{Portinari07}.\\
Although the value found for the zCOSMOS sample agrees with most of these values within errors, it implies a stronger evolution in luminosity with redshift. There are several possible explanations for this. One reason could be that the sample used in this thesis consists of objects with lower maximum rotation velocities as only a small subsample of zCOSMOS galaxies was selected for the kinematic analysis. The comparison between the $M_{B}$, $r_{d}$ and $v_{max}$ distributions of the zCOSMOS sample and the FDF sample from BZ16 (figure \ref{fig:distributions}) shows that the latter has more galaxies with luminosities brighter than $M_{B}=-21$ mag and disc scale lengths larger than $r_{d}=3.5$ kpc. This would explain the differences in the $v_{max}$ distributions and the smaller number of galaxies with high $v_{max}$ to some extent. Another possibility is that $v_{max}$ may be underestimated for some objects. Furthermore, the errors of $v_{max}$, which also contribute to the mean offset, are likely estimated too small. Finally, the different approaches for the $M_{B}$-determination performed in the case of the zCOSMOS and the FDF galaxies could also be a reason. Systematic differences between the luminosity determinations could explain the different mean TF offsets.\\
Compared to other studies, the number of galaxies used for the kinematic analysis in this thesis is smaller. So enlarging the sample would allow to make more accurate statements about the evolution of the B-band TFR.\\

\noindent
\textbf{The velocity-size relation:} By means of the VSR, which links the rotation velocity $v_{max}$ to the disk scale length $r_{d}$, a possible evolution of the disk sizes of spiral galaxies with look-back time can be explored.\\
The distant VSR was constructed by fitting a line with the slope fixed to the value of the local VSR to the zCOSMOS sample (figure \ref{fig:VSR}). Compared to the local relation, the resulting distant VSR is offset toward smaller sizes by $\Delta log(r_{d})=-0.02\pm 0.02$ dex. This agrees with the prediction of the hierarchical scenario of galaxy formation that distant galaxies are smaller than local galaxies. However the indicated evolution is very small. With $\Delta log(r_{d})=-0.10\pm 0.05$ dex, BZ16 find for example a significantly larger offset of the distant VSR~\cite{Boehm16}. If however only high quality objects of the zCOSMOS sample are taken into account, the mean offset changes to $\langle\Delta log(r_{d})\rangle=-0.08\pm 0.02$ dex which is comparable to the value found by BZ16. Also, the majority of the high-z objects are located on the "right" side of the local relation, i.e. toward smaller sizes. This agrees with the assumption that objects with higher redshifts have smaller sizes.\\
Investigating the offsets $\Delta log(r_{d})$ against the redshift (figure \ref{fig:Deltard_z}) implies that spiral galaxies at z=1 are on average by $\Delta log(r_{d})=-0.09\pm 0.09$ dex smaller than their local counterparts. It is noteable that the error is of the same order as the change in size. This predicted evolution in size is smaller than the evolution found by other studies. For example, BZ16 and Dutton et al. (2011) find $\Delta log(r_{d})=-0.16$ dex and $\Delta log(r_{d})=-0.19$ dex at z=1, respectively~\cite{Boehm16}~\cite{Dutton11}. In contrast, Vergani et al. (2012) also find a weaker decrease in size ($\Delta log(r_{d})=-0.12$ dex at z=1.2)~\cite{Vergani12}. This is still slightly larger than the evolution found in this thesis, however the difference is not very large.\\
Two problems related to the computation of the disk scale length $r_{d}$ have to be considered. First, the disk scale length was calculated from the effective radius $r_{1/2}$ with the equation $r_{d}=r_{1/2}/1.7$ which is valid only for galaxies with an exponential profile with the Sersic index n=1. It is thus assumed that all galaxies used for the construction of the distant VSR have a Sersic index of n=1, which may not be the case. The second point is that the $r_{1/2}$ were all measured in one HST filter (F814W) and that the correction for the wavelength dependence of the size was not taken into account.\\
Computing the disk scale lengths by taking into account the correct restframe wavelength and Sersic index, would allow to make more precise statements regarding the evolution of the VSR. However it should be noted that for redshifts up to $z \sim 1$, the effect of the wavelength dependence is assumedly very weak in the F814W filter (maximum overestimate of $r_{d}:$ $\sim 11\%$)~\cite{Boehm16}. Also, the majority of the available Sersic indices scatter around the value $n=1$ (see figure \ref{fig:Sersic}). Therefore it can be said that the $r_{d}$'s used in this thesis are still representative and that the results implying an ongoing disc growth with cosmic time, hold true. For example, BZ16 write that if they omit the correction for the wavelength dependence of $r_{d}$, their sample yields a weaker size evolution: $\Delta log(r_{d})=-0.11$ dex at z=1~\cite{Boehm16}. This is very similar to the evolution found in this thesis.\\

\noindent
\textbf{$\mathbf{\Delta M_{B}}$ vs. $\mathbf{\Delta logr_{d}}$:} In order to investigate whether the evolution of the TFR and the evolution of the VSR are linked, the respective offsets were plotted against each other (figures \ref{fig:deltaMB_deltard}). The local sample of Haynes et al. (1999) is taken as a comparison~\cite{Haynes99}. The TF and VS offsets $\Delta M_{B}$ and $\Delta log(r_{d})$ are obviously correlated which can be explained by means of the fundamental plane of spiral galaxies. If all 70 objects with available offsets are taken into account, the offset in terms of $\Delta log(r_{d})$ from the local relation is $-0.26\pm 0.02$ dex. If only the low-z or high-z objects are used, the offset is $-0.20\pm 0.02$ dex and $-0.39\pm 0.03$ dex, respectively. It can be said that the correlation between $\Delta M_{B}$ and $\Delta log(r_{d})$ obviously holds at least up to $z\sim 1$. As expected, the offset from the local $\Delta M_{B}-\Delta log(r_{d})$ relation is larger for high-z galaxies than for low-z galaxies, but the shape of the distribution seems to be the same as for local galaxies. It is simply shifted towards smaller sizes and higher luminosities. In addition, it can be seen that galaxies with the strongest evolution in luminosity need not have also the strongest evolution in size, and vice versa.\\
For comparison, BZ16 find similar results for a sample of FDF galaxies by dividing the sample into three redshift bins. Using fixed-slope fits they find the following offsets from the local relation in terms of $\Delta log(r_{d})$: $-0.08\pm 0.06$ dex for $z<0.36$, $-0.19\pm 0.08$ dex for $z<0.59$ and $-0.29\pm 0.07$ dex for $z>0.59$~\cite{Boehm16}. Their offset for galaxies with $z<0.59$ is in good agreement with the offset computed for the zCOSMOS low-z sample. In contrast, the offset of the high-z galaxies found in this thesis is significantly higher than the one found by BZ16 for galaxies with $z>0.59$. But it should be kept in mind that with 47 objects the low-z sample of the zCOSMOS galaxies is twice as large as the high-z sample. So statistically speaking, the fit to the low-z sample is also more reliable.\\
In addition, BZ16 find that galaxies with the strongest evolution in luminosity have usually $\Delta log(r_{d})>0$, and that objects with the strongest decrease in size scatter around $\Delta M_{B}\sim 0$. These two findings are also true for the zCOSMOS sample, although they are not as pronounced due to the smaller number of objects.\\

\noindent
\textbf{The Stellar Tully-Fisher relation:} The stellar TFR is another popular variant of the TFR. It links the stellar masses $M_{*}$ of spiral galaxies to their maximum rotation velocities $v_{max}$.\\
The stellar masses were available for 51 zCOSMOS galaxies (32 low-z and 19 high-z). First, the positions of these 51 objects relative to the main sequence of spiral galaxies were examined in more detail. The main sequence of star-forming galaxies	is a relation that links the stellar mass of spirals to their specific star formation rate (sSFR). The sSFR was calculated from the available star formation rates and stellar masses. As this relation depends on redshift, two main sequences were computed, for the low-z and the high-z sample, respectively (see figure \ref{fig:main_seq}). Objects that were more than $\pm 0.3$ dex away from the scatter area of the main sequence, were excluded. The remaining 41 objects were used for the construction of the distant stellar TFR (figure \ref{fig:TFRMstar}).\\
As a local reference, the relation from Reyes et al. (2011) was used who derived the local stellar TFR using $v_{80}$ as the measure for the rotation velocity ~\cite{Reyes11}. However, $v_{80}$ can be seen as a good approximation for $v_{max}$.\\
In order to investigate the evolution of the stellar TFR, a line with the slope fixed to the local value was fitted to the low-z and the high-z objects, respectively. It is expected that $v_{max}$, representing the total mass of a galaxy, does not really change up to $z\sim 1$. The stellar mass $M_{*}$, however, should increase with decreasing redshift because over time more and more gas is converted into stars. This is partially true for the zCOSMOS sample. The lines fitted to the two different redshift-samples are indeed both offset from the local relation towards smaller $M_{*}$. The offset of the low-z sample is larger than the one from the high-z sample. However, the offsets are consistent within scatter.\\
If a fixed-slope line is fitted to all 41 objects, the resulting offset from the local relation is $\Delta M_{*}=-0.37\pm 0.05$ dex. Puech et al. (2008) found an evolution of $\Delta M_{*}=-0.36$ dex for spiral galaxies up to $z\sim 0.6$~\cite{Puech08}. Tiley et al. (2016) constructed the stellar TFR for galaxies with redshifts $0.8<z<1.0$ and found an offset of $\Delta M_{*}=-0.41\pm 0.08$ dex~\cite{Tiley16}. The offset for all 41 objects together found in this thesis is in good agreement with results from other studies.\\
Finally, the positions of the excluded objects in the stellar Tully-Fisher diagram were examined (figure \ref{fig:TFRMstar_bad}). The majority of these ten objects lie below the local stellar TFR, i.e. towards smaller $M_{*}$. Only one high-z and two low-z galaxies have larger stellar masses than expected from the local relation. Although they have too high or too low sSFR according to the main sequence, they fit well into the stellar Tully-Fisher diagram.\\

\noindent
\textbf{The Stellar mass-Halo mass relation:} For the zCOSMOS galaxies with available stellar masses $M_{*}$, the relation between $M_{*}$ and $M_{h}$, that is the mass of a galaxy's dark matter halo, was studied. It is assumed that due to formation processes, $M_{*}$ and $M_{h}$ are tightly correlated.\\
The halo mass $M_{h}$ was estimated from the dynamical mass $M_{dyn}$, which was calculated from the maximum rotation velocity $v_{max}$ and the half-light radius $r_{1/2}$. 51 zCOSMOS galaxies were used to study the $M_{*}$-$M_{h}$ relation. In order to compare the results with numerical simulations, the predicted local stellar-to-halo mass (SHM) ratio derived by Moster et al. (2010) was used (see equation \ref{Moster10})~\cite{Moster10}. Moster et al. (2010) found that in halos with different masses, baryons are not converted into stars with the same efficiency. The SHM ratio is not constant, but rather increases with increasing $M_{h}$. After reaching the maximum at the characteristic mass $log(M_{1}/M_{\odot})=11.899$, it decreases again.\\
In order to compare the SHM ratios of the zCOSMOS galaxies with the predicted local ratio from Moster et al. (2010), three diagrams were made: first the stellar mass $M_{*}$ was plotted against the halo mass $M_{h}$ (figure \ref{fig:MsMh}), and then the SHM ratio was plotted against $M_{h}$ (figure \ref{fig:MsMh_ratio1}) and $M_{*}$ (figure \ref{fig:MsMh_ratio2}), respectively.\\
In all three diagrams the objects are in quite good agreement with the fit from Moster et al. (2010), in particular in the lower-mass regime. Especially in the $M_{*}$-$M_{h}$ and the SHM-$M_{*}$ plots, the distributions of objects with lower masses agree very well with the fit, whereas in the SHM-$M_{h}$ plot it is more difficult to say whether the distribution of the galaxies is consistent with the numerical predictions. It seems that there is a slight increase of the SHM ratio up to the characteristic mass $M_{1}$, however it is difficult to see a clear trend due to the large scatter of the objects in this plot. Moreover, it is not possible to make reliable statements about the behaviour of objects in the higher-mass regime. This applies to all three diagrams and is because of the lack of objects with halo masses larger than $M_{1}$ and stellar masses larger than $log(M_{*}/M_{\odot})\sim 10.5$, respectively.\\
To see how well the zCOSMOS galaxies follow the expected SHM ratio, a line was fitted through the objects in the $M_{*}$-$M_{h}$ plot. The resulting slope was compared to the slope derived by Moster et al. (2010). Because of the small number of objects with masses larger than $M_{1}$, this was done only for the lower-mass regime. A linear fit yields a slope of $\beta=1.0995$ ($\beta=1.094$ for only high-quality objects). This is in good agreement with Moster et al. (2010) who use $\beta=1.068$.\\
In all three diagrams there is no clear difference between low-z and high-z galaxies visible. If however a line is fitted through the low-z and high-z samples separately, the resulting slopes are $\beta=0.984$ and $\beta=1.321$, respectively. This indicates that the star formation efficiency increased stronger with $M_{h}$ at $z>0.5$ than at $z<0.5$. This result is in agreement with the results from Moster et al. (2010) who also studied the evolution of the parameters in eq. \ref{Moster10} and found that the low-mass slope increases with increasing redshift.\\
It should be kept in mind that the computation of $M_{h}$ is based on rough estimates and for given assumptions. Normally it should be calculated for each galaxy separately. Nevertheless the results from this section are in line with predictions from abundance matching models based on simulations. This shows that this new approach can in fact be used to get new insights in the SHM ratio of individual galaxies.\\

\noindent
In conclusion, it can be said that the results found in this master thesis are mostly in good agreement with the results from other studies. One possible explanation for the discrepancies found regarding the evolution of the TFR and the VSR, are underestimated maximum rotation velocities or incorrect B-band magnitudes $M_{B}$. Another possibility is that the number of objects used for the kinematic analysis is too small. Enlarging the sample would allow to make more precise statements about the evolution of scaling relation and thus spiral galaxies with redshift, and consequently enable more significant comparisons with other studies and simulations.

\begin{appendices}
\chapter{Tables}
The tables on the following pages list the most important parameters for the 76 objects that have been used for the kinematic analysis.\\
The columns of table \ref{tab:table_1} are: (1) the ID, (2) the redshift \textit{z}, (3) the emission line used for the RC extraction, (4) \& (5) the maximum rotation velocity $v_{max}$ and its error $v_{err}$, (6) the effective radius $r_{12}$, (7) the inclination \textit{i} and (8) the position angle of the apparent major axis $\theta$. Columns (9) and (10) give information about whether the turn-over radius $r_{t}$ has been held fixed during the simulation or not, and about the quality class of the object, respectively.\\
Table \ref{tab:table_2} lists (1) the ID, (2) the absolute B-band magnitude $M_{B}$, (3) the B-band magnitude corrected for intrinsic dust absorption $M_{Bkorr}$, (4) the Sersic index $n_{Sersic}$, (5) \& (6) the star formation rate SFR and the specific star formation rate sSFR, (7) the stellar mass $M_{*}$ and (8) the halo mass $M_{h}$. Note that the parameters listed in table \ref{tab:table_2} are not available for all 76 objects.
\newpage

\begin{table} [h!]
\begin{center}
\small
\caption{\small The ID, the redshift \textit{z}, the emission line used for the RC extraction, the maximum rotation velocity $v_{max}$, its error $v_{err}$, the effective radius $r_{12}$, the inclination \textit{i} and the position angle of the apparent major axis $\theta$ of the 76 objects used for the kinematic analysis. The last two columns give information about whether the turn-over radius $r_{t}$ has been held fixed or not, and about the quality class of the object.
\emph{The complete table is available on request.}}
	\begin{tabular}{ c  c  c  c  c  c  c  c  c  c}
ID	&	z	&	Line	&	$v_{max}$ [km/s]	&	$v_{err}$ [km/s]	&	$r_{12}$ [kpc] &	i [$^{\circ}$]	&	$\theta$	 [$^{\circ}$]	&	$r_{t}$ fixed	&	Q	\\	\hline
810929	&	0.3493	&	[OIII](2.)	&	103.66	&	1.71	&	3.58	&	84.37	&	-41.36	&	y	&	H	\\	
810934	&	0.4401	&	[OIII](2.)	&	93.36	&	6.46	&	6.92	&	49.75	&	57.02	&	n	&	L	\\
811102	&	0.4404	&	[OIII](2.)	&	91.41	&	3.56	&	6.67	&	79.43	&	-16.52	&	n	&	L	\\
\end{tabular}
\label{tab:table_1}
\end{center}
\end{table}

%%%%%%%%%%%%%%%%%%%%%%%%%%%%%%%%%%%%

\begin{table} [h!]
\begin{center}
\small
\caption{\small The ID, the absolute B-band magnitude $M_{B}$, the B-band magnitude corrected for intrinsic dust absorption $M_{Bkorr}$, the Sersic index $n_{Sersic}$, the star formation rate SFR, the specific star formation rate sSFR, the stellar mass $M_{*}$, and the halo mass $M_{h}$ of the 76 objects used for the kinematic analysis. \emph{The complete table is available on request.}}
	\begin{tabular}{ c  c  c  c  c  c  c  c }

ID	&	$M_{B}$	&	$M_{Bkorr}$	&	$n_{Sersic}$	&	log(SFR/$M_{\odot}/yr$)&	log(sSFR/$Gyr^{-1}$)	&	log($M_{*}/M_{\odot}$)	&	log($M_{h}/M_{\odot}$)	\\	\hline
810929	&	-18.25	&	-18.94	&	0.87	&	-0.74	&	-0.92	&	9.18	&	11.43	\\	
810934	&	-20.72	&	-20.89	&	0.83	&	-0.29	&	-1.29	&	10.00	&	11.62	\\	
811102	&	-19.27	&	-19.94	&	0.67	&	-0.75	&	-1.72	&	9.97	&	11.59	\\	
\end{tabular}
\label{tab:table_2}
\end{center}
\end{table}

\end{appendices}

\backmatter

\chapter{Abstract English}
Scaling relations, which link important physical properties of galaxies to one another, are a very powerful tool to analyse the nature and evolution of galaxies. In case of spiral galaxies, the Tully-Fisher-Relation (TFR), which links the luminosity of the stellar population to the maximum velocity $v_{max}$, and the velocity-size relation (VSR) are very important.\\
In this master thesis, the evolution of the B-band TFR, the stellar TFR and the VSR up to a redshift of $z\sim 1$ is investigated by means of a sample of zCOSMOS galaxies. Furthermore, the stellar-to-halo mass ratio (SHM) is studied and compared to predictions from simulations.\\
For the computation of $v_{max}$, spectra recorded with the spectrograph VIMOS (Visible Multi-Object Spectrograph) were used. Five masks, containing together 160 spectra, were reduced with the software package VIPGI. From the reduced spectra, rotation curves (RCs) were extracted from sufficiently bright emission lines. These observed RCs were subsequently modelled, taking into account geometrical and observational effects, which yielded the $v_{max}$ of the individual objects. Regarding their shape, the modeled RCs were divided into three quality classes. The objects of the third group were not used for the kinematic analysis.\\
The other required parameters like the B-band magnitudes $M_{B}$, the stellar masses $M_{*}$ and the disk scale lengths $r_{d}$ were taken from the zCOSMOS catalogues. By means of these parameters the scaling relations for 76 zCOSMOS galaxies were constructed and compared to the corresponding local relations.\\
Regarding the B-band TFR, the zCOSMOS sample used in this thesis is on average by $\Delta M_{B} = -0.83\pm 0.09$ mag brighter than local galaxies. The found offset is larger than in the case of other studies. If however only high quality (HQ) RCs are used, the resulting offset agrees within margins with the results from other studies. According to the VSR of the zCOSMOS sample, galaxies were smaller in the past which is in agreement with the hierarchical scenario of galaxy formation. Compared to local galaxies, the stellar masses of the zCOSMOS galaxies are on average by $\Delta M_{*}=-0.37\pm 0.05$ dex smaller.\\
In summary, we find that at a given $v_{max}$ the intermediate zCOSMOS galaxies are brighter, smaller and less massive (in terms of stellar mass) compared to local galaxies, a trend qualitatively in agreement with other studies.\\
The halo masses $M_{h}$  were estimated for a part of the zCOSMOS galaxies by means of $v_{max}$ and the effective radius $r_{1/2}$. The resulting $M_{*}/M_{h}$ ratios (SHM) are in good agreement with predictions from simulations, according to which the SHM ratio increases up to a characteristic mass and then decreases again. It is interesting that our $M_{*}$ vs. $M_{h}$ relation is in agreement with abundance matching models based on numerical simulations, given the simple derivations of $v_{circ}$ from $v_{max}$ and of $r_{vir}$ from $r_{1/2}$. This shows that this new approach can be used complementary to abundance matching techniques to get new insights in the stellar content of dark matter halos for individual galaxies.\\

\chapter{Abstract Deutsch}
So genannte Skalenrelationen verbinden bestimmte physikalische Eigenschaften von Galaxien und sind daher ein wichtiges Hilfsmittel um Aussagen {\"u}ber die Natur und Entwicklung von Galaxien zu machen. Im Fall von Spiralgalaxien, sind v.a. die Beziehungen zwischen der Leuchtkraft und der maximalen Rotationsgeschwindigkeit $v_{max}$ (\textit{Tully-Fisher relation}, TFR) sowie zwischen der Gr{\"o}\ss e und $v_{max}$ (\textit{velocity-size relation}, VSR) von Bedeutung.\\
Ziel dieser Masterarbeit ist es, die Entwicklung der B-Band TFR (B-Band Magnituden $M_{B}$ vs. $v_{max}$), der stellaren TFR (stellare Masse $M_{*}$ vs. $v_{max}$) und der VSR bis zu einer Rotverschiebung von $z\sim 1$ anhand eines Samples von zCOSMOS-Galaxien zu untersuchen. Weiters soll auch die Beziehung zwischen $M_{*}$ und der Halo-Masse $M_{h}$ betrachtet und mit Vorhersagen von Simulationen verglichen werden.\\
F{\"u}r die Berechnung von $v_{max}$ wurden Spektren verwendet, die mit dem Spektrographen VIMOS (Visible Multi-Object Spectrograph) aufgenommen wurden. F{\"u}nf Masken mit insgesamt 160 Spektren wurden mit dem Software-Programm VIPGI reduziert. Aus ausreichend hellen Emissionslinien in den reduzierten Spektren wurden Rotationskurven (\textit{rotation curves}, RCs) extrahiert. Die anschlie\ss ende Modellierung dieser beobachteten RCs lieferte die $v_{max}$-Werte der einzelnen Galaxien. Dabei wurden geometrische Effekte und Beobachtungseffekte ber{\"u}cksichtigt. Die modellierten RCs wurden anhand ihrer Form in drei Qualit{\"a}sgruppen unterteilt. Die Objekte der dritten Gruppe wurden dabei nicht f{\"u}r die kinematischen Analysen verwendet.\\
Die anderen ben{\"o}tigten Parameter $M_{B}$, $M_{*}$ und die Skalenl{\"a}nge $r_{d}$ wurden den zCOSMOS-Katalogen entnommen. Mithilfe dieser Parameter wurden schlie\ss lich die Skalenrelationen f{\"u}r 76 zCOSMOS Galaxien konstruiert und mit den entsprechenden Relationen f{\"u}r das lokale Universum verglichen.\\
Was die B-Band TFR betrifft, ist das in dieser Arbeit verwendete zCOSMOS Sample durchschnittlich um $\Delta M_{B}=-0.83\pm 0.09$ mag heller als lokale Galaxien. Der gefundene Unterschied in den Helligkeiten ist gr{\"o}\ss er als bei anderen Studien. Wenn jedoch nur Objekte der ersten Qualit{\"a}tsgruppe (high quality, HQ) verwendet werden, stimmt der Wert innerhalb der Fehlerbalken mit den Werten anderer Studien {\"u}berein. Die VSR des zCOSMOS-Samples zeigt, dass Galaxien in der Vergangenheit kleiner waren, was mit der Vorhersage des hierarchischen Modells der Galaxienentstehung {\"u}bereinstimmt. Die stellaren Massen der zCOSMOS-Galaxien sind im Mittel um $\Delta M_{*}=-0.37\pm 0.05$ dex kleiner als Galaxien des lokalen Universums.\\
Zusammenfassend sind die zCOSMOS-Galaxien bei bestimmten $v_{max}$ also heller und kleiner und haben weniger stellare Masse als lokale Galaxien. Diese Ergebnisse stimmen qualitativ mit den Resultaten anderer Studien {\"u}berein.\\
Anhand von $v_{max}$ und des Effektivradius $r_{1/2}$ wurden schlie\ss lich die Halo-Massen $M_{h}$ f{\"u}r einen Teil der zCOSMOS-Galaxien abgesch{\"a}tzt. Es zeigt sich, dass die berechneten Verh{\"altnisse $M_{*}/M_{h}$ gut mit den Vorhersagen von Simulationen {\"u}bereintreffen, wonach $M_{*}/M_{h}$ bis zu einer bestimmten charakteristischen Masse ansteigt, danach jedoch wieder abnimmt. Unter Ber{\"u}cksichtigung der einfachen Herleitung der Geschwindigkeit des Halos $v_{circ}$ von $v_{max}$ und des Virialradius $r_{vir}$ von $r_{1/2}$, ist es interessant, dass die in dieser Arbeit bestimmte $M_{*}$ vs. $M_{h}$-Relation mit Modellen von numerischen Simulationen gut {\"u}bereinstimmt. Das zeigt, dass dieser neue Ansatz verwendet werden kann um neue Erkenntnisse bez{\"u}glich der stellaren Materie von Dunkle-Materie-Halos individueller Galaxien zu bekommen.

\phantomsection 
\addcontentsline{toc}{chapter}{Bibliography} 
\bibliographystyle{abbrv}
\bibliography{Masterarbeit}

\end{document}